\documentclass[
              reprint,
              superscriptaddress,
              nofootinbib,
              showpacs,
              amsmath,amssymb,aps,prd,
              ]{revtex4-1}
\pdfoutput=1 			

\usepackage{graphics}
\usepackage{epsfig}
\usepackage{graphicx}
\usepackage{dcolumn}
\usepackage{bm}
\usepackage{longtable}
\usepackage{ifpdf}
\ifpdf
\usepackage[pdfencoding=auto]{hyperref}
\usepackage{bookmark}
\fi

\begin{document}

\vspace*{0.5cm}

\title{\boldmath
	Measurement of $\sin^2\theta^{\rm lept}_{\rm eff}$ using $e^+e^-$
	pairs from $\gamma^*/Z$ bosons produced in $p\bar{p}$ collisions at
	a center-of-momentum energy of 1.96 TeV
       \unboldmath}

\input{cdf_auth_050516.itex}

\date{\today}

\begin{abstract}
At the Fermilab Tevatron proton-antiproton ($p\bar{p}$)
collider, Drell-Yan lepton pairs are produced in the
process $p \bar{p} \rightarrow e^+e^- + X$ through an
intermediate $\gamma^*/Z$ boson. The forward-backward
asymmetry in the polar-angle distribution of the $e^-$ as a
function of the $e^+e^-$-pair mass
is used to obtain  $\sin^2\theta^{\rm lept}_{\rm eff}$,
the effective leptonic determination of the electroweak-mixing
parameter $\sin^2\theta_W$.
The measurement sample, recorded by the Collider Detector at
Fermilab (CDF), corresponds to 9.4~fb$^{-1}$ of integrated
luminosity from $p\bar{p}$ collisions at a center-of-momentum
energy of 1.96~TeV, and is the full CDF Run II data set.
The value of $\sin^2\theta^{\rm lept}_{\rm eff}$ is found
to be $0.23248 \pm 0.00053$. The combination with the previous
CDF measurement based on $\mu^+\mu^-$ pairs yields 
$\sin^2\theta^{\rm lept}_{\rm eff} = 0.23221 \pm 0.00046$.
This result, when interpreted within the specified context
of the standard model assuming
$\sin^2 \theta_W = 1 - M_W^2/M_Z^2$ and that the $W$- and
$Z$-boson masses are on-shell, yields
$\sin^2\theta_W = 0.22400 \pm 0.00045$, or equivalently a
$W$-boson mass of $80.328 \pm 0.024 \;{\rm GeV}/c^2$.
\end{abstract}


\maketitle

\section{Introduction}

In this paper, the angular distribution of charged leptons
$(\ell^\pm)$ from the Drell-Yan~\cite{DrellYan} process is used
to measure the electroweak-mixing parameter $\sin^2\theta_W$
\cite{*[{}] [{, and 2015 update.}] PDGreviews}.
At the Fermilab Tevatron collider, Drell-Yan pairs are
produced by the process
$p\bar{p} \rightarrow \ell^+\ell^-  + X$, where the $\ell^+\ell^-$
pair is produced through an intermediate $\gamma^*/Z$ boson,
and $X$ is the final state associated with the production of
the boson. In the standard model, the production of Drell-Yan
lepton pairs at the Born level proceeds through two parton-level
processes,
\begin{eqnarray*}
  q\bar{q} & \rightarrow & \gamma^* \rightarrow \ell^+\ell^- \; {\rm and} \\
  q\bar{q} & \rightarrow & Z \rightarrow \ell^+\ell^- ,
\end{eqnarray*}
where the $q$ and $\bar{q}$ are the quark and antiquark, respectively,
from the colliding hadrons. The virtual photon couples the vector
currents of the incoming and outgoing fermions $(f)$, and the
spacetime structure of a photon-fermion interaction vertex is
$\langle \bar{f} | Q_f \gamma_\mu |f\rangle$,
where $Q_f$, the strength of the coupling, is the fermion charge
(in units of $e$), and $|f\rangle$ is the spinor for fermion $f$.
An interaction vertex of a fermion with a $Z$ boson contains both
vector $(V)$ and axial-vector $(A)$ current components, and its
structure is
$\langle \bar{f} | g_V^f \gamma_\mu + g_A^f \gamma_\mu\gamma_5 |f\rangle$.
The Born-level coupling strengths are
\begin{eqnarray*}
  g_V^f & = & T_3^f - 2Q_f \: \sin^2\theta_W \; {\rm and} \\
  g_A^f & = & T_3^f ,
\end{eqnarray*}
where $T_3^f$ is the third component of the fermion weak-isospin,
which is $T_3^f = \frac{1}{2}$ $(-\frac{1}{2})$
for positively (negatively) charged fermions.
At the Born level, and in all orders of the on-shell
renormalization scheme~\cite{OnShellScheme}, the $\sin^2\theta_W$
parameter is related to the $W$-boson mass $M_W$ and the $Z$-boson
mass $M_Z$ by the relationship $\sin^2\theta_W =  1 - M_W^2/M_Z^2$.
Radiative corrections alter the strength
of the Born-level couplings into {\rm effective} couplings.
These effective couplings have been investigated at the
Tevatron
\cite{CDFIIsw2e,zA4ee21prd,*zA4ee21prdE,cdfAfb9mmprd,D0sw2e10,*D0sw2e},
at the LHC \cite{ATLASsw2eff,CMSsw2eff1,LHCBsw2eff}, and
at LEP-1 and SLC~\cite{LEPfinalZ,LEPfinalZ2}. The on-shell
$\sin^2\theta_W$ coupling has been investigated
with neutrino-nucleon collisions at the Tevatron
\cite{NuTev1,*NuTev2} and with electron-proton collisions
at HERA \cite{h1HERAsw2Mw}.
\par
The effective $\sin^2 \theta_W$ coupling at the lepton vertex,
denoted as $\sin^2\theta^{\rm lept}_{\rm eff}$, has been
accurately measured at the LEP-1 and SLC $e^+e^-$ colliders
\cite{LEPfinalZ,LEPfinalZ2}. The combined average of six individual
measurements yields a value of $0.23149 \pm 0.00016$. However,
there is tension between the two most precise individual
measurements: the combined LEP-1 and SLD $b$-quark
forward-backward asymmetry ($A_{\rm FB}^{0,{\rm b}})$ yields
$\sin^2\theta^{\rm lept}_{\rm eff} = 0.23221 \pm 0.00029$,
and the SLD left-right polarization asymmetry of $Z$-boson
production $({\cal A}_\ell)$
yields $\sin^2\theta^{\rm lept}_{\rm eff} = 0.23098 \pm 0.00026$.
They differ by 3.2 standard deviations.
\par
The Drell-Yan process at hadron-hadron colliders is
also sensitive to the $\sin^2\theta^{\rm lept}_{\rm eff}$
coupling. Measurements of the forward-backward asymmetry
in the $\ell^-$ polar-angle distribution as a function of
the lepton-pair invariant mass are used to extract the
coupling. This paper presents a new measurement of the
$\sin^2\theta^{\rm lept}_{\rm eff}$ coupling and an inference
of the $\sin^2\theta_W$ parameter using a sample of
$e^+e^-$ pairs corresponding to an integrated $p\bar{p}$
luminosity of 9.4~fb$^{-1}$ collected at the Tevatron $p\bar{p}$
collider.
Innovative methods for
the calibration of the electron energy and the measurement of
the forward-backward asymmetry are used. Electroweak
radiative corrections used for the extraction of
$\sin^2\theta^{\rm lept}_{\rm eff}$ and $\sin^2\theta_W$
are derived from an approach used by LEP-1 and SLD.
\par
An outline of the paper follows.
Section~\ref{LeptAngDistr} provides an overview of the
lepton angular distributions and the extraction of
$\sin^2\theta^{\rm lept}_{\rm eff}$.
Section~\ref{QCDEWKpred} discusses quantum chromodynamics (QCD)
calculations for the forward-backward asymmetry and the
inclusion of electroweak radiative-correction form factors used
in the analysis of high-energy $e^+e^-$ collisions. The form
factors are required for the determination of $\sin^2\theta_W$
from the measurement of $\sin^2\theta^{\rm lept}_{\rm eff}$.
Section~\ref{CDFdetector} describes the experimental apparatus.
Section~\ref{DataSelection} reports on the selection of data.
Section~\ref{AfbexpDatSim} describes the simulation of the
reconstructed data.
Sections~\ref{CorrDatSim} and \ref{AfbexpInput} present the
experimental calibrations and the measurement of the asymmetry,
respectively, along with corresponding corrections to data
and simulation.
Section~\ref{sw2scans} describes the method used to extract
$\sin^2\theta^{\rm lept}_{\rm eff}$.
Section~\ref{systUncerts} describes the systematic uncertainties.
Section~\ref{finalResults} presents the results of this
measurement using $e^+e^-$ pairs and 
Sec.~\ref{cdfemuCombination} describes the combination of
results from this measurement and a
previous CDF measurement using $\mu^+\mu^-$ pairs
\cite{cdfAfb9mmprd}.
Finally, Sec.~\ref{theEndSummary} presents the summary.
Standard units are used for numerical values of particle masses
and momenta, e.g., 40~GeV/$c^2$ and 20~GeV/$c$, respectively,
where $c$ denotes the speed of light. Otherwise,
natural units ($\hbar = c = 1$) are used.

\section{\label{LeptAngDistr}
Lepton Angular Distribution}

The angular distribution of leptons from the Drell-Yan process
in the rest frame of the boson is governed by the polarization
state of the $\gamma^*/Z$ boson. In amplitudes at higher order
than tree level, initial-state QCD interactions of the colliding
partons impart to the $\gamma^*/Z$ boson a momentum component
transverse to the collision axis,
thus affecting the polarization states.
\par
In the laboratory frame, the  $p\bar{p}$ collision axis is
the $z$ axis, with the positive direction oriented along the
direction of the proton. The transverse component of any vector,
such as the momentum vector, is defined to be relative to that
axis. The transverse component of vectors
in other reference frames is defined to be relative to
the $z$ axes in those frames.
\par
For the description of the Drell-Yan process, the rapidity, transverse
momentum, and mass of a particle are denoted as $y$, $P_{\rm T}$,
and $M$, respectively. The energy and momentum of particles are denoted
as $E$ and $\vec{P}$, respectively. In a given coordinate frame, the rapidity
is $y = \frac{1}{2} \, \ln[\,(E + P_{\rm z})/(E - P_{\rm z})\,]$,
where $P_{\rm z}$ is the component of the momentum vector along the $z$
axis of the coordinate frame. 
\par
\begin{figure}
\includegraphics
   [height=50mm]
   {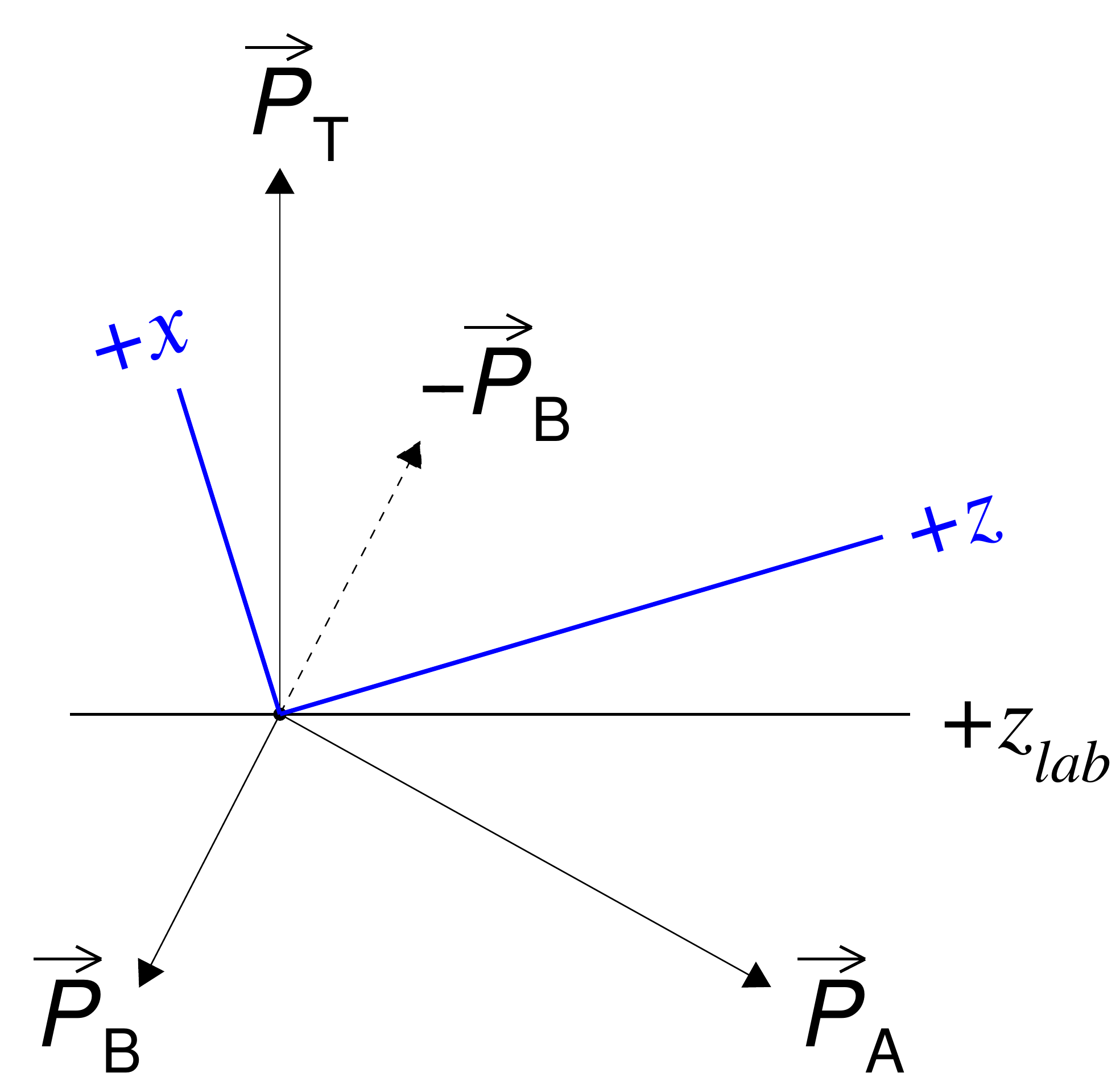}
\caption{\label{fig_CSframe}
Representation of the Collins-Soper coordinate axes $(x,z)$ in the
lepton-pair rest frame, along with the laboratory $z$ axis ($z_{lab}$).
The three axes are in the plane formed by the
proton ($\vec{P}_{\rm A}$) and antiproton ($\vec{P}_{\rm B}$)
momentum vectors in the rest frame. The $z$ axis is the
angular bisector of $\vec{P}_{\rm A}$ and $-\vec{P}_{\rm B}$.
The $y$ axis is along the direction of
$\vec{P}_{\rm B} \times \vec{P}_{\rm A}$, and the $x$ axis
is in the direction away from the transverse component of
$\vec{P}_{\rm A}+\vec{P}_{\rm B}$.
}
\end{figure}

The polar and azimuthal angles of the $\ell^-$ direction in
the rest frame of the boson are denoted as $\vartheta$ and
$\varphi$, respectively.
For this analysis, the ideal positive $z$ axis coincides with the
direction of the incoming quark so that the definition of $\vartheta$
parallels the definition used in $e^+e^-$ collisions at
LEP~\cite{LEPfinalZ,LEPfinalZ2}.
This frame is approximated by the Collins-Soper (CS) rest
frame~\cite{CollinsSoperFrame} for $p\bar{p}$ collisions.
The rest frame is reached from the laboratory frame via two
Lorentz boosts, first along the laboratory $z$-axis into a frame
where the $z$ component of the lepton-pair momentum vector is zero,
followed by a boost along the transverse component of the lepton-pair
momentum vector. 
A view of the CS frame is shown in Fig.~\ref{fig_CSframe}.
\par
The general structure of the Drell-Yan lepton angular-distribution
in the boson rest frame consists of terms from nine helicity
cross-sections that describe the polarization state of the boson,
\begin{eqnarray}
\frac{dN}{d\Omega}
  & \propto &
        \: (1 + \cos^2 \vartheta) +  \nonumber \\
  &   & A_0 \:\frac{1}{2} \:
             (1 -3\cos^2 \vartheta) + \nonumber \\
  &   & A_1 \: \sin 2\vartheta
               \cos \varphi +   \nonumber \\
  &   & A_2 \: \frac{1}{2} \:
               \sin^2 \vartheta
               \cos 2\varphi +  \nonumber \\
  &   & A_3 \: \sin \vartheta
               \cos \varphi +   \nonumber \\
  &   & A_4 \: \cos \vartheta + \nonumber \\
  &   & A_5 \: \sin^2 \vartheta
               \sin 2\varphi +  \nonumber \\
  &   & A_6 \: \sin 2\vartheta
               \sin \varphi +   \nonumber \\
  &   & A_7 \: \sin \vartheta
               \sin \varphi \: ,
\label{eqnAngDistr}
\end{eqnarray}
where each term is relative to the cross section for unpolarized
production integrated over the lepton angular distribution
\cite{MirkesA0to7a, *MirkesA0to7b}. The coefficients $A_{0-7}$ are
functions of kinematic variables of the boson and vanish when the
lepton-pair transverse momentum is zero, except for $A_4$, which
contributes to the tree-level QCD amplitude and generates the
forward-backward
$\ell^-$ asymmetry in $\cos \vartheta$. Thus, at zero transverse
momentum, the angular distribution reduces to the tree-level
form $1 + \cos^2 \vartheta + A_4\cos \vartheta$. The $A_4$ coefficient
is relatively uniform across the range of transverse momentum where
the cross section is large (at values smaller than approximately
$45$ GeV/$c$), but slowly decreases
for larger values of transverse momentum, where the cross section is
very small. The $A_0$ and $A_2$ coefficients, corresponding to the
longitudinal and transverse states of polarization, respectively,
are the most significant and have been previously measured, along
with $A_3$ and $A_4$ \cite{CSangcoef21}. The $A_1$ coefficient, from
the interference between the longitudinal and transverse states of
polarization, is small in the CS frame.
The $A_{5-7}$ coefficients appear at second
order in the QCD strong coupling, $\alpha_s$, and are small in the
CS frame \cite{MirkesA0to7a, *MirkesA0to7b}. Hereafter, the angles
$(\vartheta, \: \varphi)$ and the angular coefficients $A_{0-7}$ are
intended to be specific to the CS rest frame.
\par
The $A_4 \cos\vartheta$ term violates parity, and is due to
the interference of the amplitudes of the vector and axial-vector
currents. Its presence induces an asymmetry in the
$\varphi$-integrated $\cos \vartheta$ dependence of the cross section.
Two sources contribute: the interference between
the $Z$-boson vector and axial-vector amplitudes, and the
interference between the photon vector and $Z$-boson axial-vector
amplitudes. The asymmetric component from the $\gamma^*$-$Z$
interference cross section contains $g_A^f$ couplings
that are independent of $\sin^2 \theta_W$.
The asymmetric component from $Z$-boson self-interference
contains a product of $g_V^f$ from the lepton and quark
vertices, and thus is related to $\sin^2 \theta_W$.
At the Born level, this product is
\begin{displaymath}
   T_3^\ell \: (1 - 4|Q_\ell|\sin^2\theta_W) \;
   T_3^q    \: (1 - 4|Q_q|\sin^2\theta_W) ,
\end{displaymath}
where $\ell$ and $q$ denote the lepton and quark, respectively.
For the Drell-Yan process, the relevant quarks are predominantly
the light quarks $u$, $d$, or $s$. The coupling factor has an
enhanced sensitivity to $\sin^2\theta_W$ at the lepton-$Z$
vertex: for a $\sin^2\theta_W$ value of $0.223$, a 1\% variation
in $\sin^2\theta_W$ changes the lepton factor (containing
$Q_\ell$) by about 8\%, and it changes the quark factor (containing
$Q_q$) by about 1.5\% (0.4\%) for the $u$ ($d$ or $s$) quark.
Electroweak radiative corrections do not alter significantly
this Born-level interpretation. Loop and vertex
electroweak radiative corrections are multiplicative
form-factor corrections to the couplings that change their
values by a few percent \cite{zA4ee21prd,*zA4ee21prdE}.
\par
\begin{figure}
\includegraphics
   [width=85mm]
   {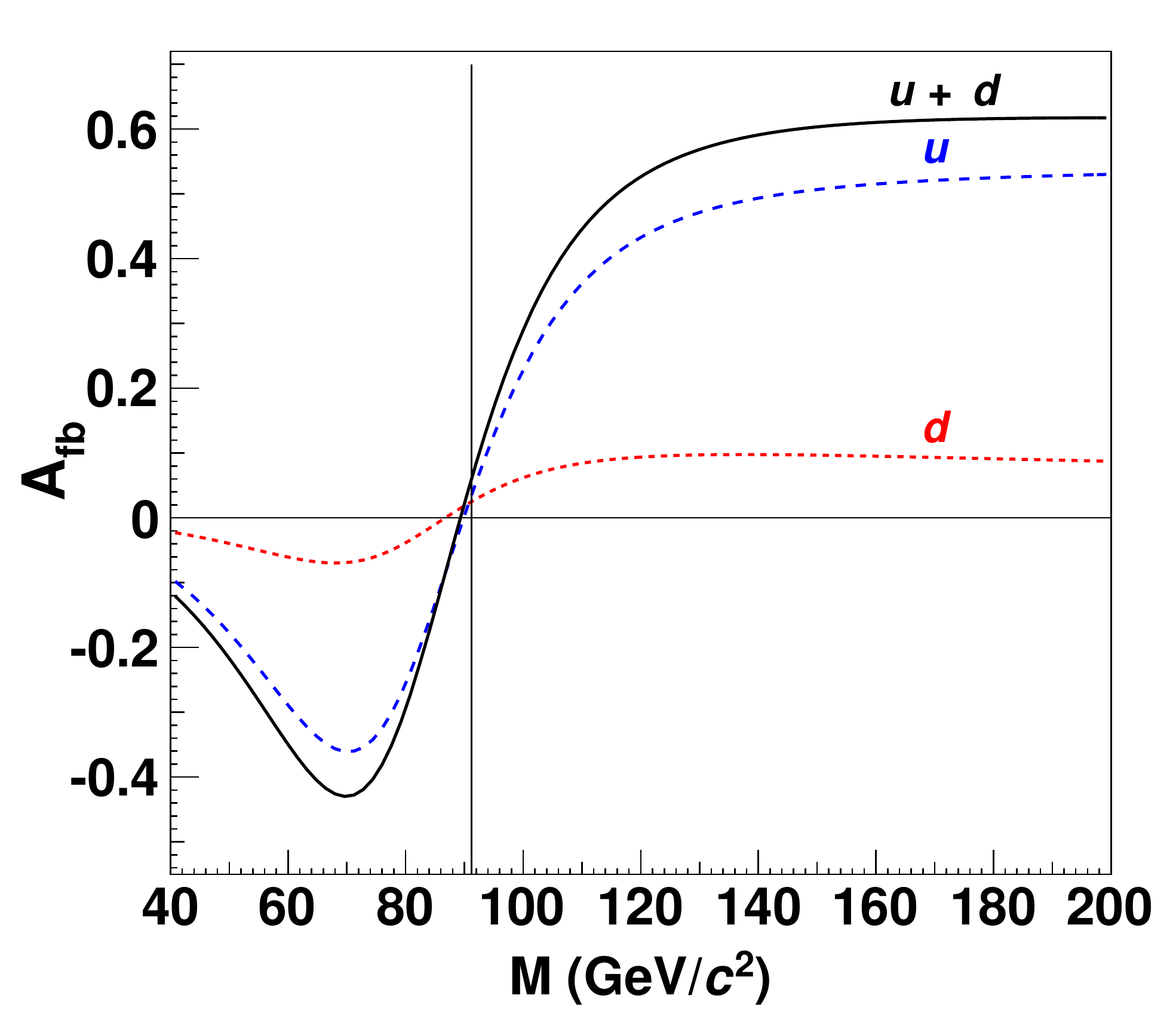}
\caption{\label{fig_loAfbVSmass}
 Typical dependence of $A_{\rm fb}$ as a function of the lepton-pair
 invariant mass $M$. The label $u$ + $d$ denotes the overall asymmetry,
 and the labels $u$ and $d$ denote the contribution to the overall
 asymmetry from quarks with charge $2/3$ and $-1/3$, respectively.
 The contribution of quarks categorized by the $u$ or $d$ label is
 $(\sigma^+_q - \sigma^-_q)/\sigma$,
 where $q = u$ or $d$, $\sigma^{+(-)}$ their forward (backward)
 cross section, and $\sigma$ the total cross section from all quarks. 
 The vertical line is at $M = M_Z$.
}
\end{figure}

The $\ell^-$ forward-backward asymmetry in $\cos \vartheta$
is defined as
\begin{equation}
  A_{\rm fb}(M) = \frac{\sigma^+(M) - \sigma^-(M)}
                       {\sigma^+(M) + \sigma^-(M)} 
                = \frac{3}{8}A_4(M) \:,
\label{eqnAfbDef}
\end{equation}
where $M$ is the lepton-pair invariant mass,
$\sigma^+$ is the total cross section for $\cos \vartheta \geq 0$,
and $\sigma^-$ is the total cross section for $\cos \vartheta < 0$.
Figure~\ref{fig_loAfbVSmass} shows the typical dependence of the
asymmetry as a function of the lepton-pair invariant mass from a
Drell-Yan QCD calculation.
The offset of $A_{\rm fb}$ from zero at $M = M_Z$ is related
to $\sin^2\theta_W$. Away from the $Z$ pole, the asymmetry is
dominated by the component from $\gamma^*$-$Z$ interference,
whose cross section is proportional to
$(M^2-M_Z^2)/M^2$, and the asymmetries in these regions are
primarily related to the flux of partons. Consequently,
the asymmetry distribution is sensitive to both $\sin^2\theta_W$
and the parton distribution functions (PDF) of the proton.

\par
The $\sin^2\theta^{\rm lept}_{\rm eff}$ coupling is derived from
the measurement of $A_{\rm fb}(M)$ and predictions
of $A_{\rm fb}(M)$ for various input values of $\sin^2 \theta_W$.
Electroweak and QCD radiative corrections are included in the
predictions of $A_{\rm fb}(M)$, with the electroweak radiative
corrections derived from an approach adopted at LEP
\cite{Dizet, *zfitter621, *zfitter642}.

\section{\label{QCDEWKpred}
Enhanced QCD Predictions}

Drell-Yan cross-section calculations with QCD radiation do not
typically include electroweak radiative corrections. However,
the QCD, quantum electrodynamic, and weak-interaction corrections
can be organized to be individually gauge invariant so that they
are applied as independent components.
\par
Quantum electrodynamic (QED) radiative corrections that result
in final-state photons are the most important for measurements
derived from the Drell-Yan process, and they are included
in the physics and detector simulation described in
Sec.~\ref{AfbexpDatSim}. The effects of QED radiation are
removed from the measured distribution of $A_{\rm fb}$ using
the simulation so that the measurement can be directly
compared with QCD calculations of $A_{\rm fb}$ that do not
include it. 
\par
The Drell-Yan process and the production of quark pairs in
high-energy $e^+e^-$ collisions are analogous processes:
   $q\bar{q} \rightarrow e^+e^-$ and
   $e^+e^-   \rightarrow q\bar{q}$.
At the Born level,
the process amplitudes are of the same form except for the
interchange of the electrons and quarks. Electroweak radiative
corrections, calculated and extensively used for precision
fits of LEP-1 and SLD measurements to the standard model
\cite{LEPfinalZ,LEPfinalZ2},
are therefore applicable to the Drell-Yan process.
\par
In the remainder of this section, the technique used to incorporate
independently
calculated electroweak radiative corrections for $e^+e^-$ collisions
into existing QCD calculations for the Drell-Yan process is
presented.

\subsection{\label{EWKradcor}
Electroweak radiative corrections}

The effects of virtual electroweak radiative corrections are
incorporated into Drell-Yan QCD calculations via form factors
for fermion-pair production according to
$e^+e^- \rightarrow Z \rightarrow f \! \bar{f}$. 
The $Z$-amplitude form factors are calculated by
\textsc{zfitter 6.43}~\cite{Dizet, *zfitter621, *zfitter642},
which is used with LEP-1 and SLD measurement inputs for precision
tests of the standard model \cite{LEPfinalZ,LEPfinalZ2}.
Corrections to fermion-pair production via a
virtual photon include weak-interaction $W$-boson loops in the photon 
propagator, and $Z$-boson propagators at fermion-photon vertices; these
corrections are not gauge-invariant except when combined with their
gauge counterparts in the $Z$ amplitude.  The \textsc{zfitter} weak
and QED corrections are organized to be separately gauge-invariant.
Consequently, weak corrections to fermion-pair production via the
virtual photon are included through the  $Z$-amplitude form factors.
\textsc{zfitter} uses the on-shell scheme~\cite{OnShellScheme},
where particle masses are on-shell, and
\begin{equation}
  \sin^2 \theta_W = 1 - M_W^2/M_Z^2
\label{baseSW2}
\end{equation}
holds to all orders of perturbation theory by definition.
Since the $Z$-boson mass is accurately known
(to $\pm 0.0021$ GeV/$c^2$ \cite{LEPfinalZ,LEPfinalZ2}),
the inference of $\sin^2 \theta_W$ is equivalent to an indirect
$W$-boson mass measurement.
\par
Form factors calculated by \textsc{zfitter} are tabulated for later
use in QCD calculations. The specific standard-model assumptions
and parameters used in the form-factor calculation are presented
in the appendix, as well as their usage in the scattering
amplitude $A_q$. The calculated form factors are
$\rho_{eq}$, $\kappa_e$, $\kappa_q$, and $\kappa_{eq}$, where
the label $e$ denotes an electron and $q$ denotes a quark. As the
calculations use the massless-fermion approximation, 
the form factors only depend on the charge and weak isospin
of the fermions. Consequently, the tabulated form factors are
distinguished by three labels, $e$ (electron type),
$u$ (up-quark type), and $d$ (down-quark type). The form factors
are complex valued, and are functions of the $\sin^2\theta_W$
parameter and the Mandelstam $\hat{s}$ variable of the
$e^+e^- \rightarrow Z \rightarrow f \! \bar{f}$ process.
The $\rho_{eq}$, $\kappa_e$, and $\kappa_q$ form factors of the
amplitude can be reformulated as corrections to the
Born-level $g_A^f$ and $g_V^f$ couplings,
\begin{eqnarray*}
  g_V^f & \rightarrow & \sqrt{\rho_{eq}}\,
                       ( T_3^f - 2Q_f \kappa_f \: \sin^2\theta_W )
		       \: {\rm and} 
                       \nonumber \\
  g_A^f & \rightarrow & \sqrt{\rho_{eq}} \, T_3^f ,
\end{eqnarray*}
where $f$ represents $e$ or $q$. 
\par
The products $\kappa_f \sin^2 \theta_W$, called
effective-mixing terms, are directly accessible from measurements
of the asymmetry in the $\cos \vartheta$ distribution. However,
neither the $\sin^2 \theta_W$ parameter nor the $\hat{s}$-dependent
form factors can be inferred from measurements
without assuming the standard model. The effective-mixing terms
are denoted as $\sin^2 \theta_{\rm eff}$ to distinguish them from
the on-shell definition of the $\sin^2 \theta_W$ parameter of
Eq.~(\ref{baseSW2}). The Drell-Yan process is most sensitive
to the $\sin^2 \theta_{\rm eff}$ term of the lepton vertex,
$\kappa_e \sin^2 \theta_W$. At the $Z$ pole, $\kappa_e$ is
independent of the quark flavor, and the flavor-independent value
of $\kappa_e \sin^2 \theta_W$ is commonly denoted as
$\sin^2 \theta^{\rm lept}_{\rm eff}$. 
For comparisons with other measurements, the value of
$\sin^2 \theta^{\rm lept}_{\rm eff}$ at the $Z$ pole is taken to be
$\operatorname{Re}[\kappa_e(M_Z^2)] \sin^2 \theta_W$.

\subsection{\label{QCDcalcs}
QCD calculations}

The \textsc{zfitter} form factors $\rho_{eq}$, $\kappa_e$, and
$\kappa_q$ are inserted into the Born $g_A^f$ and $g_V^f$
couplings of the Drell-Yan process. The $\kappa_{eq}$ form factor
is incorporated as an amplitude correction. This provides an
enhanced Born approximation (EBA) to the electroweak terms of the
amplitude. The form factor for the QED self-energy correction to
the photon propagator is also part of the EBA.
Complex-valued form factors are used in the amplitude. Only the
electroweak-coupling factors in the QCD cross sections are
affected. The standard LEP $Z$-boson resonant line-shape and
the total decay width calculated by \textsc{zfitter} are used.
\par
Both leading-order (LO) and next-to-leading-order (NLO) QCD
calculations of $A_{\rm fb}$ for the process
$p\bar{p} \rightarrow \gamma^*/Z \rightarrow \ell^+\ell^-$
are performed with form factors provided by \textsc{zfitter}.
Two sets of PDFs are used to
provide the incoming parton flux used in all QCD calculations
discussed in this section, except where specified otherwise.
They are the NLO CTEQ6.6~\cite{cteq66pdf} PDFs and the 
next-to-next-to-leading order (NNLO)
NNPDF-3.0~\cite{ nnpdf301, *nnpdf302, *nnpdf303, *nnpdf304,
		*nnpdf305, *nnpdf306, *nnpdf306e,*nnpdf307}
PDFs. For consistency with the \textsc{zfitter} calculations,
the NNPDFs selected are derived with a value of the strong-interaction
coupling of 0.118 at the $Z$ mass.
\par
Two NLO calculations, \textsc{resbos}
\cite{ResBos1, *ResBos2, *ResBos3, *ResBosc221}
and the \textsc{powheg-box} implementation \cite{Powheg-Box}
of the Drell-Yan process \cite{PowhegBoxVBP}, are modified to be
EBA-based QCD calculations. For both calculations, the cross
section is finite as $P_{\rm T}^2$ vanishes.
The \textsc{resbos} calculation combines a NLO fixed-order
calculation at high boson $P_{\rm T}$ with the
Collins-Soper-Sterman resummation formalism
\cite{methodCSS, *wfactorCSS1, *wfactorCSS2, *wfactorCSS3}
at low boson $P_{\rm T}$, which is an all-orders summation of
large terms from gluon emission calculated to
next-to-next-to-leading log accuracy.
The \textsc{resbos} calculation uses CTEQ6.6 NLO PDFs.
The \textsc{powheg-box} calculation uses
the NNLO NNPDF-3.0 PDFs, and is a fully unweighted
partonic event generator that implements Drell-Yan production
of $\ell^+\ell^-$ pairs at NLO. The NLO production implements a
Sudakov form factor \cite{Sudakov-FFeng, *Sudakov-FFrus}
that controls the infrared diverence
at low $P_{\rm T}$, and is constructed to be
interfaced with parton showering to avoid double counting.
The \textsc{pythia}~6.41~\cite{pythia64} parton-showering
algorithm is used to produce the final hadron-level event.
The combined implementation has next-to-leading log
resummation accuracy. The LO calculations of $A_{\rm fb}$ are
based on numerical integrations of the LO cross section using
NNPDF-3.0 PDFs, and  are used for direct comparisons with the
\textsc{powheg-box} calculations.

\par
The \textsc{powheg-box} NLO program, in conjunction with the
NNPDF-3.0 NNLO PDFs, is chosen as the default
EBA-based QCD calculation of $A_{\rm fb}$ with various input
values of $\sin^2 \theta_W$. The \textsc{resbos}
calculation is used as a reference for resummed calculations.
The LO calculation serves as a reference calculation
for the sensitivity of $A_{\rm fb}$ to QCD radiation.

\section{\label{CDFdetector}
Experimental Apparatus}

The CDF II apparatus is a general-purpose detector~\cite{refCDFII} at the
Fermilab Tevatron, a $p\bar{p}$ collider with a 
center-of-momentum (cm) energy of 1.96~TeV. The positive-$z$ axis of the
detector coordinate system is directed along the proton
direction. For particle trajectories, the polar angle $\theta_{\rm cm}$ is
relative to the proton direction and the azimuthal angle $\phi_{\rm cm}$ is
oriented about the beamline axis with $\pi/2$ being vertically upwards.
The pseudorapidity of a particle is
$\eta = -\ln \tan (\theta_{\rm cm}/2)$. Detector coordinates
are specified as $(\eta_{\rm det}, \phi_{\rm cm})$, where $\eta_{\rm det}$
is the pseudorapidity relative to the detector center~($z=0$).
\par
The momentum $\vec{P}$ of a charged particle is measured in the magnetic
spectrometer, which consists of charged-particle tracking detectors
(trackers) immersed in a magnetic field. The energy of a particle
is measured in the calorimeters surrounding the magnetic spectrometer.
The component of momentum transverse to the beamline is
$P_{\rm T} = |\vec{P}| \sin \theta_{\rm cm}$. The component of energy
transverse to the beamline is $E_{\rm T} = E \sin \theta_{\rm cm}$.
\par
The tracking detectors consist of a central tracker and an inner
silicon tracker. The central tracker is a 3.1~m long, open-cell drift
chamber~\cite{refCOT}
that extends radially from 0.4 to 1.4~m. Between the Tevatron beam pipe
and the central tracker is a 2~m long silicon tracker~\cite{refSVXII}.
Both trackers are immersed in a 1.4~T axial magnetic field produced by a
superconducting solenoid just beyond the outer radius of the drift
chamber. Combined, these two trackers provide efficient, high-resolution
tracking and momentum measurement over $|\eta_{\rm det}|<1.3$.
\par
Outside the solenoid is the central
barrel calorimeter~\cite{refCEM,refChad} that covers the region
$|\eta_{\rm det}|<1.1$. The forward end-cap regions,
$1.1<|\eta_{\rm det}|<3.5$, are covered
by the end-plug calorimeters~\cite{refPEM,refPES,refPHA}.
The calorimeters are scintillator-based sampling calorimeters,
which are segmented along their depth into electromagnetic (EM) and
hadronic (HAD) sections, and transversely into projective towers.
The EM calorimeter energy resolutions measured in test beams with
electrons are $\sigma/E = 13.5\%/\sqrt{E_{\rm T}}$ for the central
calorimeter, and $\sigma/E = 16\%/\sqrt{E} \oplus 1\%$ for the plug
calorimeter, where the symbol $\oplus$ is a quadrature sum,
and $E_{\rm T}$ and $E$ are in units of GeV. Both the
central and plug EM calorimeters have preshower and shower-maximum
detectors for electromagnetic-shower identification and
centroid measurements. The combination of the plug shower-maximum
detector and silicon tracker provides enhanced tracking coverage
to $|\eta_{\rm det}| = 2.8$. However, as $|\eta_{\rm det}|$
increases for plug-region tracks, the transverse track-length
within the magnetic field decreases, resulting in increasingly
poor track-curvature resolution. Within the plug shower-maximum
detector, $|\eta_{\rm det}| = 2.8$ corresponds to a radial extent
from the beamline of 23 cm.

\section{\label{DataSelection}
Data Selection}

The data set, collected over 2002--2011, is the full
CDF Run~II sample and consists of $p\bar{p}$ collisions
corresponding to an integrated luminosity of 9.4~fb$^{-1}$. 
Section~\ref{ElectronTriggers} reports on the online selection of
events (triggers) for the $A_{\rm fb}$ measurement.
Section~\ref{OfflineEleSelection} describes the offline selection
of electron candidates, and Sec.~\ref{ElePairSelection} describes
the selection of electron pairs.

\subsection{\label{ElectronTriggers}
Online event selection}

Electron candidates
are selected from two online triggers, \textsc{central-18}, and
\textsc{$Z$-no-track}. The \textsc{central-18} selection
accepts events containing at least one electron candidate
with $E_{\rm T} > 18$~GeV in the central calorimeter region.
Candidates are required to have electromagnetic shower clusters
in the central calorimeters that are geometrically matched to
tracks from the central tracker. Electron candidates for the
\textsc{$Z$-no-track} selection have no track requirement and
are only required to be associated with an electromagnetic shower
cluster with $E_{\rm T} > 18$~GeV. The selection, which accepts
events containing at least one pair of candidates located in any
calorimeter region, is primarily for dielectrons in the
plug-calorimeter region where online tracking is not available. 
It also accepts the small fraction of dielectron events that fail
the track requirements of the \textsc{central-18} trigger.

\subsection{\label{OfflineEleSelection}
Offline electron selection}

After offline event reconstruction, the purity of the sample is
improved with the application of CDF standard-electron identification
and quality requirements \cite{refCDFII}. Fiducial constraints are
applied to ensure that the electrons are in well-instrumented
regions, thus ensuring good-quality and predictable reconstruction
performance. Each electron candidate is required to be
associated with a track, to significantly reduce backgrounds.
The track-vertex position along the beamline $(z_{\rm vtx})$ is
restricted to be within the luminous region,
$|z_{\rm vtx}| < 60$~cm. Overall, 3\% of the $p\bar{p}$ luminous
region along the beamline is outside this fiducial region. 

\par
Electron identification in the central calorimeter region is
optimized for electrons of $P_{\rm T}>10$~GeV/$c$ \cite{refCDFII}.
It uses information from the central and silicon trackers, the
longitudinal and lateral (tower) segmentation of the electromagnetic
and hadronic calorimeter compartments, and the shower-maximum strip
detector (CES) within the electromagnetic calorimeter.
The highest quality of signal selection and background rejection
is provided by the trackers in combination with the CES.
An electron candidate must have shower
clusters within the electromagnetic calorimeter towers and CES
signals compatible with the lateral profile of an electromagnetic
shower. A candidate must also be associated to a track that
extrapolates to the three-dimensional position of the CES shower
centroid. The transverse momentum of the particle associated with
the track must be consistent with the associated electron shower
$E_{\rm T}$ via an $E/P$ selection when $P_{\rm T} < 50$~GeV/$c$
\cite{refCDFII}. For both the
track matching in the CES and the $E/P$ selection, allowances are 
included for bremsstrahlung energy-loss in the tracking volume,
which on average is about 20\% of a radiation length. The fraction
of shower energy in the hadronic-calorimeter towers behind the
tower cluster of the electromagnetic calorimeter must be consistent
with that for electrons through an $E_{\rm HAD}/E_{\rm EM}$ requirement.
These selections are more restrictive than those applied in the
online selections described in Sec.~\ref{ElectronTriggers}.

\par
Such an offline selection has high purity
and is called the tight central electron (TCE) selection.
To improve the selection efficiency of central-electron pairs,
a looser selection,
called the loose central electron (LCE) selection, is applied
on the second electron candidate. The looser variant does not use
transverse shower-shape constraints, the $E/P$ constraint, or
track matching in the CES. For track associations, the track is
only required to project into the highest-energy calorimeter tower
within the cluster of towers associated with the electromagnetic
shower.

\par
Electron identification in the plug calorimeter also uses
tracker information, the longitudinal and lateral (tower) segmentation
of the electromagnetic and hadronic calorimeter compartments, and the
shower-maximum strip detector (PES) within the electromagnetic
calorimeter. As the plug-calorimeter geometry differs from
the central geometry, the details of the selection requirements differ.

\par
The end-plug calorimeters, with sampling planes perpendicular to the
beamline, have projective towers that are physically much smaller
than the central calorimeter towers and vary in size \cite{refPEM}.
The electromagnetic
showers in the plug calorimeter are clustered into ``rectangular''
$3 \times 3$ detector-tower clusters centered on the highest-energy tower.
Good radial containment of these showers is achieved.
The preshower detector is the
first layer of the plug-electromagnetic calorimeter and it is instrumented
and read out separately. As there are approximately 0.7 radiation lengths
of material in front of it, its energy is always included in the
electromagnetic-cluster shower energy.

\par
Tracks entering the plug calorimeters have limited geometrical
acceptance in the central tracker for $|\eta_{\rm det}| > 1.3$.
The forward coverage of the silicon tracker is exploited with a
dedicated calorimetry-seeded tracking algorithm called ``Phoenix''.
It is similar to the central tracking-algorithm, where tracks found
in the central tracker are projected into the silicon tracker and hits
within a narrow road of the trajectory initialize the silicon track
reconstruction. With the Phoenix algorithm, the track helix in the
magnetic field is specified by the position of the $p\bar{p}$
collision vertex, the three-dimensional exit position of the
electron into the PES, and a helix curvature. The collision
vertex is reconstructed from tracks found by
the trackers. The curvature is derived from the $E_{\rm T}$ of the
shower in the electromagnetic calorimeter. Two potential helices
are formed, one for each charge. The algorithm projects each helix into
the silicon tracker to initialize the track reconstruction. If both
projections yield valid tracks, the higher-quality one is selected.
Depending on its vertex location along the beamline, a track
traverses zero to eight layers of silicon. A Phoenix track is
required to traverse at least three silicon layers and have at 
least three silicon hits. Eighty percent of the tracks traverse
four or more silicon layers, and the average track acceptance
is 94\%.

\par
An electron candidate in a plug calorimeter must have a shower
cluster within the electromagnetic calorimeter towers, an associated
PES signal compatible with the lateral profile of an electromagnetic
shower, and a longitudinal profile, measured using
$E_{\rm HAD}/E_{\rm EM}$, that is consistent with that expected for
electrons. The candidate must also be associated with a
Phoenix track. Neither a $P_{\rm T}$ nor an $E/P$ selection
requirement is applied because the track momentum determined by the
Phoenix algorithm is correlated with the calorimeter energy. Charge
misidentification is significant at large $|\eta_{\rm det}|$
because of the reduced track-helix curvature resolution. The
resolution is inversely proportional to the track-exit radius
at the PES, which varies from 23 to 129~cm.

\par
As Drell-Yan high-$E_{\rm T}$ leptons are typically produced in
isolation, the electron candidates are required to be isolated
from other calorimetric activity. The isolation energy,
$E_{\rm iso}$, is defined as the sum of $E_{\rm T}$ over towers
within a 0.4 isolation cone in $(\eta,\phi)$ surrounding the
electron cluster. The towers of the electron cluster are not
included in the sum. For central-electron candidates, the
isolation requirement is $E_{\rm iso}/E_{\rm T} < 0.1$, and
for plug-electron candidates, it is $E_{\rm iso} < 4$~GeV.

\subsection{\label{ElePairSelection}
Offline electron-pair event selection}

Events are required to contain two electron candidates in
either the central or plug calorimeters. These events are
classified into three topologies, CC, CP, and PP,
where C (P) denotes that the electron is detected in the
central (plug) calorimeter.
The electron kinematic variables are based on the electron
energy measured in the calorimeters and on the track direction.
The kinematic and fiducial regions of acceptance for electrons
in the three topologies are described below.
\begin{enumerate}
 \item Central--central (CC)
    \begin{enumerate}
      \item $E_{\rm T} > 25$ (15)~GeV for electron 1 (2);
      \item $0.05 < |\eta_{\rm det}| < 1.05$.
    \end{enumerate}
 \item Central--plug (CP)
    \begin{enumerate}
      \item $E_{\rm T} > 20$~GeV for both electrons;
      \item Central electron: $0.05 < |\eta_{\rm det}| < 1.05$;
      \item Plug electron: $1.2 < |\eta_{\rm det}| < 2.8$.
    \end{enumerate}
 \item Plug--plug (PP)
    \begin{enumerate}
      \item $E_{\rm T} > 25$~GeV for both electrons;
      \item $1.2 < |\eta_{\rm det}| < 2.8$.
    \end{enumerate}
\end{enumerate}
The CC topology consists of TCE-LCE combinations with
asymmetric $E_{\rm T}$ selections on electrons 1 and 2, the
electrons in the pair with the higher and lower $E_{\rm T}$,
respectively. Either electron can be the TCE candidate, and
its LCE partner can also be a TCE candiate because they are
a subset of the LCE candidates.
The asymmetric selection, an optimization from the previous
measurement of electron angular-distribution coefficients
\cite{CSangcoef21}, improves the acceptance.
For the CP topology, the central
electron candidate must pass the TCE selection.
The PP-topology electron candidates are both required to be
in the same end of the CDF~II detector; Drell-Yan electrons
of the PP topology on opposite ends of the CDF~II detector
are overwhelmed by QCD dijet backgrounds at low $P_{\rm T}$.
In addition, the longitudinal separation of vertex positions
of the associated tracks of the candidates is required to be
within 4~cm of each other.

\par
The measurement of $A_{\rm fb}$ is based on the direction of
the $e^-$ in the CS frame, and any charge misidentification
dilutes the result. Charge misidentification is
small for central tracks and significant for plug tracks.
Consequently, only CC- and CP-topology pairs are used in the
measurement. For the CP-topology, the central electron is
used to identify the $e^-$. Electron pairs of the PP topology
are only used for plug-calorimeter calibrations and cross
checks. The same-charge pairs of the CC topology are also
not used in the measurement, but they are used for calibrations,
simulation tuning, and consistency checks.

\par
Signal events intrinsically have no imbalance in the total energy
in the transverse plane from undetected particles except for those
within uninstrumented regions of the detector or from semileptonic
decays of hadrons. The transverse energy imbalance
$E\!\!\!/_{\rm T}$ is the magnitude of $-\sum_i E_{\rm T}^i \hat{n}_i$,
where the sum is over calorimeter towers, $\hat{n}_i$ is the unit
vector in the azimuthal plane that points from the $p\bar{p}$
collision vertex to the center of the calorimeter tower $i$, and
$E_{\rm T}^i$ is the corresponding transverse energy in that
tower. Events with $E\!\!\!/_{\rm T} < 40$~GeV are selected,
therefore  poorly reconstructed signal events, characterized by large
$E\!\!\!/_{\rm T}$, are removed. Only a very small fraction of
signal events is removed. About half of the background events
containing leptonically decaying $W$-bosons are removed because
they have large intrinsic $E\!\!\!/_{\rm T}$ from neutrinos, which
are undetected.

\section{\label{AfbexpDatSim}
Signal simulation}

Drell-Yan pair production is simulated using the
\textsc{pythia}~\cite{Pythia621} Monte Carlo event generator and CDF~II
detector-simulation programs. \textsc{pythia} generates the hard,
leading-order QCD interaction $q\bar{q} \rightarrow \gamma^*/Z$,
simulates initial-state QCD radiation via its parton-shower algorithms,
and generates the decay $\gamma^*/Z \rightarrow \ell^+\ell^-$.
The CTEQ5L~\cite{Cteq5pdf} PDFs are used in the calculations.
The underlying-event and
boson-$P_{\rm T}$ parameters are derived from the \textsc{pythia}
configuration \textsc{pytune} 101 (AW), which is a
tuning to previous CDF data~\cite{Pythia621,run1CDF-Z,PyTuneAW}.
\par
Generated events are first processed by the event simulation, and then
followed by the CDF~II detector simulation based on \textsc{geant}-3
and \textsc{gflash}~\cite{nimGflash}. The event simulation includes 
\textsc{photos} 2.0~\cite{Photos20a, *Photos20b, Photos20c},
which adds final-state QED radiation (FSR) to decay vertices with
charged particles (e.g., $\gamma^*/Z \rightarrow ee$).
The default implementation of \textsc{pythia} plus \textsc{photos}
(\textsc{pythia+photos}) QED radiation in the simulation
has been validated in a previous 2.1~fb$^{-1}$
measurement of $\sin^2\theta^{\rm lept}_{\rm eff}$ using Drell-Yan
electron pairs \cite{zA4ee21prd,*zA4ee21prdE}.
\par
The \textsc{pythia+photos} calculation is adjusted using the
data and the \textsc{resbos} calculation. The generator-level
$P_{\rm T}$ distribution of the boson is adjusted so that the
shape of the reconstruction-level, simulated $P_{\rm T}$ distribution
matches the data in two rapidity bins: $0 < |y| < 0.8$ and
$|y| \ge 0.8$. For this adjustment, reconstructed $ee$ pairs of all
topologies (CC, CP, and PP) in the 66--116~GeV/$c^2$ mass region
are used. The generator-level boson-mass distribution is
adjusted with a mass-dependent K-factor. The K-factor is the
ratio of the \textsc{resbos} boson-mass distribution calculated
using CTEQ6.6 PDFs relative to the \textsc{pythia}~6.4 \cite{pythia64}
boson-mass distribution calculated using CTEQ5L PDFs. No kinematic
restrictions are applied.
\par
Standard time-dependent beam and detector conditions are
incorporated into the simulation, including the $p$ and $\bar{p}$
beamline parameters; the luminous region profile;
the instantaneous and integrated luminosities per data-taking period;
and detector component calibrations, which include
channel gains and malfunctions. The simulated events
are reconstructed, selected, and analyzed in the same way as the
experimental data.

\section{\label{CorrDatSim}
Data and Simulation Corrections}

In this section, time- and position-dependent corrections and
calibrations to the experimental and simulated data are presented.
They include event-rate normalizations of the simulation to the
data, energy calibrations of both the data and simulation,
and modeling and removal of backgrounds from the data.
The detector has 1440 EM calorimeter towers, each with
different responses over time and position within the
tower. Many instrumental effects are correlated, and the overall
correction and calibration process is iterative.

\subsection{\label{RateNormSim}
Event rate normalizations}

The simulation does not model
the trigger and reconstruction efficiences observed in the
data with sufficient precision. Time-, detector-location-,
and luminosity-dependent differences are observed.
To correct the observed differences in rate
between the data and simulation, a scale-factor event weight
is applied to simulated events. The scale factor is the ratio
of the measured offline-selection efficiencies observed in data
to the simulation versus time, detector location, and instantaneous
luminosity. 
\par
The base correction described above using measured efficiencies
is inadequate for the $A_{\rm fb}$ measurement for two reasons:
1) due to the more stringent selection requirements
for the efficiency measurements, the bin sizes for the time,
position, and luminosity dependence are wide, and a finer
resolution is needed; and 2) the Tevatron $p\bar{p}$ luminosity
profile is difficult to simulate. The second-level correction uses
event-count ratios between the data and simulation, or scale
factors, as event weights. Events are required to pass all
standard selection requirements and the $ee$-pair mass is required
to be within the 66--116 GeV/$c^2$ range. Events are separated into
the CC, CP, and PP topologies and corrected separately.
\par
The time and luminosity dependencies are related. The distributions
of the number of $p\bar{p}$ collision vertices in each event
$(n_{\rm vtx})$ and the location of these vertices along the
beamline~$(z_{\rm vtx})$ changed significantly with improvements
to the Tevatron collider. These distributions are inadequately
simulated and are corrected separately. For the $n_{\rm vtx}$
correction, the data and simulation are grouped into 39 calibration
periods, and the distribution corrected on a period-by-period basis.
The correction of the $z_{\rm vtx}$ distribution is organized into a
smaller set of seven time intervals corresponding to improvements in
the Tevatron collider. The $z_{\rm vtx}$ distribution has an rms
spread of 30~cm, and it needs to be simulated accurately because at large
$|z_{\rm vtx}|$ the electron acceptance as a function of $E_{\rm T}$
changes significantly.
\par
The second-level correction to remove detector-location
dependencies is a function of $|\eta_{\rm det}|$.
In the central calorimeter, the corrections are small.
In the plug calorimeters, the corrections are larger and
they correct the effects of tower-response differences between data
and simulation near tower boundaries.

\subsection{\label{EnergyCalibrations}
Energy calibrations}

The energy calibrations are relative to the standard
calibrations for time-dependent beam and detector conditions.
Energy calibrations are multidimensional, and since it is not
feasible to calibrate all components simultaneously, they are
iteratively calibrated with a sequence of four steps using groups
of lower dimension.
\par
\begin{figure}
\includegraphics
   [width=85mm] 
   {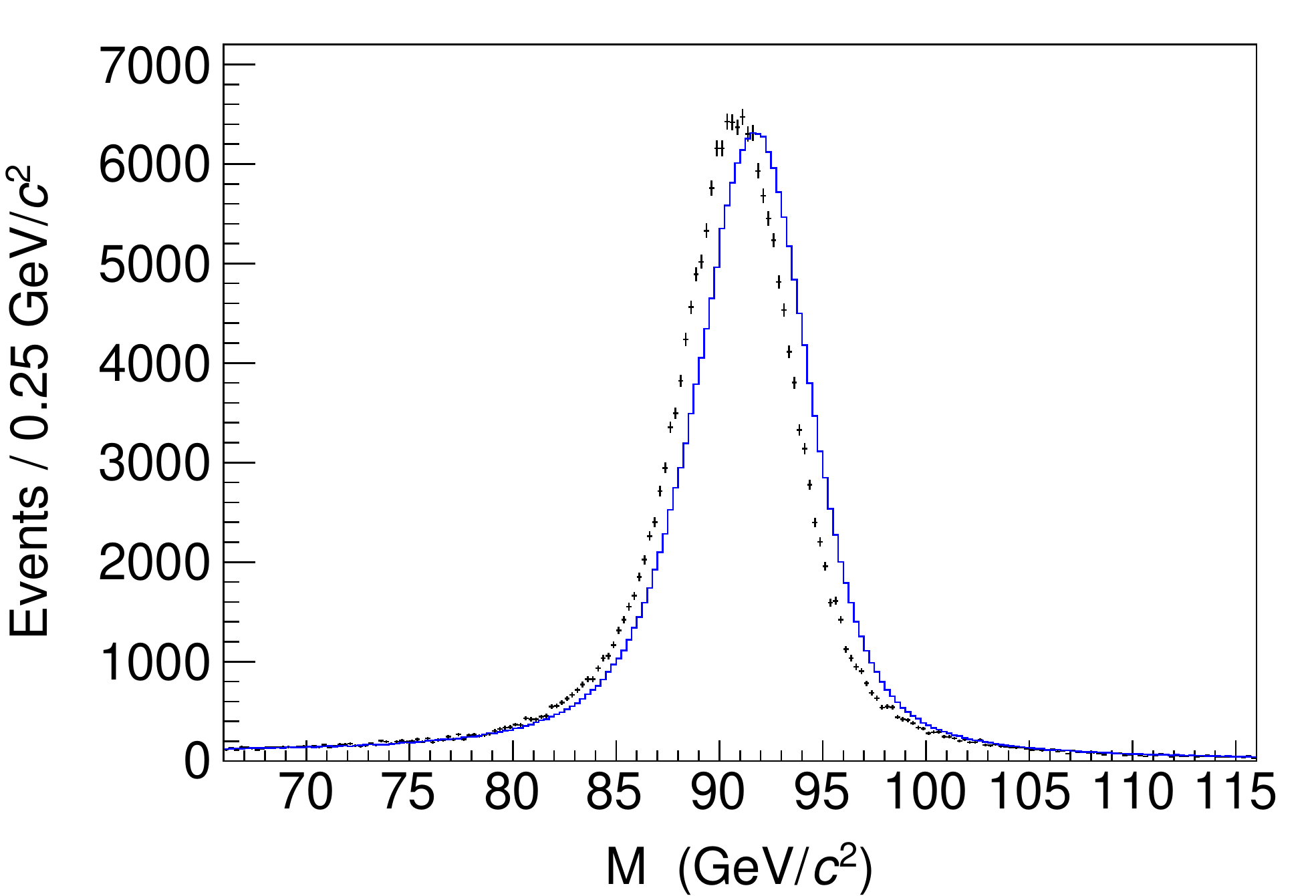}
\caption{\label{fig_rawmee1CCos}
Invariant $ee$-mass distribution for opposite-charged CC events
prior to the calibration and background subtractions.
The crosses are the data and the solid histogram is the simulation.
}
\end{figure}
\begin{figure}
\includegraphics
   [width=85mm]
   {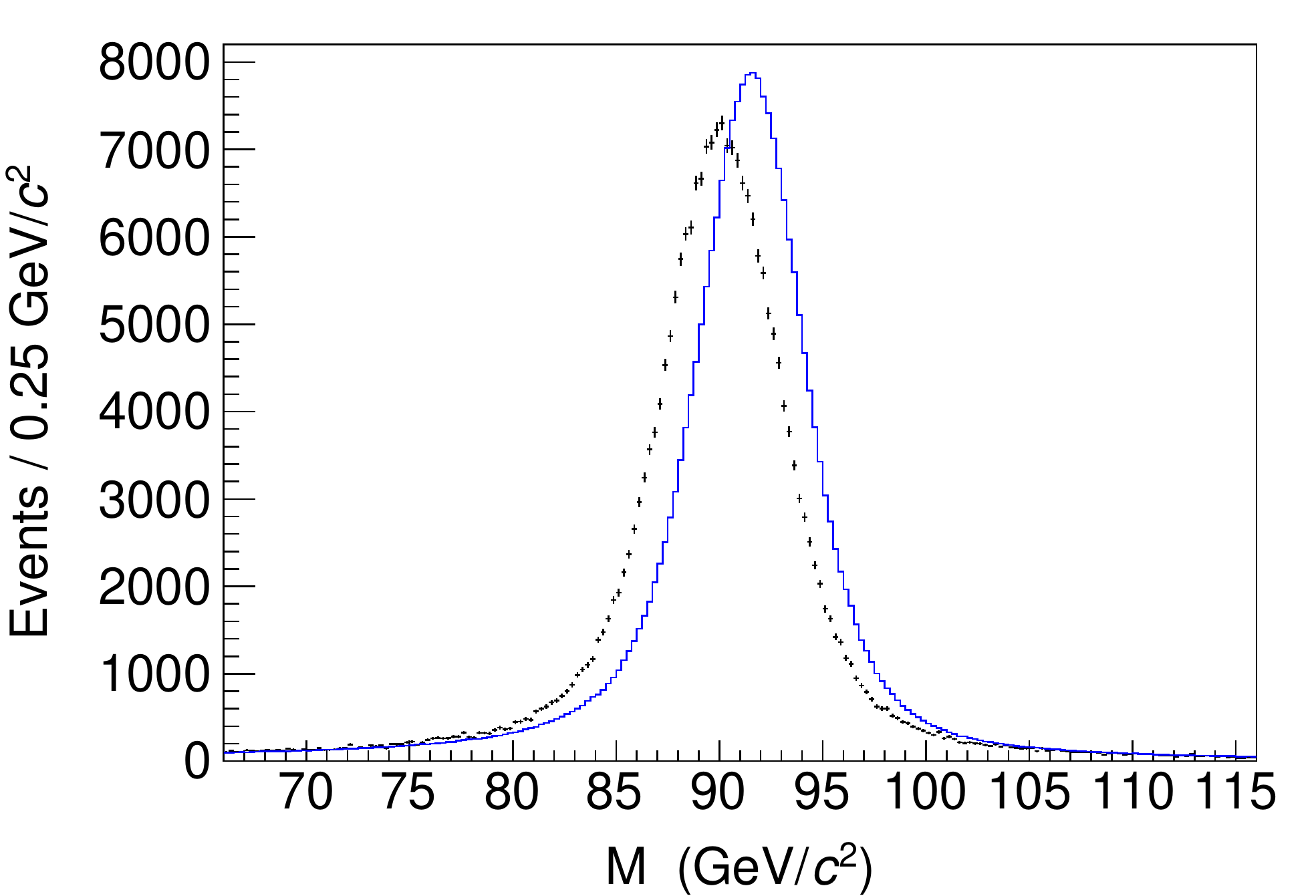}
\caption{\label{fig_rawmee1CP}
Invariant $ee$-mass distribution for CP events
prior to the calibration and background subtractions.
The crosses are the data and the solid histogram is the simulation.
}
\end{figure}

The standard calibrations for the calorimeter have energy-scale
miscalibrations that depend on time and detector location,
and range up to 5\% in magnitude. The miscalibrations differ for
the data and the simulation, and are larger at the edges of the
plug calorimeter. The energy resolution of the simulation
also needs additional tuning. Without any adjustments, the mass
distributions of CC- and CP-topology electron pairs are as
shown in Figs. \ref{fig_rawmee1CCos} and \ref{fig_rawmee1CP},
respectively.
\par
Adjustments to correct the
miscalibrations are constrained using the mass distributions
of $e^+e^-$ pairs about the $Z$~pole.
Calibration adjustments are based on three electron-pair
mass distributions: 1) generator level, 2) simulated data, and
3) data. All three mass distributions are fit to a line shape
that includes the $Z$-pole mass as a fit parameter. The $Z$-pole
mass values obtained from fits to the experimental and simulated
data are separately aligned to the corresponding generator-level
value~\cite{muPcorrMethod}.
\par
The generator-level mass is evaluated using clustered energies and
includes the effects of QED FSR. The FSR electrons and photons are
clustered about the seed tower in a manner similar to the clustering
of electron reconstruction. The seed tower is based on the
reconstructed electron, and the projection from the $p\bar{p}$
collision vertex to the tower is achieved by extrapolating the
track helix.
Since the detector acceptance slightly alters the line-shape of
the mass distribution, generator-level events are selected by
requiring that their kinematic properties after detector simulation
meet all selection criteria.
\par
The generator-level mass distribution is fit to the standard LEP
$Z$-boson resonant line shape. The data and simulation mass
distributions are fit to the standard LEP $Z$-boson resonant
line shape convoluted with the resolution functions of the
calorimeters, which are Gaussian. Typically, the fit range
is $\pm 5$~GeV/$c^2$ around the $Z$ peak.
The $Z$-pole mass and resolution
width values are allowed to vary but the resonant width is fixed
to the corresponding generator-level fit value. With this method,
the resolution width values of the simulation and data are
directly comparable and are used to calibrate the
energy resolution of the simulation to the data.
\par
Electron pairs of all topologies that satisfy the selection
requirements are used in the calibration. The set of
CC+PP events, and separately, of CP events, provides two
independent sets of calibrations for all calorimeter
components, such as towers.
The electrons used to calibrate the energy scale of a
component are denoted as reference electrons.
The partners to these electrons can be anywhere in the
detector so that miscalibrations of the current iteration are
averaged out, and they also serve as references elsewhere.
Energy-scale adjustments require the constraint of the sharp,
nearly symmetric peak at the $Z$ pole of the mass distribution.
The energy distribution of the electrons is not as suitable
because it is broad and asymmetric, and sensitive to the
boson transverse-momentum and rapidity distributions, as well
as the $e^+e^-$ angular distribution. 

\par
The first step in an iteration is the time-dependent calibration
of the overall energy scales for the central and plug
calorimeters. Corrections are determined for each of the
39~calibration periods introduced in the previous section.
\par
The next step is the relative calibration of calorimeter towers
and the response maps within their boundaries. In this step,
the bins are small and do not have enough events for accurate
mass fits. Consequently, the energy response for each bin
is quantified using the statistically more accurate mean of the
scaled electron-pair mass $M/(91.15\;{\rm GeV}/c^2)$
over the range 0.9--1.1, and is normalized to the overall average
of the central or plug calorimeters.
Tower-response corrections are important in the high
$|\eta_{\rm det}|$ region of the plug calorimeters where
standard calibrations are difficult. Corrections are
determined for two time periods: calibration periods 0--17 and
18--39. Period 18 is the start of consistently efficient
high-luminosity Tevatron operations, which commenced from
April~2008. Both the central and plug calorimeter towers require
additional response-map tuning at the periphery of the towers.
\par
The third step calibrates the energy scales of the $\eta$-tower
rings of the calorimeter. A ring consists of all towers in the
$\phi$ dimension with the same $|\eta_{\rm det}|$ dimension. The
adjustments from this step isolate the systematic variation of the
energy scale in the $\eta$ dimension of the standard calibration
relative to the underlying physics.
There are 22 $\eta$-tower rings, 12 of which are in the plug
calorimeter. The lowest and highest $\eta$-tower rings
of the plug calorimeter are not in the acceptance region.
Separate calibrations for the CC+PP and CP data are iteratively
determined using two passes, with corrections determined
for two time periods, 0--17 and 18--39. First,
the central and plug rings are calibrated with events from the
CC+PP data. These calibrations are used only for CC- and
PP-topology pairs. Then the CP data set calibration is derived
from CP events, using the CC+PP calibrations as initial values
for the calibration. The calibrations
from the CC+PP and CP sets are expected to be slightly
different due to the wide $z_{\rm vtx}$ distribution of
$p\bar{p}$ collision vertices at the Tevatron. The geometry
of an electron shower within the CDF calorimeters depends on the
position of the collision vertex. Away from  $z_{\rm vtx} = 0$,
the transverse segmentation of the calorimeter is less
projective, and the fraction of the shower energy sampled by
the calorimeter is different. As the magnitude of
$z_{\rm vtx}$ increases, the electron energy reconstructed
in the calorimeter can change.
\par
Accompanying the $\eta$-ring correction is the extraction
of the underlying-event energy contained within an electron-shower
cluster. The electron-pair mass distributions show an observable
dependence on the number of $p\bar{p}$ collision
vertices in an event. Assuming that the underlying-event
energy per shower cluster increases linearly with $n_{\rm vtx}$,
these mass distributions are used to extract the associated
underlying-event energy of a shower cluster per vertex for
each $\eta$ ring. For the central calorimeter, the value is
approximately constant at 35~MeV. For the plug calorimeters,
the value is approximately
150~MeV for $|\eta_{\rm det}| < 2$ and increases to 1.5~GeV at
$|\eta_{\rm det}| \approx 2.8$. The expected underlying-event energy
is subtracted from the measured electron energy.
\par
The fourth step removes residual miscalibrations in both
$\eta_{\rm det}$ and $\phi$. The energy scales on a grid with
16 $\eta_{\rm det}$ and 8 $\phi_{\rm cm}$ bins are calibrated,
along with determinations of the corresponding energy
resolutions. The
$\eta_{\rm det}$ bins span both ends of the detector, with
eight bins each for the central and plug calorimeters. Events
in each $(\eta_{\rm det},\phi_{\rm cm})$ bin are further
divided into electron pairs with $\eta_{\rm det}$ values of
the same sign (SS) and pairs with opposite-sign values (OS).
There are differences
of a few tenths of a percent between the SS- and OS-pair
calibrations. The electrons of SS and OS pairs also have
differing showering geometries within the calorimeters due to 
the wide $z_{\rm vtx}$ distribution of $p\bar{p}$ collisions.
The fraction of SS pairs for the CC topology is approximately
50\%. For the CP topology, the fraction varies with the 
$\eta$-bin index, and the range is approximately 50\% to
80\%. The PP-topology sample consists entirely of SS pairs. 
\par
The energy resolution of the calorimeter simulation is also
adjusted for each calibration bin of the fourth step.
Line-shape fits to the mass distributions of the data and
the simulation provide the resolution-smearing parameters
$\sigma^2_{\rm d}$ and $\sigma^2_{\rm s}$, respectively.
The fit values of $\sigma_{\rm d}$ are approximately 2~GeV/$c^2$
for all bins. For most bins, the simulation resolution is
adjusted with an additional Gaussian rms deviation of
$\sigma^2_{\rm d} - \sigma^2_{\rm s}$.
For 24\% of the central bins, this value is negative, and
the alternative is to rescale the simulation energy bias
$\Delta E_{\rm bias} \equiv E_{\rm gen} - E_{\rm rec}$
of each event,
where $E_{\rm gen}$ is the generator-level clustered energy
and $E_{\rm rec}$ is the reconstruction-level energy. The
resolution is modified by scaling the event-by-event bias
with the factor $f_{\rm bias}$ so that the new
reconstruction-level energy is
$E_{\rm gen} - f_{\rm bias} \Delta E_{\rm bias}$.
The value of $f_{\rm bias}$ does not deviate from its expected
value of unity by more than 17\%.

\par
\begin{figure}
\includegraphics
   [width=85mm]
   {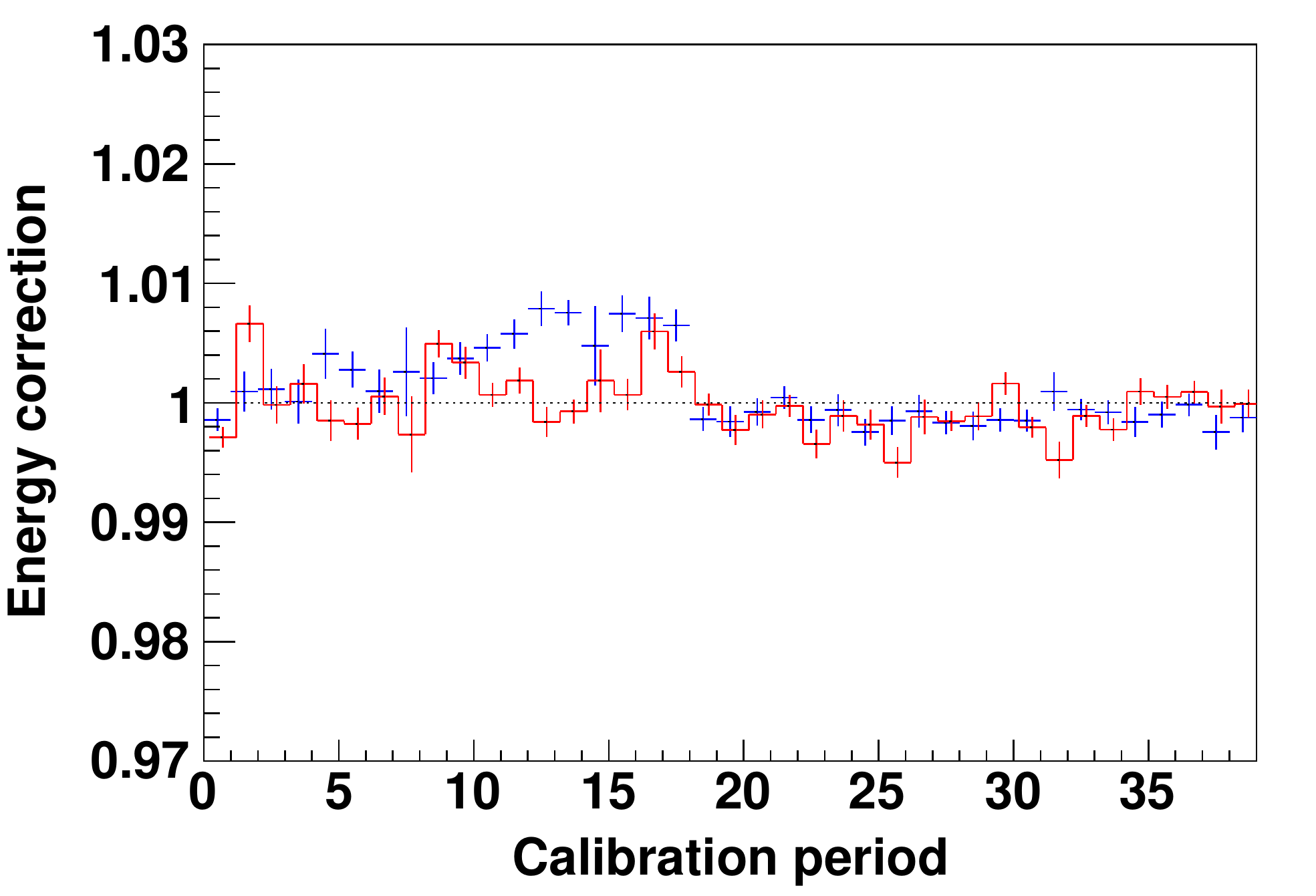}
\caption{\label{fig_ecalibRefE}
Corrections to the global energy scales as functions
of the calibration period for the data. The central
calorimeter corrections are the crosses (blue), and
for the plug calorimeter, the histogram (red) gives
the corrections.
}
\end{figure}

The energy calibration stabilizes after three iterations.
The time-dependent global corrections to the energy
scales of the central and plug calorimeters from step one
are shown in Fig.~\ref{fig_ecalibRefE}.
Approximately 20\% of the data is contained in time periods
0--10, and 68\% in time periods 18-38.
The energy calibrations over $\eta$-tower rings from
step three have the largest effect.
\begin{figure}
\includegraphics
   [width=85mm]
   {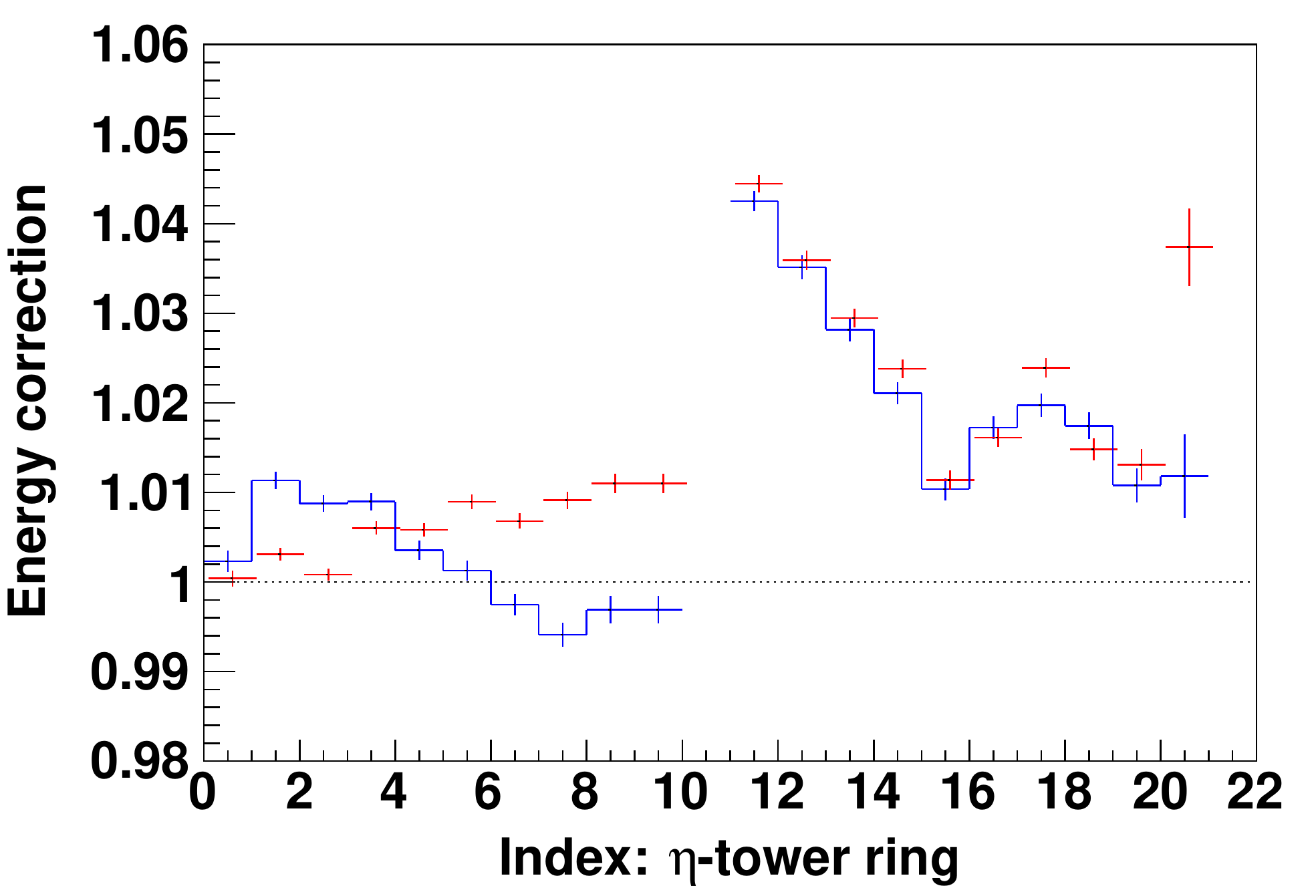}
\caption{\label{fig_ecalibSimE}
Corrections to the energy scales as functions of the
calorimeter $\eta$-tower ring index for the CP-topology data.
The corrections from time periods 0 to 17 are represented by
the histograms (blue), and those from time periods 18 to 38
by the crosses (red). The central calorimeter region
is index 0 to 9. Index 21 towers are about 23~cm from the
beamline.
}
\end{figure}
Figure \ref{fig_ecalibSimE} shows the corrections derived
from the CP calibration set for the two time periods, 0--17 and
18--38. The corrections derived from the CC+PP calibration set
are similar. For the central calorimeter, the corrections
from periods 0 to 17 and 18 to 38 are different because its
standard calibration procedure was modified prior to the start of
period 18. The tower-gain calibrations include an additional
$\eta$-dependent correction that ranges from 0 to 2\%.
For periods 0--10 and 11--17, the central-calorimeter corrections
are close to and compatible with the combined corrections shown
in Fig.~\ref{fig_ecalibSimE}. 
The mass distributions of CC- and CP-topology electron pairs
after the energy calibration adjustments, and other corrections
presented next, are shown later in Figs. \ref{fig_mee1CCos} and
\ref{fig_mee1CP}, respectively.

\subsection{\label{EEBackgrounds}
Backgrounds}

The backgrounds are negligible in the $Z$-peak region used for
the energy calibration but they are detectable in the low- and
high-mass regions of the mass distributions.
In this section, the level and shapes of the backgrounds in
the $ee$-pair mass distribution are determined separately for
each of the CC, CP, and PP topologies.
\par
The backgrounds are from the production of QCD dijets,
$Z \rightarrow \tau^+\tau^-$, $W+$jets, dibosons
(\textit{WW}, \textit{WZ}, \textit{ZZ}), and $t\bar{t}$ pairs. 
All backgrounds except for QCD are derived from
\textsc{pythia}~\cite{Pythia621} samples
that are processed with the detector simulation, and in which the
integrated luminosity of each sample is normalized to that of the
data. The diboson and $t\bar{t}$ sample normalizations use total
cross sections calculated at NLO \cite{MCFM345}.
The $W+$jets and $Z \rightarrow \tau^+\tau^-$ sample normalizations
use the total cross sections calculated at LO multiplied by an
NLO-to-LO K-factor of 1.4. Sample normalizations based on these
calculated cross sections are referenced as default normalizations.
Simulated events are
required to pass all selections required of the data.
\par
The QCD background is primarily from dijets that are misidentified
as electrons. This background is extracted from the data assuming
that its combination with the sum of the simulated signal
and other backgrounds matches the observed mass distribution.
The QCD background distribution, parametrized with level and
shape parameters, is determined in a fit of the data to the sum
of all backgrounds in conjunction
with the simulated signal. The mass range for the fit is 42--400
GeV/$c^2$ with 50 equally spaced bins in $\ln M$, and the
minimization statistic is the $\chi^2$ between the data and the sum
of predicted components over all bins. The normalizations of the
simulated signal and backgrounds are also allowed to vary from
their default values via scale factors. However,
as most simulated backgrounds are very small, they are only allowed
to vary within their normalization uncertainties. The constraint is
implemented with an additional $\chi^2$ term $(f_{\rm norm}-1)^2/0.085^2$,
where $f_{\rm norm}$ is the scale factor of the background
calculation. The uncertainty of the measured luminosity is 6\%
\cite{cdfR2CLC}; the prediction uncertainty is taken to be equally as
large; and their combination gives the estimate for the constraint
uncertainty of 0.085.
The $t\bar{t}$, diboson, and $W+$jets backgrounds are always constrained.
The $Z \rightarrow \tau^+\tau^-$ background is the second largest, and
for CC-topology events, the scale factor is determined with the data.
However, for CP- and PP-topology events, the $Z \rightarrow \tau^+\tau^-$
background scale factors are constrained to their default normalizations.
\par
For the QCD-background analysis, two independent data samples are used:
events passing the selection criteria and events failing them.
The first sample, denoted as the signal sample, is used to determine the
level of the QCD background and its shape over the mass distribution.
The second sample, denoted as the QCD-background sample, is
derived from events failing the selection criteria, and is for the
event-by-event background subtractions from kinematic distributions.
\par
Electron-like candidates for the QCD-background samples are selected
by reversing criteria that suppress hadrons and QCD jets. One
candidate passes all electron selection requirements except the
isolation criterion. The other is required to be ``jet-like'' by
reversing the isolation and $E_{\rm HAD}/E_{\rm EM}$ selection
criteria. Since there is a small fraction of $\gamma^*/Z$ events
in the initial background sample, the reverse selections are
optimized for each $ee$-pair topology to remove them. As the
reversed selection criteria bias the mass distributions, events of
these QCD-background samples are reweighted so that the overall
normalization and the shapes of the mass distributions match
those extracted from the signal samples.
\par
\begin{figure}
\includegraphics
   [width=85mm]
   {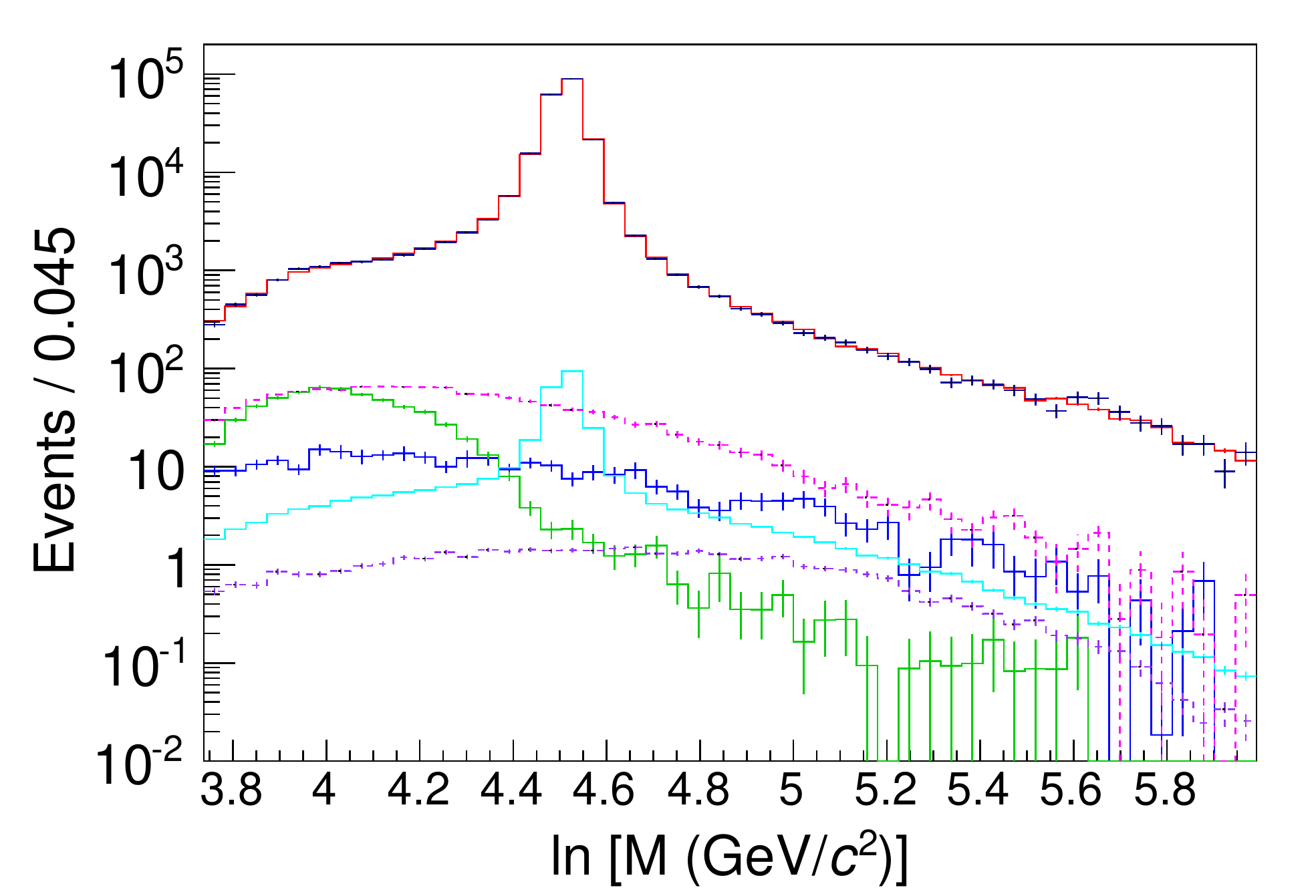}
\caption{\label{fig_bkgrCCos}
Logarithmically binned mass distributions for oppositely
charged $ee$-pair candiates in CC-topology events. The (black)
crosses are the data, the (red) histogram overlapping the data is the
sum of all components, the (green) histogram concentrated at lower
masses is the $Z \rightarrow \tau^+\tau^-$ component, and the (cyan)
histogram in the middle with the $Z$ peak is the diboson component.
The remaining broad distributions, from top to bottom, are
QCD (magenta), $W+$jets (blue), and $t\bar{t}$ (purple).
The comparison of the data with the sum of the components yields
a $\chi^2$ of 56 for 50 bins.
}
\end{figure}

For the CC topology, same-charge pairs passing the selection
criteria are also used to determine the QCD background
parameters, because 50\%-60\% of the events in the low- and
high-mass regions are from QCD. The first step in the background
determination is the extraction of the shape and default level
of the QCD background from the same-charge distribution.
Then, the mass distributions of both same-charge and
opposite-charge pairs passing the selection criteria are fit
simultaneously for the background level parameters. The large
fraction of QCD events in the same-charge distribution
constrains the QCD background level parameter.
Consequently, the scale factor for the normalization of the
$Z \rightarrow \tau^+\tau^-$ background is determined using the
data, but the default normalization is not accommodated. If the
$Z \rightarrow \tau^+\tau^-$ normalization is allowed to vary, the
fit determines a scale factor value of $0.53 \pm 0.11$. However,
if the normalization is restricted via the constraint to the
default value, the fit pulls
the scale factor away from its default value of unity to a value
of $0.83 \pm 0.07$, and the $\chi^2$ increases by six units
relative to the unconstrained fit.
The detector simulation and event normalizations
for the $Z \rightarrow \tau^+\tau^-$ sample, consisting of
lower-$E_{\rm T}$ secondary electrons from $\tau$ decays, are
not tuned. Consequently, the 0.53 value is chosen for the $A_{\rm fb}$
measurement and the 0.83 value is used as a systematic variation.
The CC-topology opposite-charge mass distributions for the
data, the simulated data plus backgrounds, and the backgrounds
are shown in Fig.~\ref{fig_bkgrCCos}.
\par
\begin{figure}
\includegraphics
   [width=85mm]
   {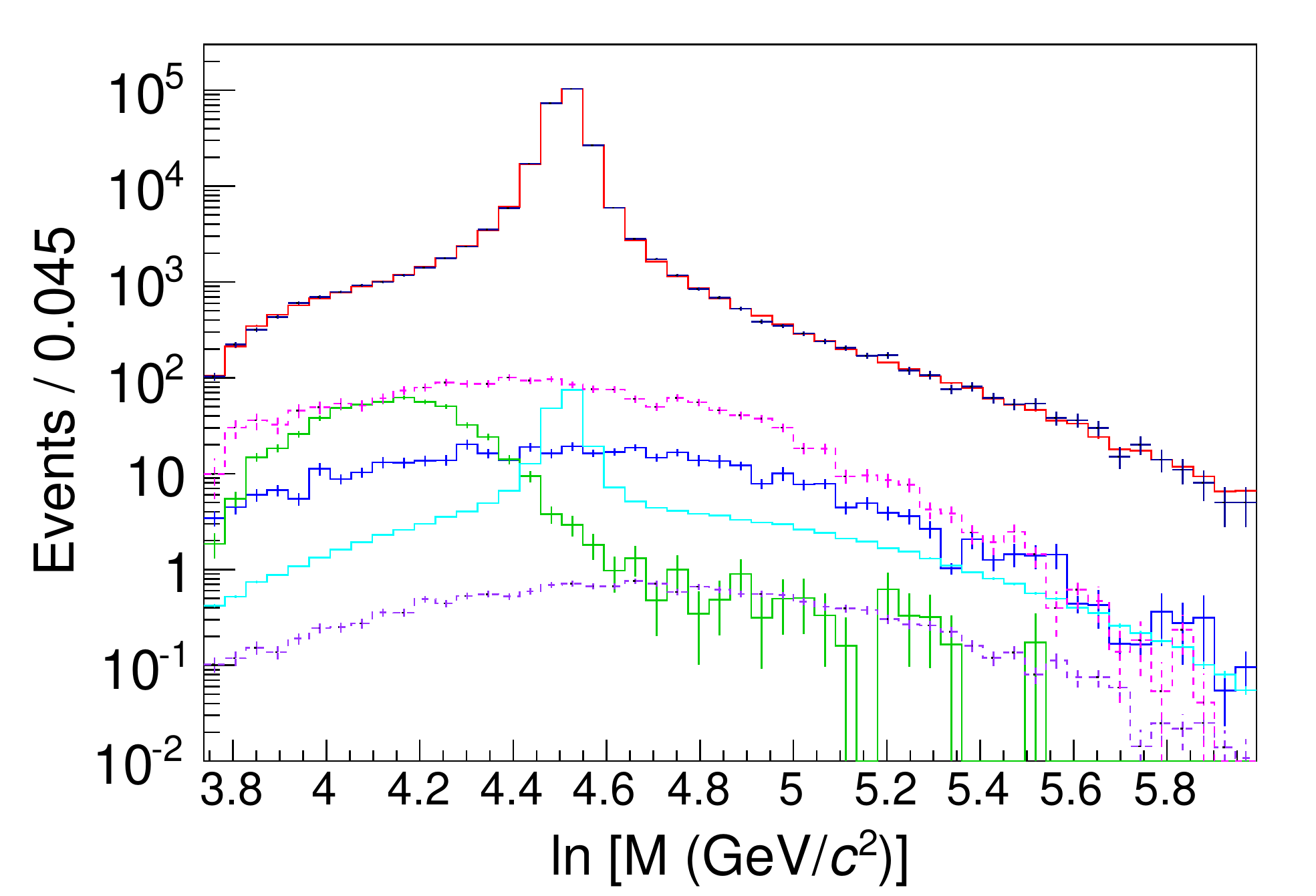}
\caption{\label{fig_bkgrCPee}
Logarithmically binned mass distributions for
CP-topology electron-pair candidates. The (black) crosses
are the data, the (red) histogram overlapping the data is the
sum of all components, the (green) histogram concentrated at
lower masses is the $Z \rightarrow \tau^+\tau^-$ component,
and the (cyan) histogram in the middle with the $Z$ peak is
the diboson component. The remaining broad distributions, from
top to bottom are:
QCD (magenta), $W+$jets (blue), and $t\bar{t}$ (purple).
The comparison of the data with the sum of the components yields
a $\chi^2$ of 50 for 50 bins.
}
\end{figure}
\begin{figure}
\includegraphics
   [width=85mm]
   {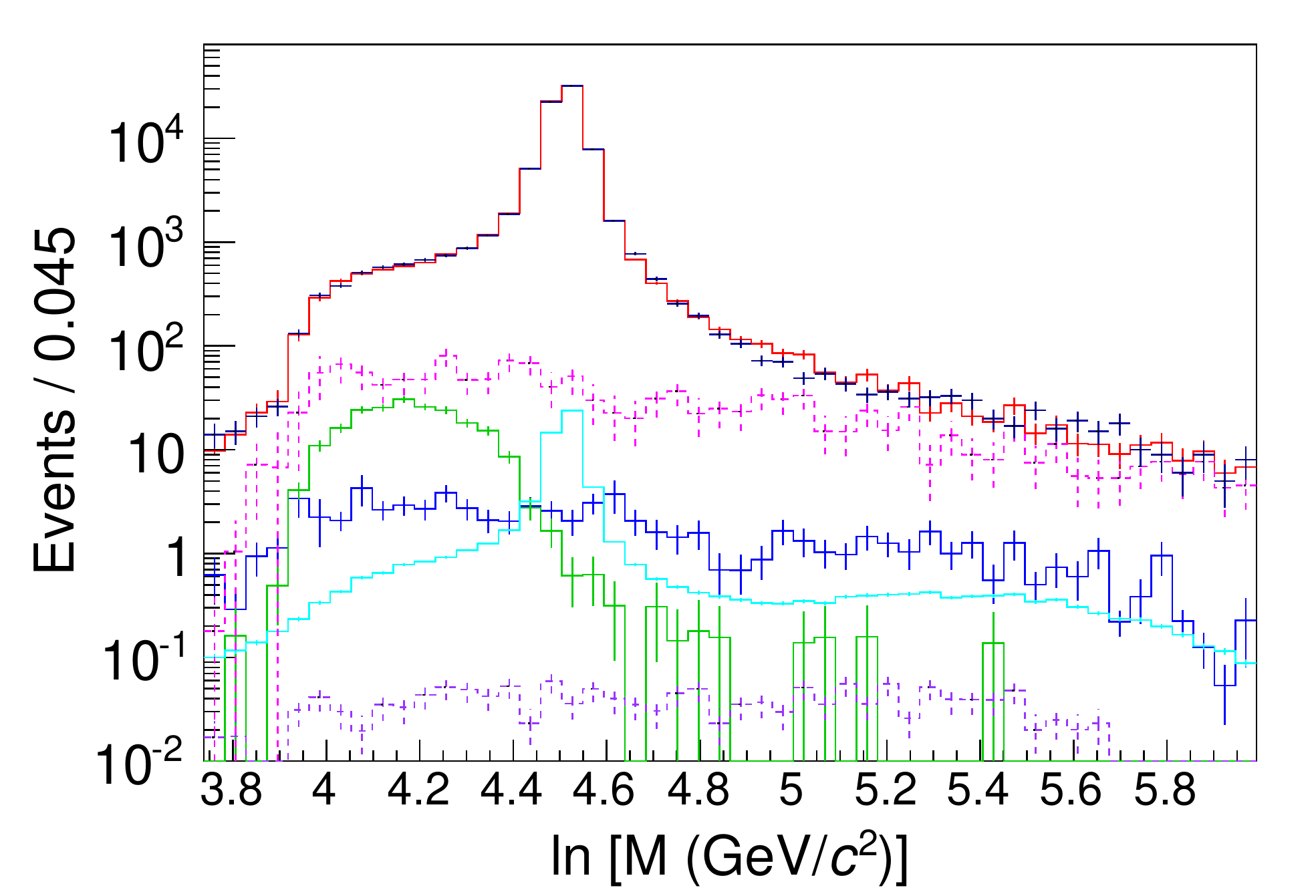}
\caption{\label{fig_bkgrPPee}
Logarithmically binned mass distributions for
PP-topology $ee$-pair candidates. The (black) crosses
are the data, the (red) histogram overlapping the data is the
sum of all components, the (green) histogram concentrated at
lower masses is the $Z \rightarrow \tau^+\tau^-$ component,
and the (cyan) histogram in the middle with the $Z$-peak is
the diboson component. The remaining broad distributions, from
top to bottom are:
QCD (magenta), $W+$jets (blue), and $t\bar{t}$ (purple).
The comparison of the data with the sum of the components yields
a $\chi^2$ of 69 for 50 bins.
}
\end{figure}

For the CP and PP topologies, the signal samples consist of both
same- and opposite-charge electron pairs. Charge separation 
is not useful because of the significant charge misidentification
rate for electrons in the plug region.
The largest background in each topology is from
QCD. However, the sum of all backgrounds is still small in relation
to the signal. If all backgrounds are allowed to vary in the fits,
the minimizations are underconstrained. Consequently,
the simulated backgrounds are constrained to their default
normalizations, and only the levels and shapes of the QCD
backgrounds are varied. The shape of the QCD background
for each topology is parametrized with an asymmetric-Gaussian
function that consists of two piecewise continuous
Gaussians joined at their common mean but with different widths.
One of the function parameters is empirically tuned in the high- or
low-mass region. As these regions have the largest level of
backgrounds, it is important to control the fit within these
regions. For the CP topology, the width on the
high-mass side is first optimized in the region
$M > 127$ GeV/$c^2$, and then the backgrounds
and simulated signal are fit to the data. For the PP topology,
the mean of the asymmetric-Gaussian is first optimized in
the low-mass region in the vicinity of the mass threshold,
and then the backgrounds and
simulated signal are fit to the data. The CP- and PP-topology
mass distributions for the data, the simulated data plus
backgrounds, and the backgrounds are shown in
Figs.~\ref{fig_bkgrCPee} and \ref{fig_bkgrPPee}, respectively.

\par
The CC-, CP-, and PP-topology samples contain approximately
\mbox{227~000}, \mbox{258~000}, and \mbox{80~000} events,
respectively, within the 42--400 GeV/$c^2$ mass region.
Table~\ref{tblBkgrFrac} summarizes the overall background levels
for these samples.
\begin{table}
\caption{\label{tblBkgrFrac}
Background fractions within the 42--400 GeV/$c^2$ mass region.
The values with uncertainties are derived from the data.
}
\begin{ruledtabular}
\begin{tabular}{cccc}
Component     & \multicolumn{3}{c}{\text{Background fraction (\%)}}  \\
              & \multicolumn{1}{c}{\text{CC}} &
	        \multicolumn{1}{c}{\text{CP}} & 
	        \multicolumn{1}{c}{\text{PP}}         \\ \hline
QCD           & 0.55$\pm$0.03 &  0.69 $\pm$ 0.13 & 1.64 $\pm$ 0.28  \\
$Z\rightarrow \tau\tau$
              & 0.26$\pm$0.06 &  0.21      &   0.27   \\
$W+$jets      &   0.13      &    0.16      &   0.10   \\
Diboson       &   0.14      &    0.10      &   0.08   \\
$t\bar{t}$    &   0.02      &    0.01      &   0.01   \\
\end{tabular}
\end{ruledtabular}
\end{table}
The total backgrounds for CC-, CP-, and PP-topology samples are
1.1\%, 1.2\%, and 2.1\%, respectively. For the CC- and CP-topology
samples shown in Figs.~\ref{fig_bkgrCCos} and \ref{fig_bkgrCPee}
respectively, the background fractions in the vicinity of the
$Z$-pole mass are small, but away from the pole mass, the levels
are larger and range from about 0.1\% to about 10\%.

\section{\label{AfbexpInput}\boldmath
The $A_{\rm fb}$ Measurement}\unboldmath

The Collins-Soper frame angle,
$\cos \vartheta$~\cite{CollinsSoperFrame}, is reconstructed
using the following laboratory-frame quantities: the lepton energies,
the lepton momenta along the beam line, the
dilepton invariant mass, and the dilepton transverse momentum.
The angle of the negatively charged lepton is
\begin{displaymath}
  \cos \vartheta = \frac{ l^-_+l^+_- - l^-_-l^+_+ }
                 { M \sqrt{M^2 + P_{\rm T}^2} }  \; ,
\end{displaymath}
where $l_\pm = (E \pm P_z)$ and
the $+$ $(-)$ superscript specifies that $l_\pm$ is for
the positively (negatively) charged lepton. Similarly, the
Collins-Soper expression for $\varphi$ in terms of
laboratory-frame quantities is
\begin{displaymath}
  \tan \varphi = \frac{\sqrt{M^2 + P_{\rm T}^2}}{M} \;
	\frac{\vec{\Delta} \cdot \widehat{R}_{\rm T}}
	     {\vec{\Delta} \cdot \widehat{P}_{\rm T}} \: ,
\end{displaymath}
where $\vec{\Delta}$ is the difference between the $\ell^-$ and
$\ell^+$ momentum vectors; $\widehat{R}_{\rm T}$ is the
transverse unit vector along $\vec{P}_p \times \vec{P}$, with
$\vec{P}_p$ being the proton momentum vector and $\vec{P}$ the
lepton-pair momentum vector; and $\widehat{P}_{\rm T}$ is the
unit vector along the transverse component of the lepton-pair
momentum vector. At $P_{\rm T} = 0$, the angular distribution
is azimuthally symmetric. The defintions of $\cos \vartheta$
and $\tan \varphi$ are invariant under Lorentz boosts along
the laboratory $z$ direction.
\par
$A_{\rm fb}$ is measured in 15 mass bins distributed over the
range $50 < M < 350$~GeV/$c^2$. This section details the
measurement method and presents the fully corrected measurement.
Section~\ref{AfbevtWtmethod} describes a newly developed
event-weighting technique.
Section \ref{AfbFinalCalib} describes final calibrations and
presents comparisons of the data and simulation.
Section \ref{AfbReslUnfold} describes the resolution-unfolding
technique and the corresponding covariance matrix of the
unfolded $A_{\rm fb}$ measurement.
Section \ref{AfbBiasCorr} describes the final corrections to
the measurement and presents the fully corrected measurement
of $A_{\rm fb}$.

\subsection{\label{AfbevtWtmethod}
Event-weighting method}

The forward-backward asymmetry $A_{\rm fb}$ of
Eq.~(\ref{eqnAfbDef}) is typically determined in terms of the
measured cross section $\sigma = N / ({\cal L}\,\epsilon A)$,
where $N$ is the number of observed events after background subtraction,
${\cal L}$ the integrated luminosity,
$\epsilon$ the reconstruction efficiency,
and $A$ the acceptance within the kinematic and fiducial
restrictions. The expression is
\begin{displaymath}
  A_{\rm fb} = \frac{ N^+/(\epsilon A)^+ - N^-/(\epsilon A)^- }
                    { N^+/(\epsilon A)^+ + N^-/(\epsilon A)^- } \: ,
\end{displaymath}
where the terms $N^{+(-)}$ and $(\epsilon A)^{+(-)}$ respectively
represent $N$ and $\epsilon A$ for $e^+e^-$ pairs with
$\cos \vartheta \geq 0$ ($\cos \vartheta < 0$), and the common
integrated luminosity is factored out. Systematic uncertainties
common to $(\epsilon A)^+$ and $(\epsilon A)^-$ cancel out.
\par
The asymmetry in this analysis is measured using the
{\it event-weighting} method~\cite{evtwtAFBmethod}, which is
equivalent to measurements of $A_{\rm fb}$ in $|\cos \vartheta|$
bins with these simplifying assumptions:
$(\epsilon A)^+ = (\epsilon A)^-$ in each $|\cos \vartheta|$ bin,
and Eq.~(\ref{eqnAngDistr}) describes the angular distributions.
As the interchange of the charge labels of the electrons
transforms $\cos \vartheta$ to $-$$\cos \vartheta$, the detector
assumption is equivalent to the postulate of a charge-symmetric
detector for single electrons. For high $P_{\rm T}$ electrons with
the same momenta, regions of the detector with charge-asymmetric
acceptances and efficiencies are small.
Thus, to first order, the acceptance and efficiency cancel out
with the event-weighting method, and the small portions that do
not cancel out are later corrected with the simulation
(Sec.~\ref{AfbBiasCorr}).
\par
The measurement of $A_{\rm fb}$ within a $|\cos \vartheta|$ bin
only depends on the event counts $N^\pm$ and is
\begin{equation}
  A_{\rm fb}^\prime = \frac{N^+ - N^-}{N^+ + N^-} =
	\frac{8}{3} A_{\rm fb}
	\left(  \frac{|\cos \vartheta|}
		     {1 + \cos^2 \vartheta + \cdots} \right) ,
\label{eqnAfbBinned}
\end{equation}
where $1 + \cos^2 \vartheta + \cdots$ denotes symmetric terms
in Eq.~(\ref{eqnAngDistr}). The event difference is proportional
to $2 A_4 |\cos \vartheta|$, and the event sum
to $2 (1 + \cos^2 \vartheta + \cdots)$.
Each bin is an independent measurement of $\frac{8}{3} A_{\rm fb}$,
or equivalently, $A_4$, with an uncertainty of $\sigma^\prime / \xi$,
where $\sigma^\prime$ is the statistical uncertainty for
$A_{\rm fb}^\prime$, and $\xi$ the angular factor in the
parentheses of Eq.~(\ref{eqnAfbBinned}). When the measurements
are combined, the statistical weight of each bin is proportional
to $\xi^2$.
\par
The binned measurements are reformulated into
an unbinned, event-by-event weighted expression
\begin{equation}
  A_{\rm fb} = \frac{N_n^+ - N_n^-}{N_d^+ + N_d^-} \, .
\label{eqnAfbWeighted}
\end{equation}
The $N_n^\pm$ and $N_d^\pm $ terms represent weighted event
counts, and the subscripts $n$ and $d$ signify the numerator
and denominator sums, respectively, which contain the same
events but with different event weights. Consider the $N^+$ and
$N^-$ events of the binned measurement of $A_{\rm fb}^\prime$
with a specific value of $|\cos \vartheta|$. In the unbinned
measurement, their numerator and denominator weights contain:
1) factors to cancel the angular dependencies of their
   event difference $(N^+ - N^-)$ and sum $(N^+ + N^-)$,
   respectively, and
2) the $\xi^2$ factor for the statistical combination of these
   events with events from other angular regions.
The method is equivalent to using
a maximum-likelihood technique, and for an ideal detector the
statistical precision of $A_{\rm fb}$ is expected to be
about 20\% better relative to the direct counting method
\cite{evtwtAFBmethod}. However, detector resolution and limited
acceptance degrade the ideal gain.
\par
While the discussion of events weights illustrates an
asymmetry measurement, the event weights presented in
Ref.~\cite{evtwtAFBmethod} and used in this analysis are for
the measurement of the $A_4$ angular coefficient.
The numerator and
denominator event weights for the measurement of $A_4$ are
$0.5 \: |\cos \vartheta| / \omega^2$ and
$0.5 \cos^2 \vartheta / \omega^3$, respectively, where $\omega$
is the symmetric $1 + \cos^2 \vartheta + \cdots$ term of
Eq.~\ref{eqnAfbBinned}.
\par
The event weights are functions of the reconstructed kinematic
variables $\cos \vartheta$, $\varphi$, and the lepton-pair
variables $M$ and $P_{\rm T}$. Only the $A_0$ and $A_2$
terms of Eq.~(\ref{eqnAngDistr}) are used in the denominator
of the angular factor of Eq.~(\ref{eqnAfbBinned}), and the
angular coefficients are parametrized with
\begin{displaymath}
  A_0 = A_2 = \frac{ k P_{\rm T}^2 }
                   { k P_{\rm T}^2 + M^2 } \, ,
\end{displaymath}
where $k$ is a tuning factor for the $P_{\rm T}$ dependence
of the $A_0$ and $A_2$ coefficients. For this analysis,
$k = 1.65$, which is derived from a previous measurement
of angular coefficients~\cite{CSangcoef21}.
The exact form of these angular terms in the
event weights is not critical for $A_{\rm fb}$ because
the bulk of the events is at low boson $P_{\rm T}$.
\par
The background events are subtracted from the weighted event sums
on an event-by-event basis by assigning negative event weights when
combining with the event sums.
\par
The event-weighting method also does not compensate for the
smearing of kinematic variables due to the detector resolution,
and the restricted sampling of the asymmetry in kinematic
regions with limited acceptance.
Resolution-smearing effects are unfolded with the aid of the
simulation, and sampling limitations are separately
compensated.

\subsection{\label{AfbFinalCalib}
Final calibrations}

Relative to the expected asymmetry distribution
illustrated in Fig.~\ref{fig_loAfbVSmass},
the observed distribution is diluted by the
detector resolution and QED FSR. The dilution from the detector
resolution is visible in the vicinity of the $Z$-boson pole
mass. The dilution from QED FSR is more pronounced at low masses
because the rate of events produced in the vicinity of the
$Z$-boson pole mass that radiate and are reconstructed in this
low-mass region is more significant in relation to the intrinsic
production rate. Detector miscalibrations add further distortions.
All sources directly affect the electron-pair mass distributions
that are primary inputs to the $A_{\rm fb}(M)$ distribution.
The precision calibrations of both the data and simulation remove
the additional distortions. In conjunction, the data-driven
adjustments to the simulation remove differences between the data
and simulation that impact the fully corrected $A_{\rm fb}(M)$
measurement.
\par
The Collins-Soper $\cos\vartheta$ distribution for the
simulation is also adjusted to improve agreement with the data.
Only the symmetric part of the distribution is
adjusted. The adjustments, determined for six
electron-pair invariant mass bins whose boundaries are aligned
with those used in the measurement, are determined from the
ratios of the data-to-simulation $\cos\vartheta$ distributions. 
The ratios are projected onto the first five Legendre
polynomials: $\Sigma_{i=0}^{i=4} \; p_i P_i(\cos\vartheta)$,
where $p_i$ are projection coefficients and
$P_i(\cos\vartheta)$ are Legendre polynomials. The ratios are
normalized so that the event count in the mass bin matches that
of the data. The symmetric parts of the projections describe
the ratios well and are used as the adjustments. Separate
adjustments are applied to the CC- and CP-topology electron
pairs as event weights. The corrections are a few percent or
smaller in regions where the acceptance is large.
\par
\begin{figure}
\includegraphics
   [width=85mm]
   {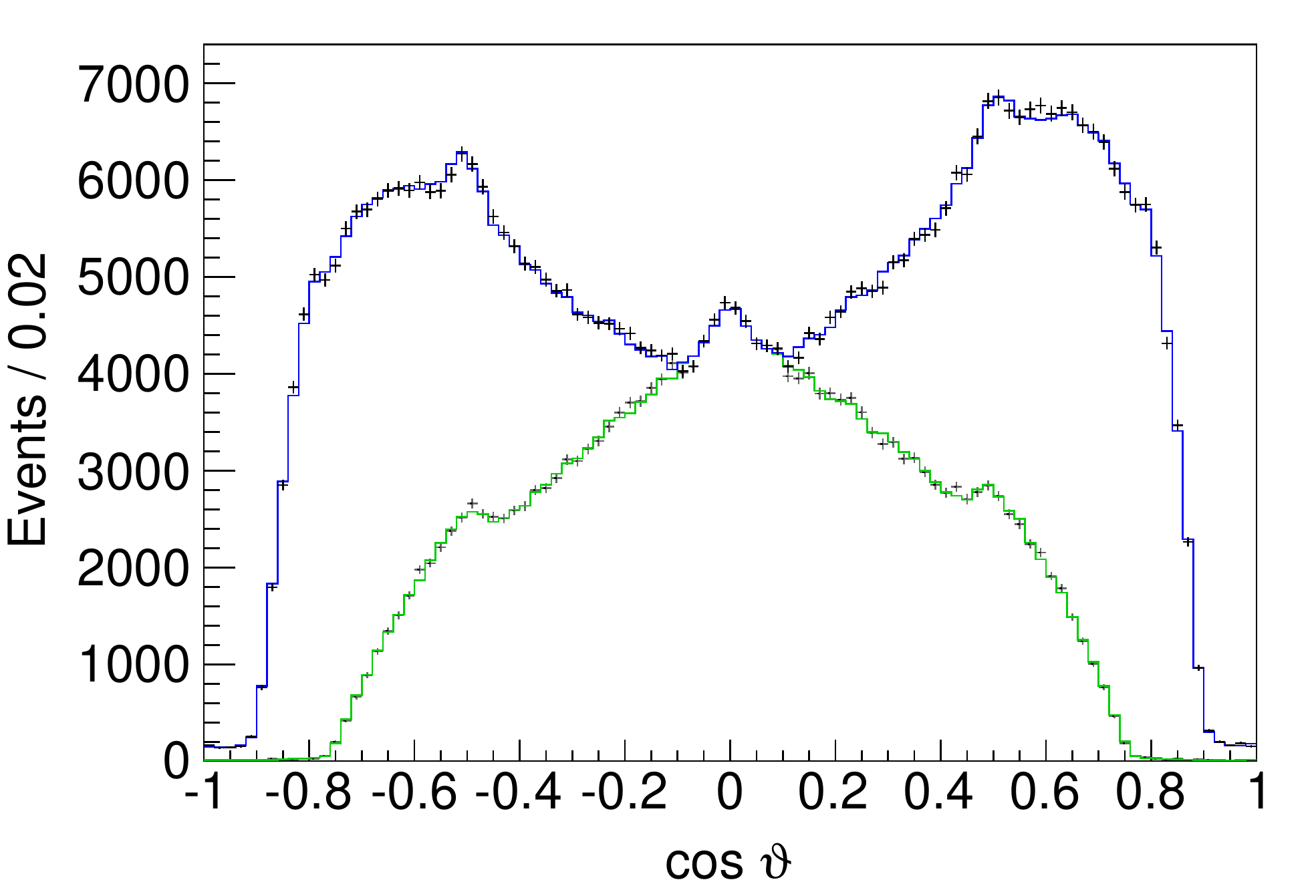}
\caption{\label{fig_tunedCoscs}
Distributions of $\cos \vartheta$ in the Collins-Soper frame
for dielectrons with $66 < M < 116$~GeV/$c^2$.
The crosses are the background-subtracted data and the
histograms are simulated data.
The upper pair of crosses and histogram is from the
combination of the CC and CP topologies, and the lower
pair is the contribution from the CC topology only.
}
\end{figure}

Figure~\ref{fig_tunedCoscs} shows the $\cos\vartheta$
distributions after all calibrations for the combination of the CC and
CP topologies and for the CC topology alone. The CP-topology dielectrons
are dominant at large $|\cos\vartheta|$ and significantly reduce the
statistical uncertainty of the measurement.
\begin{figure}
\includegraphics
   [width=85mm]
   {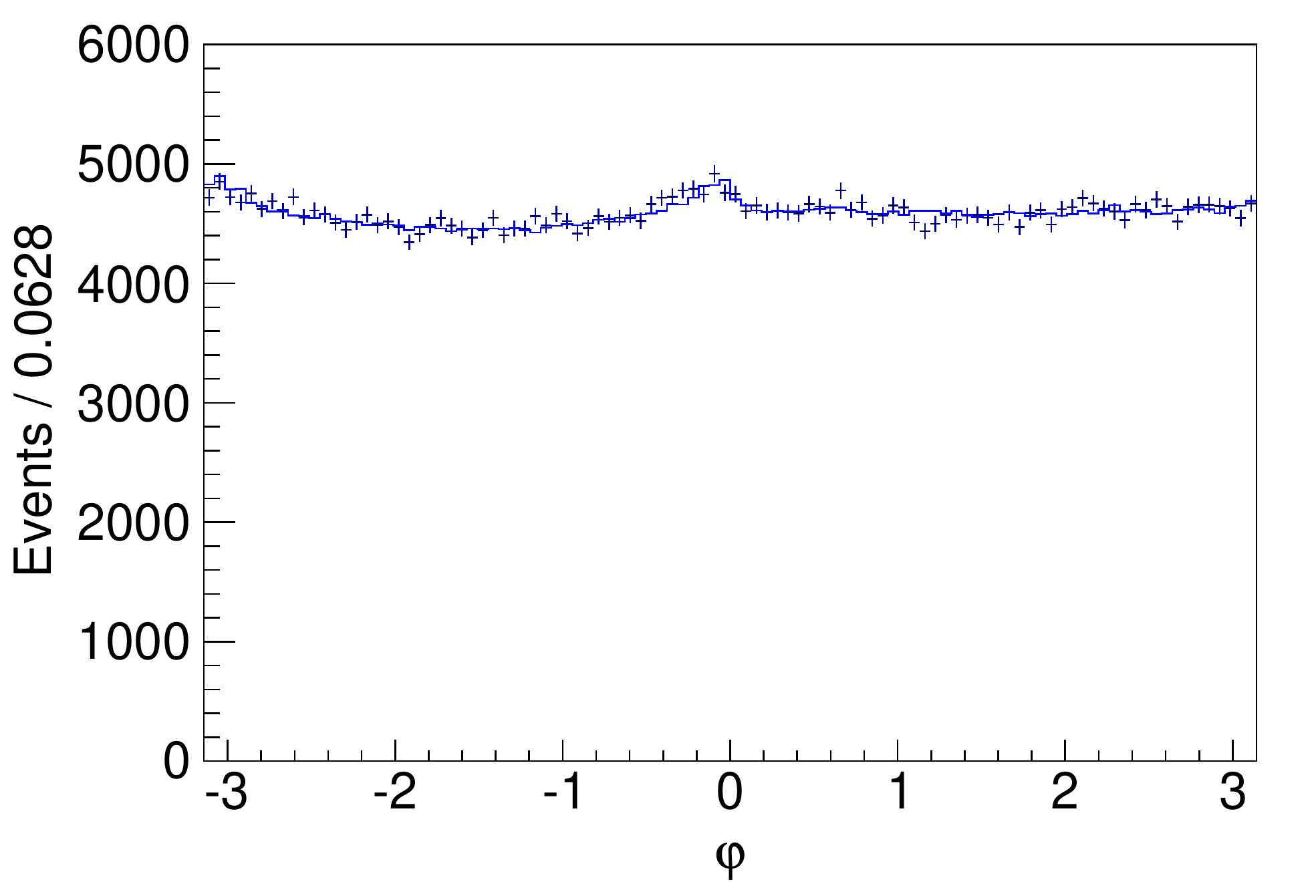}
\caption{\label{fig_defaultPhics}
Distibution of $\varphi$ in the Collins-Soper frame
for CC- and CP-topology dielectrons with
$66 < M < 116$~GeV/$c^2$. 
The crosses are the background-subtracted data and the
solid histogram is the simulation.
}
\end{figure}
Figure~\ref{fig_defaultPhics} shows the Collins-Soper
$\varphi$ distribution.

\par
\begin{figure}
\includegraphics
   [width=85mm]
   {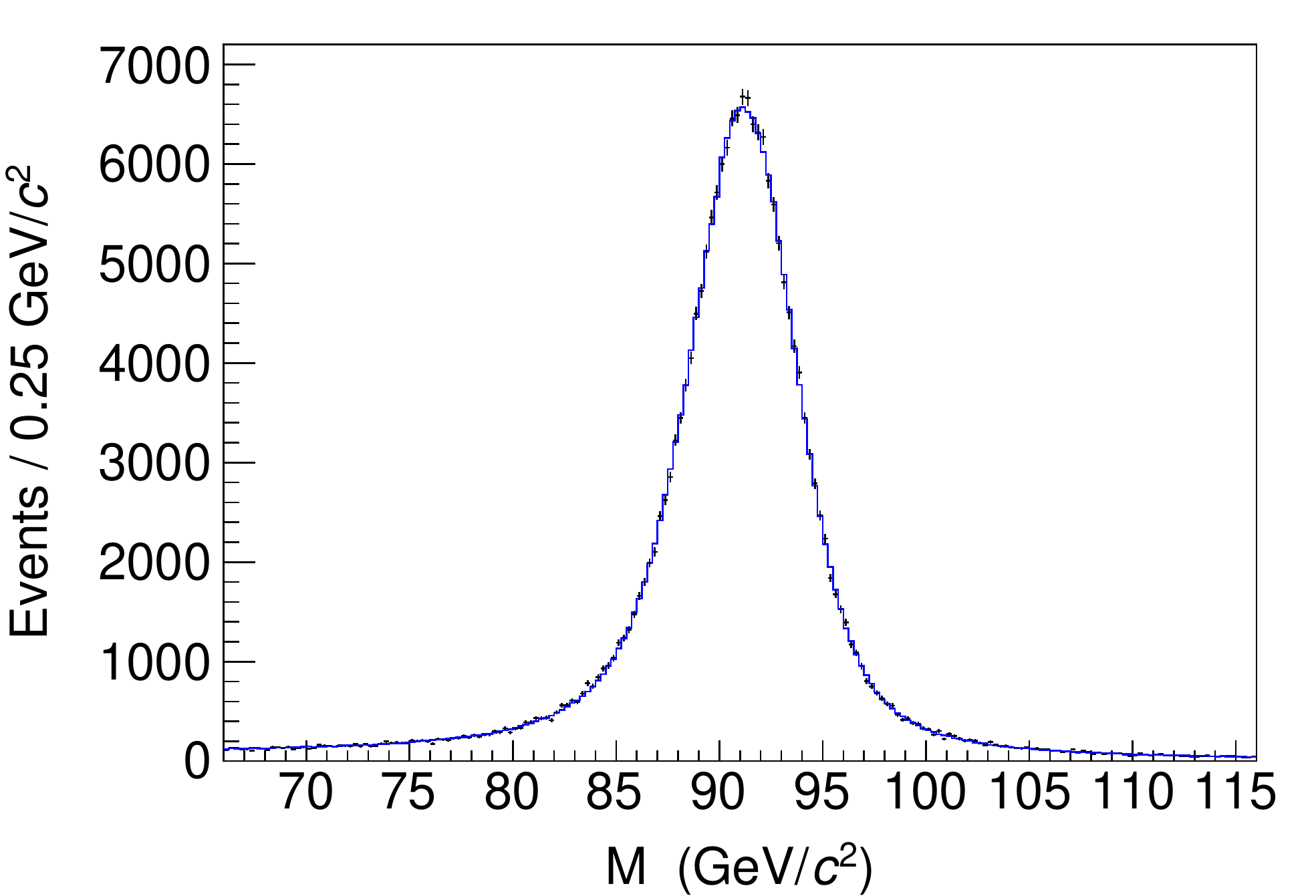}
\caption{\label{fig_mee1CCos}
Invariant $ee$-mass distribution for opposite-charged CC events.
The crosses are the background-subtracted data and the solid histogram
is the simulation. The comparison of the simulation with the
data yields a $\chi^2$ of 214 for 200 bins.
}
\end{figure}
\begin{figure}
\includegraphics
   [width=85mm]
   {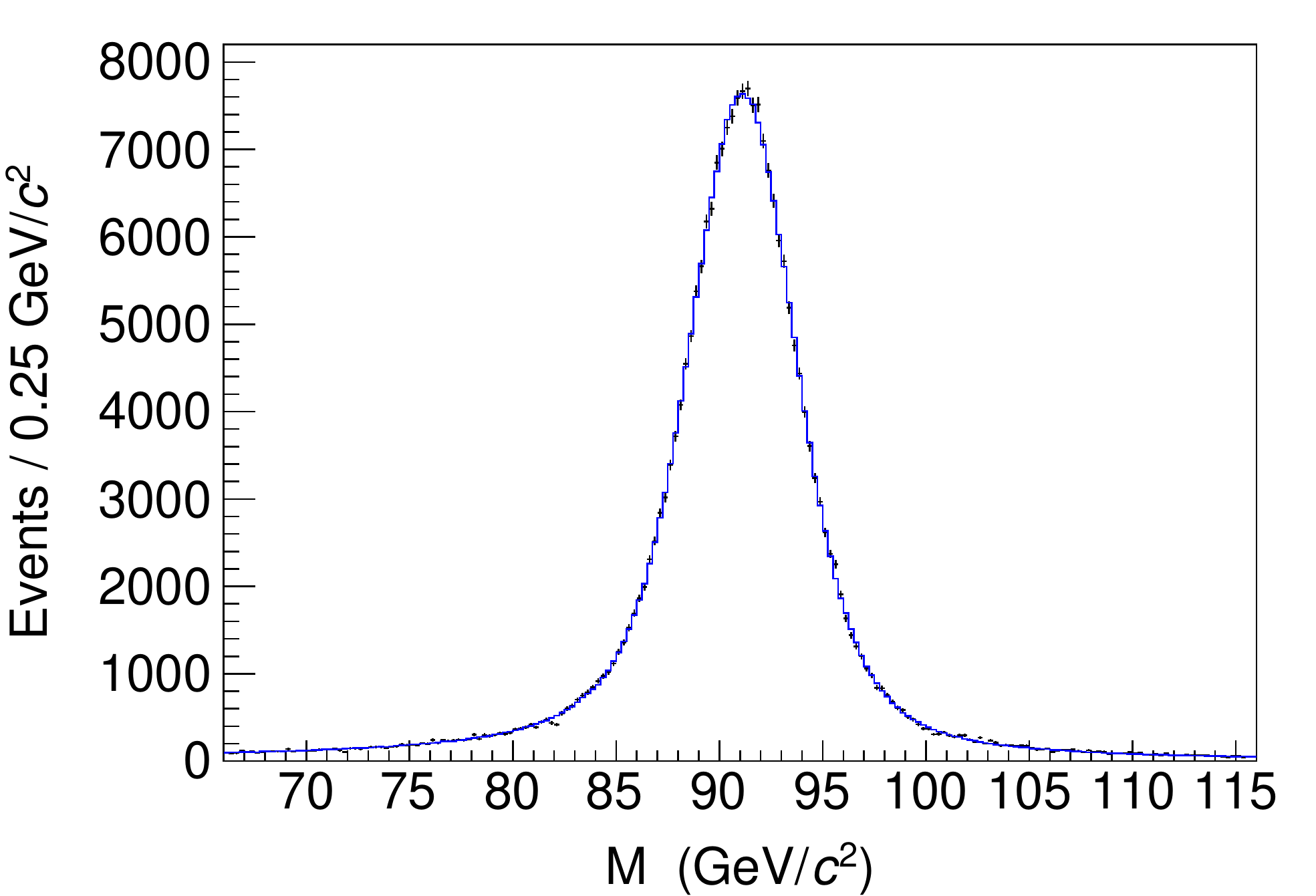}
\caption{\label{fig_mee1CP}
Invariant $ee$-mass distribution for CP events. The crosses
are the background-subtracted data and the solid histogram is
the simulation. The comparison of the simulation with the data
yields a $\chi^2$ of 235 for 200 bins.
}
\end{figure}

The CC- and CP-topology electron-pair
mass distributions in the range of 66--116~GeV/$c^2$ are shown
in Figs. \ref{fig_mee1CCos} and \ref{fig_mee1CP}, respectively.
For PP-topology electron pairs with masses in the same range,
the comparison of the simulation with the data yields a
$\chi^2$ of 232 for 200 bins.

\par
The electron $E_{\rm T}$ distributions of the data are
reasonably well described by the simulation.
\begin{figure}
\includegraphics
   [width=85mm]
   {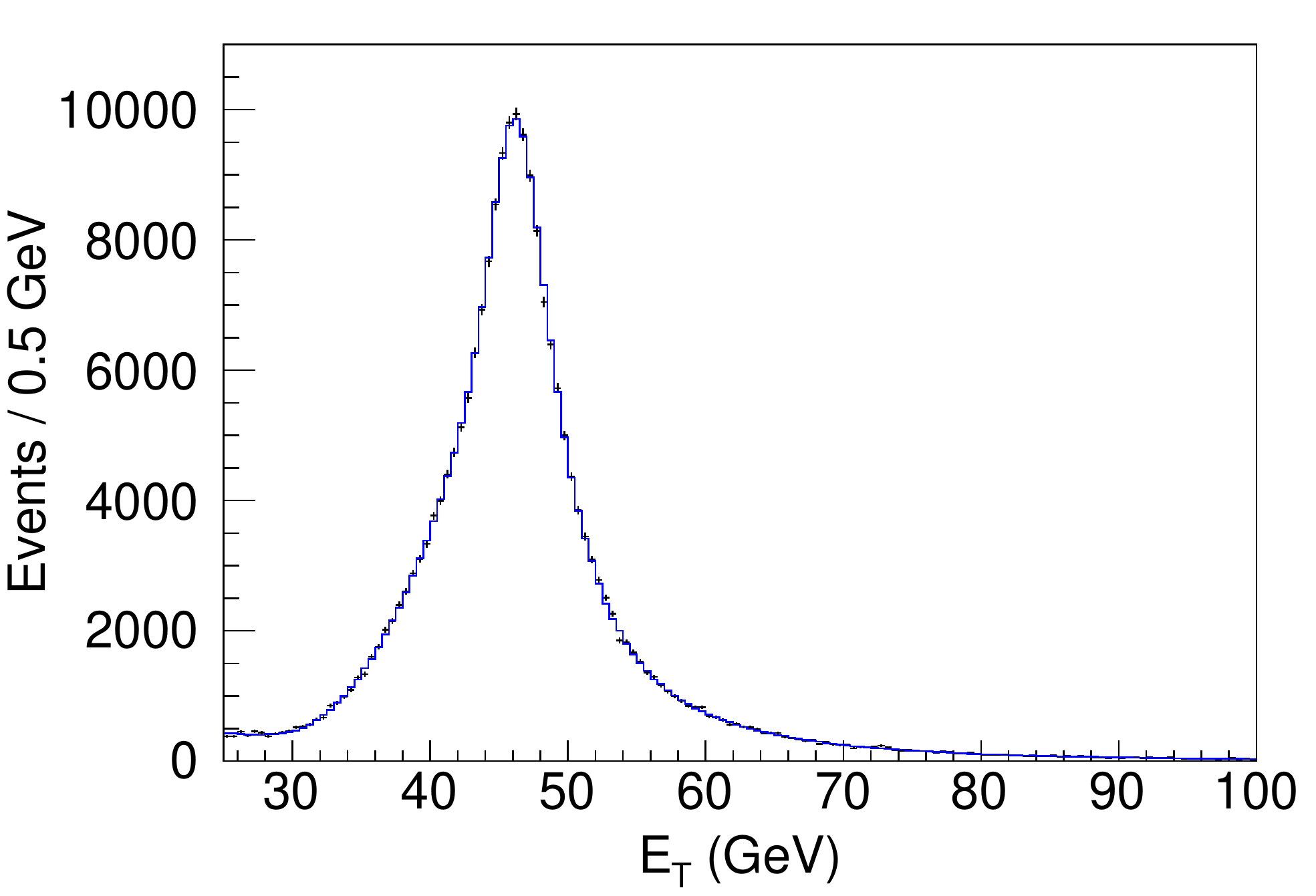}
\caption{\label{fig_erecEt0CC}
$E_{\rm T}$ distribution for the CC-topology electron
with the larger $E_{\rm T}$. The crosses
are the background-subtracted data and the solid
histogram is the simulation.
}
\end{figure}
Figure~\ref{fig_erecEt0CC} shows the $E_{\rm T}$ distribution of the
electron with the higher $E_{\rm T}$ for CC-topology dielectrons for
both the data and the simulation.
\begin{figure}
\includegraphics
   [width=85mm]
   {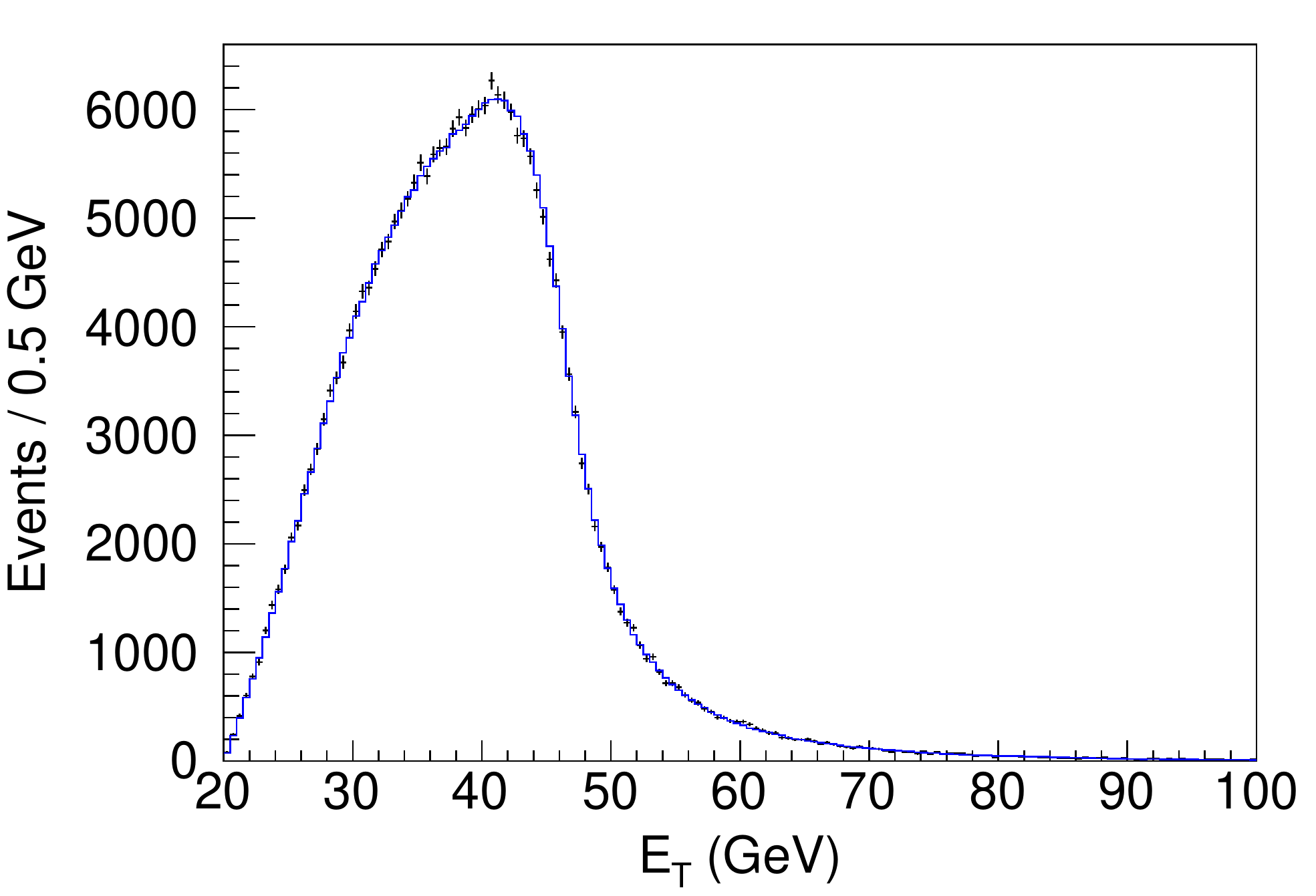}
\caption{\label{fig_erecEt0CP}
$E_{\rm T}$ distribution for the CP-topology electron
with the larger $E_{\rm T}$. The crosses
are the background-subtracted data and the solid
histogram is the simulation.
}
\end{figure}
Figure~\ref{fig_erecEt0CP} shows the equivalent distribution
for CP-topology electrons; here the electron can either
be the central or plug electron.

\par
\begin{figure}
\includegraphics
   [width=85mm]
   {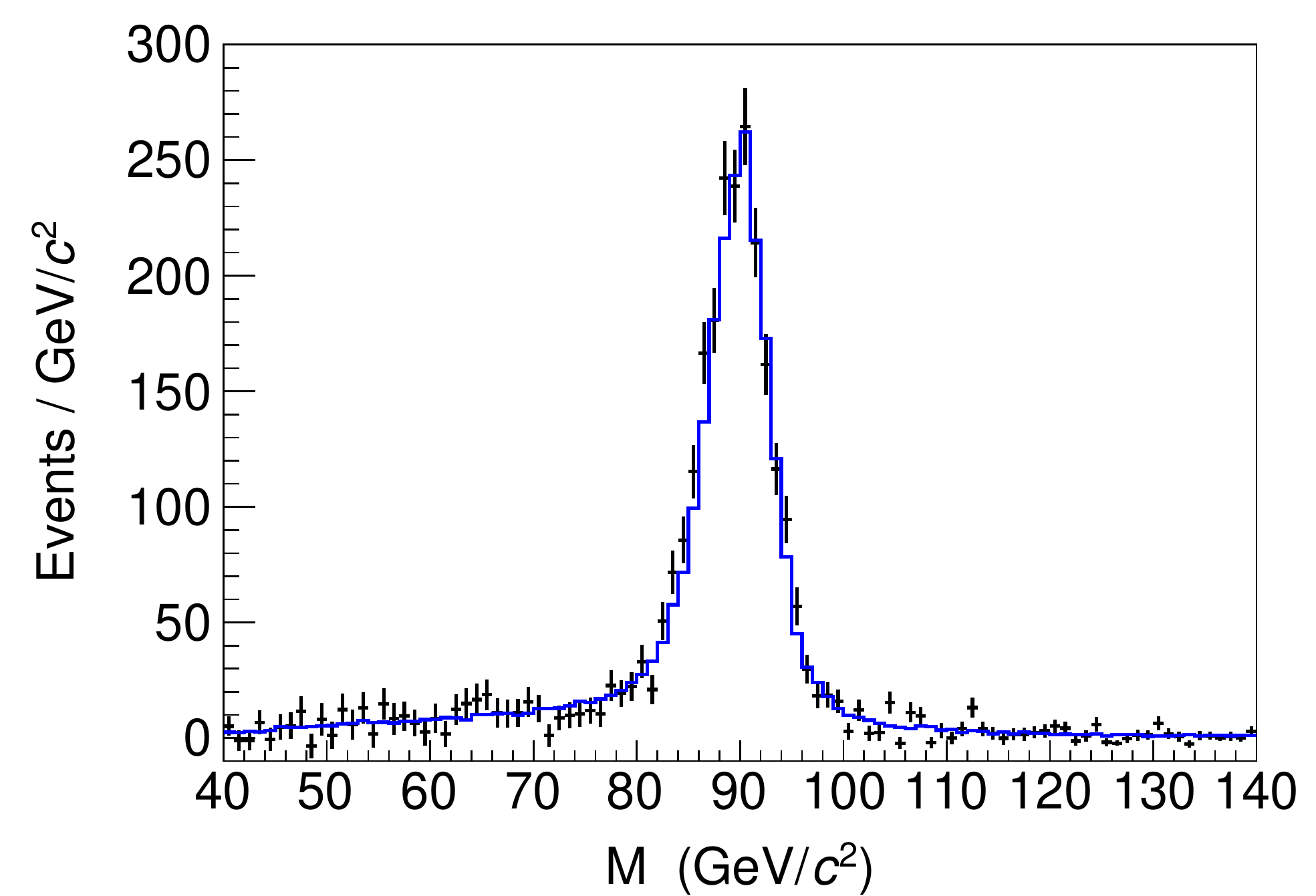}
\caption{\label{fig_meeCCss}
Invariant $ee$-mass distribution for same-charge CC events.
The crosses are the background-subtracted data and the solid
histogram is the simulation.
}
\end{figure}

The mass distribution of CC same-charge dielectrons has a
clear $Z$-boson peak from charge misidentification.
Figure~\ref{fig_meeCCss} shows the CC same-charge mass
distribution of the data and the simulation. This figure
confirms that charge misidentification
is reproduced well by the detector simulation. The
misidentification rate per central electron is 0.6\%.
Charge misidentification on the central electron of
CP pairs is thus expected to be small and properly
simulated.

\par
Charge misidentification, other categories of event
misreconstruction, and detector resolution affect
the observed value of $\cos\vartheta$. The bias of the
observed value, $\Delta \cos\vartheta$, obtained from the
simulation, is defined as the difference between its true value
prior to the application of QED FSR and the observed value.
The measurement of $A_{\rm fb}$ is in turn biased by the
fraction of events for which the sign of
the observed $\cos\vartheta$ differs from the true value;
this change of sign is denoted by sign-reversed $\cos\vartheta$.
The bias distribution consists of a narrow central core of
well reconstructed events, and a very broad distribution from
events where the electron kinematic properties are poorly reconstructed.
Charge misidentification reverses the sign of $\cos\vartheta$.
If events with charge misidentification are excluded, the
bias distributions of CC- and CP-topology events have
narrow non-Gaussian central cores centered at zero with
95\% of the events being contained within the range
$|\Delta \cos\vartheta| < 0.006$. 
For opposite-charge CC-topology events, the effects of detector
resolution dominate the bias. The fraction of events with
sign-reversed $\cos\vartheta$ is 0.3\%, with most of the events
being within the range $|\cos\vartheta|<0.1$.
For CP-topology events, the misidentification of the
central-electron charge dominates the fraction of events
with sign-reversed $\cos\vartheta$. The fraction decreases
in value from 0.6\% to 0.2\% as $|\cos\vartheta|$ increases
from 0.2 to 0.8. The measurement resolution of $\cos\vartheta$
consists of multiple components but their effects are small.

\par
\begin{figure}
\includegraphics
   [width=85mm]
   {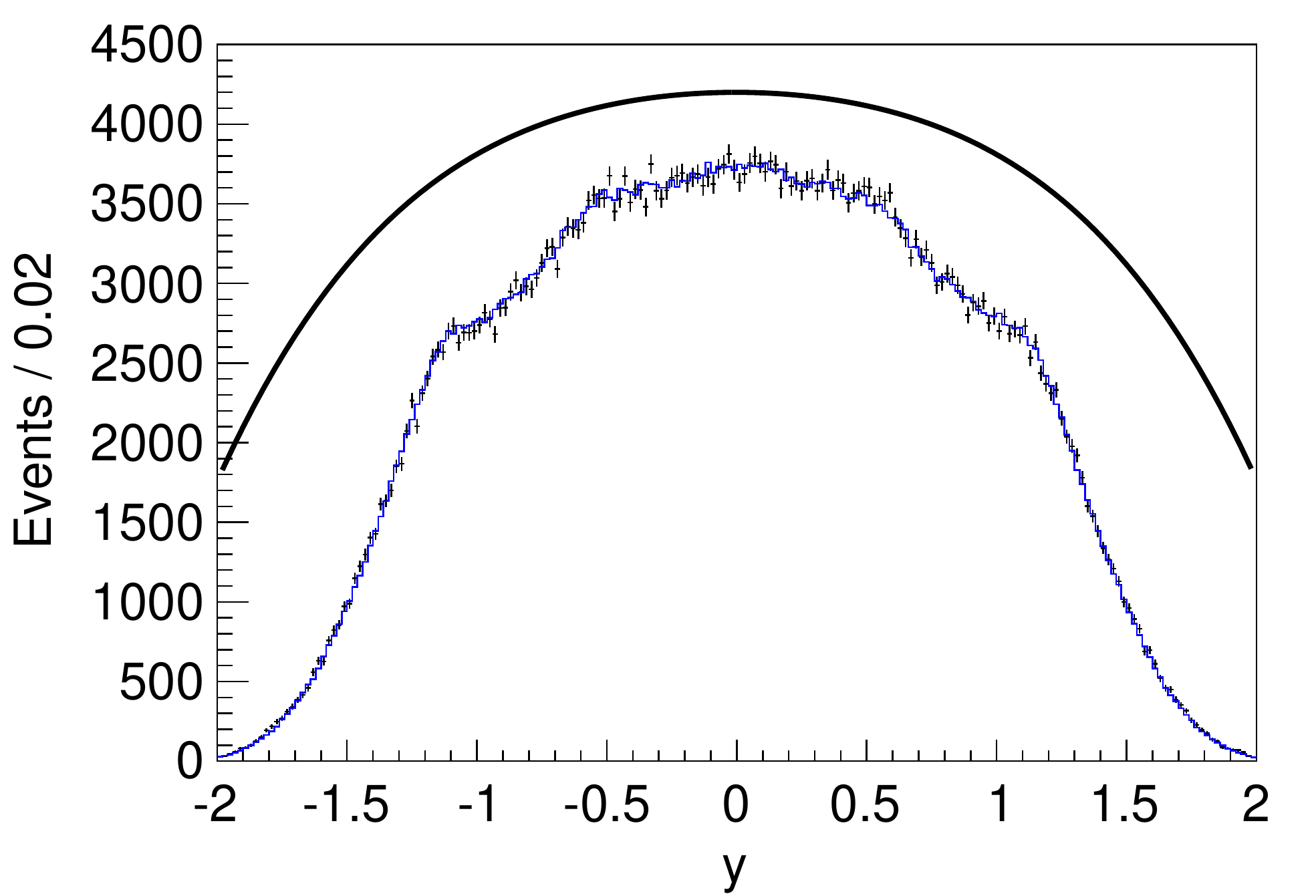}
\caption{\label{fig_eeYdist}
Rapidity distribution of electron pairs from the CC and CP
topologies with $66 < M < 116$ GeV/$c^2$. 
The crosses are the background-subtracted data and the histogram
is the simulation. The upper curve is the (arbitrarily normalized)
shape of the underlying rapidity distribution from \textsc{pythia}.
The measurement of $A_{\rm fb}$ is restricted to be within the
region $|y| < 1.7$.
}
\end{figure}

The rapidity distribution of electron pairs for the asymmetry
measurement is shown in Fig.~\ref{fig_eeYdist}, along with the
shape the underlying rapidity distribution from \textsc{pythia}.
At large values of $|y|$, the detector acceptance is significantly
reduced. For increasing values of $|y|$ in the $|y| \gtrsim 1$
region, the asymmetry slowly changes. This change
can only be tracked by event-weighting method if it has the
events to do so.
Consequently, the measurement of $A_{\rm fb}$ is restricted
to the kinematic region $|y| < 1.7$. QCD calculations
of $A_{\rm fb}$ used for comparisons with the measurement
are similarly restricted.
The electron-pair mass range of the measurement, 50 to 350~GeV/$c^2$,
corresponds to maximum $|y|$ values of 3.7 to 1.7, respectively.

\subsection{\label{AfbReslUnfold}
Resolution unfolding}

\begin{figure}
\includegraphics
   [width=85mm]
   {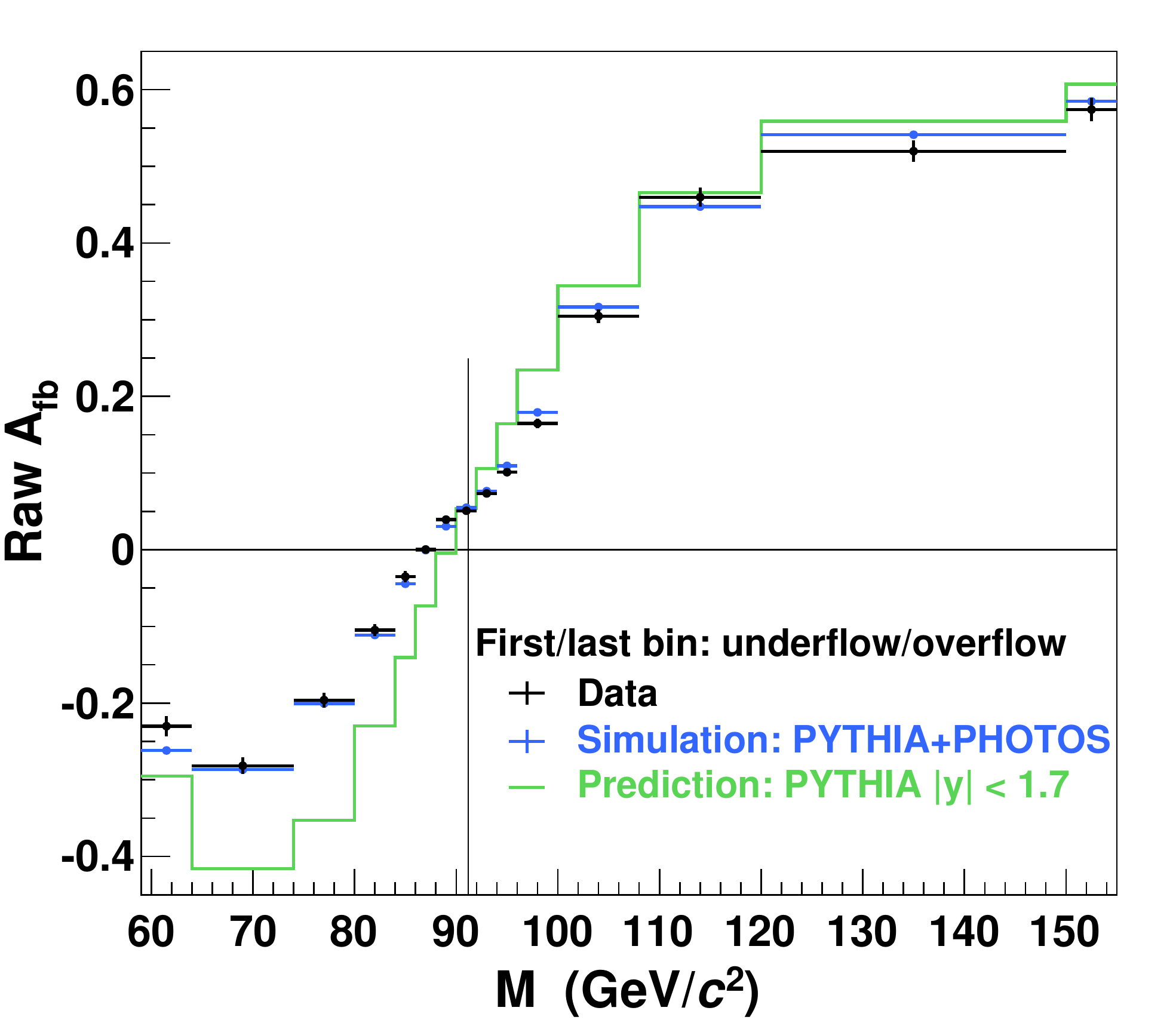}
\caption{\label{fig_eerawAfb}
Raw $A_{\rm fb}$ measurement in bins of the electron-pair
invariant mass. The vertical line is at $M = M_Z$.
Only statistical uncertainties are shown.
The \textsc{pythia} prediction for $|y|<1.7$ does not
include the effects of QED FSR.
}
\end{figure}

After applying the calibrations and corrections to the experimental
and simulated data, the asymmetry is measured in 15 bins
of the electron-pair invariant mass. The bin boundaries are
50, 64, 74, 80, 84, 86, 88, 90, 92, 94, 96, 100, 108, 120, 150,
and 350 GeV/$c^2$.
The 50--64 and 150--350 GeV/$c^2$ bins are referenced in plots as
the underflow and overflow bins, respectively, because
they include candidates reconstructed with masses outside the
range of the plot.
This measurement, denoted as raw because the effects of the
detector resolution and final-state QED radiation are not removed,
is shown in Fig.~\ref{fig_eerawAfb}.
As the mass resolution smearing of the calorimeter in the vicinity
of the $Z$-boson mass has an rms of approximately 2~GeV/$c^2$, the
calibrations and tuning of the data and simulation are important
for the resolution unfolding.
\par
The CC and CP events have different geometries and resolutions so
they are kept separate in the event-weighting phase and the
unfolding phase. They are combined for the $A_{\rm fb}$ measurement
and calculation of the measurement covariance matrix. 
\par
The unfolding of the resolution and QED FSR uses the event transfer
matrices from the simulation, denoted by $\bar{n}_{gr}$.
All data-driven corrections to the simulation are included.
The symbol $\bar{n}_{gr}$ identifies the
number of selected events that are generated in the electron-pair
$(M, \cos\vartheta)$ bin $g$ and reconstructed in the
$(M, \cos\vartheta)$ bin $r$. In addition to the 15 mass bins, the
forward-backward asymmetry has two angular regions,
$\cos\vartheta \geq 0$, and $\cos\vartheta < 0$. 
Square transfer matrices for 30-element state vectors
are implemented. The first 15 elements of the vectors are the
mass bins for the $\cos\vartheta \geq 0$ angular region, and
the remainder for the other angular region.
\par
The simulation predicts significant bin-to-bin event migration
among the mass bins when the produced and reconstructed values
of $\cos\vartheta$ have the same sign. For a mass bin, there
is very little migration of events from
one angular region to the other.  As the simulation sample size
is normalized to the integrated luminosity of the data, 
the transfer matrices provide properly
normalized estimates of event migration between bins. An
estimator for the true unfolding matrix is
$\bar{U}_{gr} = \bar{n}_{gr} / \bar{N}_r$,
where $\bar{N}_r = \sum_g \bar{n}_{gr}$ is the expected
total number of weighted events reconstructed in bin $r$.
The 30-element state vector for $\bar{N}_r$ is denoted as
$\vec{N}_r$, and the matrix $\bar{U}_{gr}$ by
$\mbox{\boldmath$U$\unboldmath}$. The estimate for the
resolution-unfolded state vector of produced events is
$\vec{N}_g = \mbox{\boldmath$U$\unboldmath} \cdot \vec{N}_r$.
\par
For the event-weighting method, there are two transfer matrices
that correspond to the weighted-event counts $N_n$ and $N_d$ of
Eq.~(\ref{eqnAfbWeighted}), and thus two separate unfolding 
matrices $\mbox{\boldmath$U$\unboldmath}$, two separate
event-weighted measurements of $\vec{N}_r$, and two separate
estimates of the resolution-unfolded $\vec{N}_g$. The CC- and
CP-event estimates of $\vec{N}_g$ for the numerator and
denominator of $A_{\rm fb}$ are summed prior to the evaluation
of $A_{\rm fb}$.  The measurements of $A_{\rm fb}$
for the 15 mass bins are collectively denoted by $\vec{A}_{\rm fb}$.
\par
The covariance matrix of the $A_{\rm fb}$ measurement,
denoted by \mbox{\boldmath$V$\unboldmath},
is calculated using the unfolding matrices, the expectation
values of $\vec{N}_r$ and $\vec{A}_{\rm fb}$ from the simulation,
and their fluctuations over an ensemble. The
per-experiment fluctuation to $\vec{N}_g$ is
$\mbox{\boldmath$U$\unboldmath} \cdot (\vec{N}_r + \delta\vec{N}_r)$,
where $\delta\vec{N}_r$ represents a fluctuation from the
expectation $\vec{N}_r$. The variation $\delta \vec{A}_{\rm fb}$
resulting from the $\vec{N}_g$ fluctuation is
ensemble-averaged to obtain the covariance matrix
\begin{displaymath}
           V_{lm} = \langle \,
                    (\delta \vec{A}_{\rm fb})_l
		    (\delta \vec{A}_{\rm fb})_m
		    \, \rangle \, ,
\end{displaymath}
where $(\delta \vec{A}_{\rm fb})_k$ ($k = l$ and $m$) denotes
the $k$th element of $\delta \vec{A}_{\rm fb}$.
Each element $i$ of $\vec{N}_r$ undergoes independent, normally
distributed fluctuations with a variance equal to the value
expected for $\bar{N}_i$. Because $\bar{N}_i$ is a sum of event
weights, fluctuations of $\bar{N}_i$ are quantified with the
variance of its event weights.
The two $\vec{N}_r$ vectors, the numerator vector and the
denominator vector, have correlations. Elements $i$ of the
numerator and denominator vectors contain the same events,
the only difference being that they have different event weights.
To include this correlation, the event-count variations of
elements $i$ of the numerator and denominator $\delta\vec{N}_r$
vectors are based on the same fluctuation from a normal
distribution with unit rms dispersion.
\par
The covariance matrix is expanded and inverted to
the error matrix using singular-value decomposition
methods. As the covariance matrix is a real-valued
symmetric
$15 \times 15$ matrix, its 15 eigenvalues and eigenvectors
are the rank-1 matrix components in the decomposition of the
covariance matrix and the error matrix
\begin{eqnarray}
\mbox{\boldmath$V$\unboldmath}      & = & \sum_n \lambda_n  \:
                      |v_n\rangle \langle v_n| 
				\; {\rm and} 
		      \nonumber \\ 
\mbox{\boldmath$V$\unboldmath}^{-1} & = & \sum_n \lambda_n^{-1} \:
                      |v_n\rangle \langle v_n| \,
		      \label{svdErrExp} , 
\end{eqnarray}
where $\lambda_n$ and $|v_n\rangle$ are the eigenvalues and
eigenvectors of \mbox{\boldmath$V$\unboldmath}, respectively,
and $|v_n\rangle \langle v_n|$ represents a vector projection
operator in the notation of Dirac bra-kets.
\par
The covariance matrix can have eigenvalues that are very small
relative to the largest eigenvalue. Their vector
projection operators select the fine structure of the resolution
model, and at a small enough eigenvalue, they become
particular to the simulation and include noise. While their
contribution to the covariance matrix is small, they completely
dominate the error matrix. The fine structures of the simulation,
measurement, and calculation are different. Consequently,
comparisons between the $A_{\rm fb}$ measurement and predictions
that use the error matrix are unstable. To alleviate these
instabilities, the decomposition of the error matrix,
Eq.~(\ref{svdErrExp}), is {\it regulated} so that the contributions
of eigenvectors with very small eigenvalues are suppressed.
A general method, as described below, is to add a regularization term
or function $r_n$ to the eigenvalues:
$\lambda_n \rightarrow \lambda_n + r_n$,
where $\lambda_n + r_n$ is the regularized eigenvalue. 

\subsection{\label{AfbBiasCorr}
Event-weighting bias correction}

After resolution unfolding, the event-weighted $A_{\rm fb}$
values have second-order acceptance and efficiency biases
from regions of limited boson acceptance,
and to a lesser extent, from detector nonuniformities resulting
in $(\epsilon A)^+ \neq (\epsilon A)^-$.
The bias is defined as the difference between the true value of
$A_{\rm fb}$ before QED FSR calculated with \textsc{pythia} and
the unfolded simulation estimate. The size of the simulation
sample is 21 times that of the data.
\begin{figure}
\includegraphics
   [width=85mm]
   {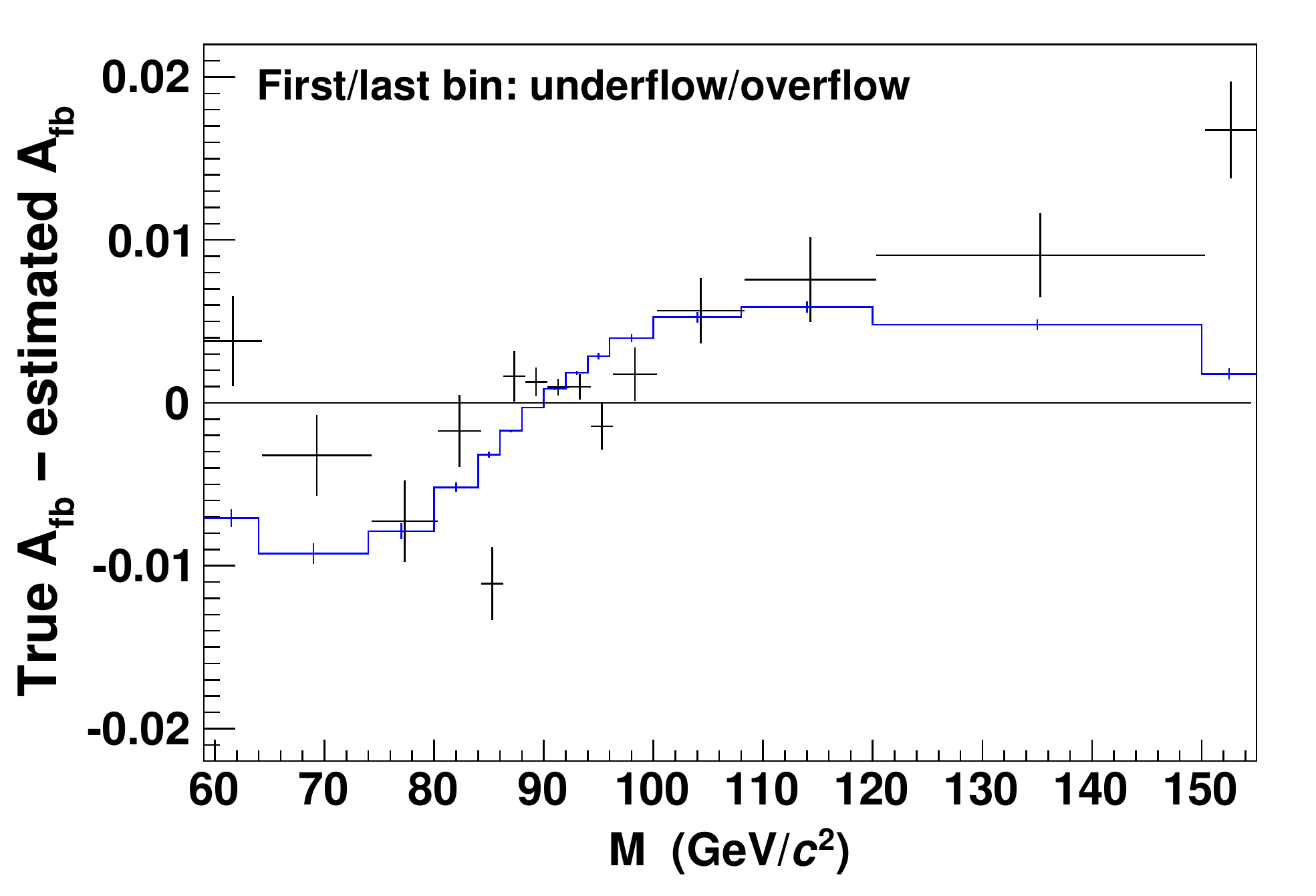}
\caption{\label{fig_afbEWbias}
Event-weighting bias in bins of the electron-pair invariant mass.
The biases are the crosses, and the uncertainties are the bin-by-bin
unfolding estimates of the simulation. The superimposed histogram is
the difference between the $A_{\rm fb}$ calculations for the rapidity
range $|y| < 1.7$ and $|y| < 1.5$, and the uncertainties are estimates
for the PDF uncertainty.
}
\end{figure}
The bias is a mass-bin-specific additive correction to the
unfolded $A_{\rm fb}$ measurement, and is shown in
Fig.~\ref{fig_afbEWbias}. All significant bias corrections
are less than 8\% of the magnitude of $A_{\rm fb}$ and most of them
are 3\% or less.
\par
Figure~\ref{fig_afbEWbias} also shows the
difference between asymmetries calculated with the measurement
rapidity range $|y| < 1.7$, and with a reduced range $|y| < 1.5$.
The difference is representative of contributions to the bias
from regions of reduced acceptance at large values of $|y|$,
and the PDF uncertainty of the
difference is specified later in Sec.~\ref{datSystUncerts}.
For increasing values of $|y|$, there is a relative increase of
the $u$- to $d$-quark flux and a decrease of the antiquark flux
from the proton.
\par
The covariance matrix of the bias-correction
uncertainties is combined with the covariance matrix for the
$A_{\rm fb}$ measurement.
\begin{figure}
\includegraphics
   [width=85mm]
   {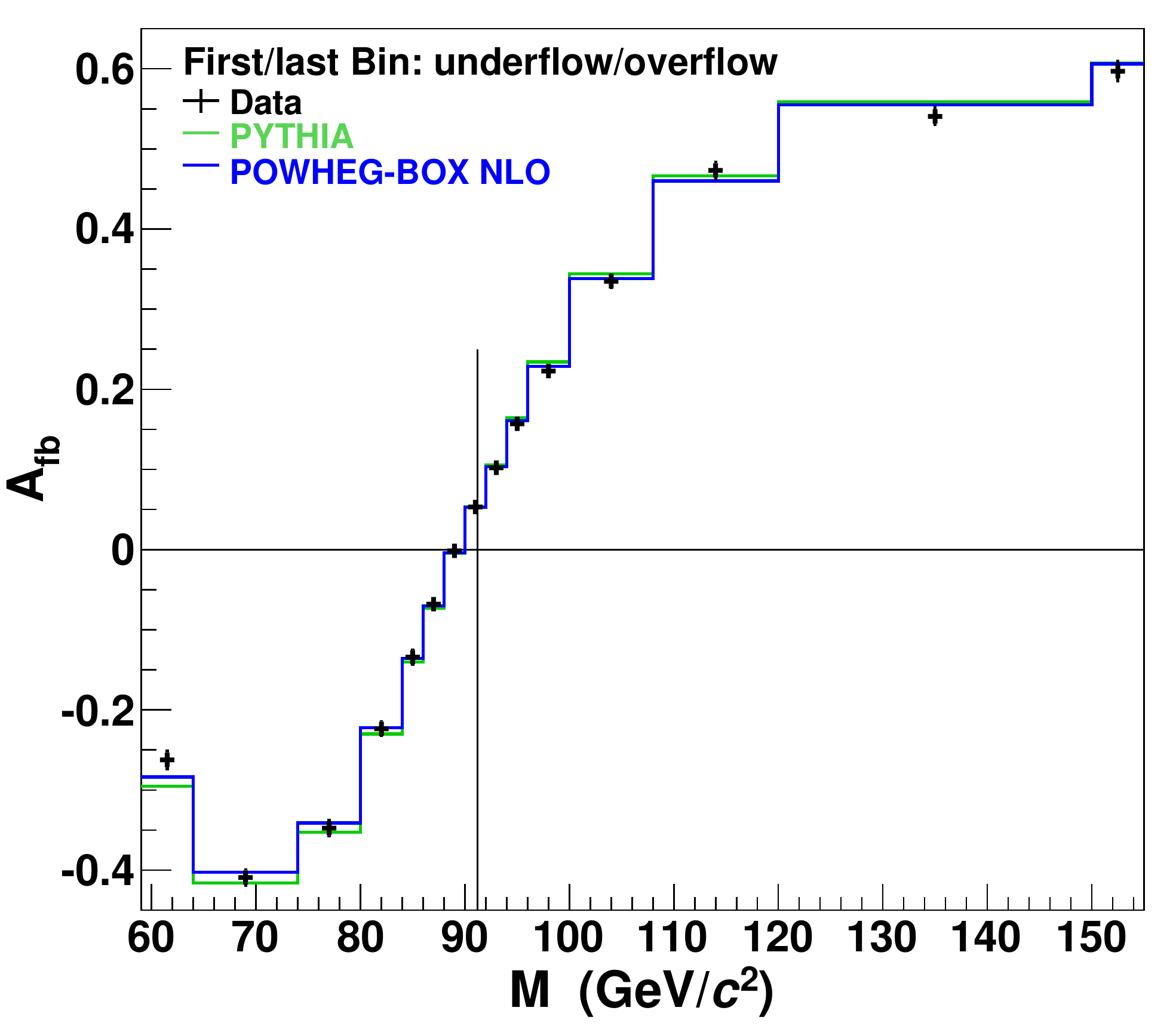}
\caption{\label{fig_correctedAfb}
Fully corrected $A_{\rm fb}$ for electron pairs with $|y| < 1.7$.
The measurement uncertainties are bin-by-bin unfolding estimates.
The vertical line is $M  = M_Z$. The \textsc{pythia} calculation uses
$\sin^2\theta^{\rm lept}_{\rm eff} = 0.232$.
The EBA-based \textsc{powheg-box} calculation uses
$\sin^2\theta_W = 0.2243$ 
$(\sin^2\theta^{\rm lept}_{\rm eff} = 0.2325)$ and
the default PDF of NNPDF-3.0.
}
\end{figure}
The fully corrected measurement of $A_{\rm fb}$, including
the bias correction, is shown in Fig.~\ref{fig_correctedAfb} 
\begin{table}
\caption{\label{tblCorrAfb}
Fully corrected $A_{\rm fb}$ measurement for electron pairs
with $|y| < 1.7$. The measurement uncertainties are
bin-by-bin unfolding estimates.
}
\begin{ruledtabular}
\begin{tabular}{cd}
Mass bin    & \multicolumn{1}{c}{$A_{\rm fb}$}   \\
(GeV/$c^2$) &                    \\ \hline
50--64    & -0.262 \pm 0.014 \\
64--74    & -0.409 \pm 0.015 \\
74--80    & -0.348 \pm 0.015 \\
80--84    & -0.224 \pm 0.014 \\
84--86    & -0.134 \pm 0.014 \\
86--88    & -0.068 \pm 0.010 \\
88--90    & -0.0015 \pm 0.0044 \\
90--92    & 0.0533 \pm 0.0017 \\
92--94    & 0.1021 \pm 0.0036 \\
94--96    & 0.1570 \pm 0.0087 \\
96--100   & 0.2228 \pm 0.0094 \\
100--108  & 0.335 \pm 0.010 \\
108--120  & 0.473 \pm 0.012 \\
120--150  & 0.541 \pm 0.012 \\
150--350  & 0.597 \pm 0.014 \\
\end{tabular}
\end{ruledtabular}
\end{table}
and tabulated in Table~\ref{tblCorrAfb}.

\section{\label{sw2scans}\boldmath
Extraction of $\sin^2\theta^{\rm lept}_{\rm eff}$}
\unboldmath

The Drell-Yan asymmetry measurement is directly sensitive to the
effective-mixing terms $\sin^2\theta_{\rm eff}$, which are
products of the form-factor functions with the static $\sin^2\theta_W$
parameter (Sec. \ref{EWKradcor}). The asymmetry is most sensitive
to the value of the effective-leptonic $\sin^2\theta_{\rm eff}$
term in the vicinity of the $Z$ pole,
or $\sin^2\theta^{\rm lept}_{\rm eff}$, and its value is derived
from the $\sin^2\theta_W$ parameter of the $A_{\rm fb}$ template
that best describes the measurement. For non-EBA calculations such
as \textsc{pythia}, the template parameter is
$\sin^2\theta^{\rm lept}_{\rm eff}$. While the value of
$\sin^2\theta^{\rm lept}_{\rm eff}$ is a direct measurement,
the interpretation of the corresponding value of $\sin^2\theta_W$
and the form factors are dependent on the details of the EBA model.
\par
The measurement and templates are compared using the $\chi^2$
statistic evaluated with the $A_{\rm fb}$ measurement error matrix.
Each template corresponds to a particular value of
$\sin^2\theta_W$ and provides a scan point for the $\chi^2$
function: $\chi^2( \sin^2\theta_W)$. The $\chi^2$ values of the scan
points are fit to a parabolic $\chi^2$ function,
\begin{equation}
  \chi^2(\sin^2\theta_W) = 
        \bar{\chi}^2 +
        (\sin^2\theta_W - \overline{\sin}^2\theta_W)^2 /
                          \bar{\sigma}^2  \, ,
\label{chiSqFit}
\end{equation}
where $\bar{\chi}^2$, $\overline{\sin}^2\theta_W$, and
$\bar{\sigma}$ are parameters. The $\overline{\sin}^2\theta_W$
parameter is the best-fit value of $\sin^2\theta_W$,
$\bar{\sigma}$ is the corresponding measurement uncertainty,
and $\bar{\chi}^2$ is the associated goodness-of-fit between
the $A_{\rm fb}$ measurement and calculation over the 15 mass
bins.
\par
\begin{figure}
\includegraphics
   [width=85mm]
   {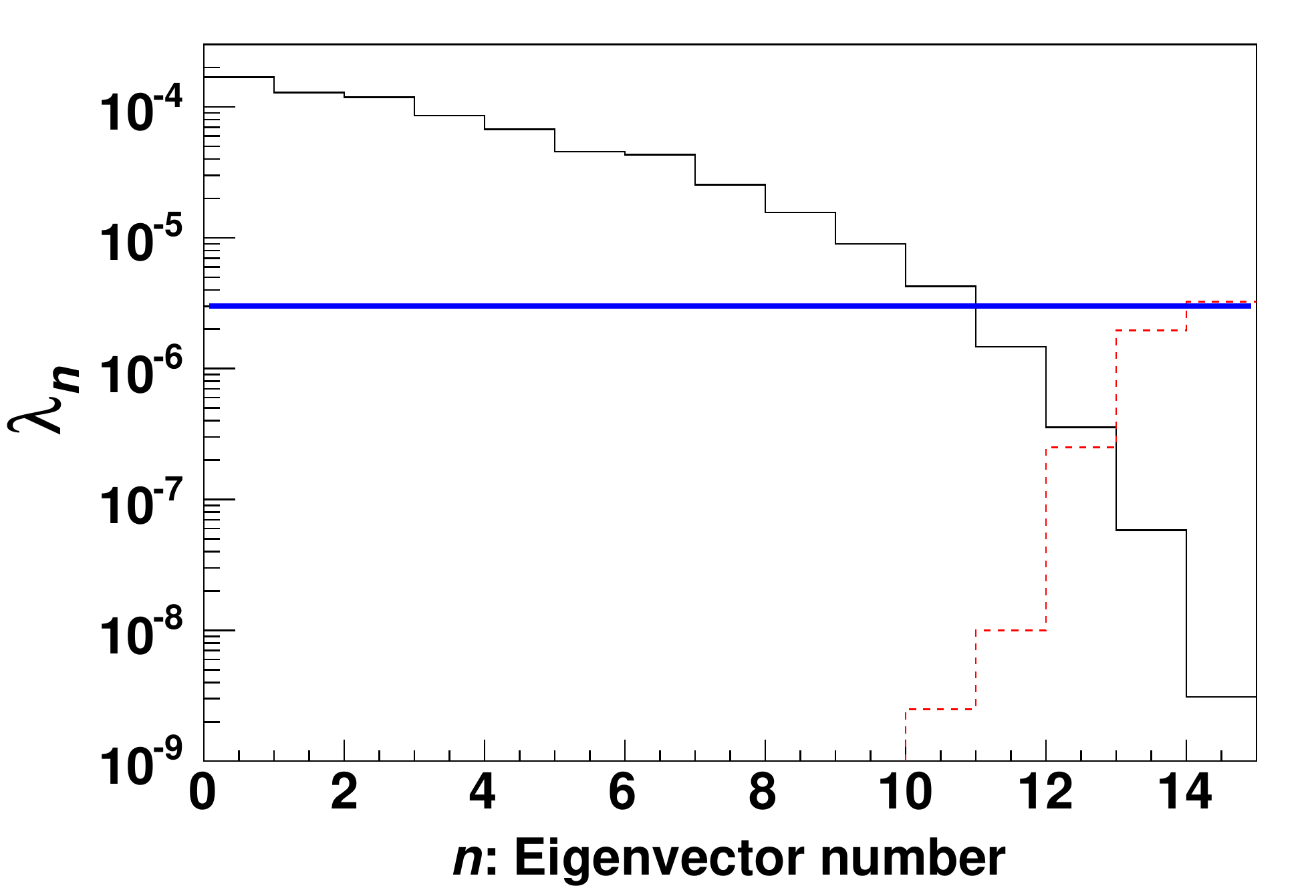}
\caption{\label{fig_covEigenval}
Eigenvalues of the error matrix (solid histogram),
and its regularization terms (dashed histogram).
The horizontal line is the square of the statistical uncertainty
of the $A_{\rm fb}$ measurement for the mass bin containing 
the $Z$ peak.
}
\end{figure}

Without regularization of the error matrix, there are large
fluctuations of the $\chi^2$ values for each scan point from the
expected parabolic form. Such fluctuations
are induced by the small eigenvalue terms in the expansion of
the error matrix, described in Eq.~(\ref{svdErrExp}). To attenuate
these fluctuations, the regularization function method described at
the end of Sec.~\ref{AfbReslUnfold} is used. The eigenvalues and
regularization terms are shown in Fig.~\ref{fig_covEigenval}.
The horizontal line of Fig.~\ref{fig_covEigenval} is an estimate,
detailed next, of
the resolving power of the measurement. The eigenvectors whose
eigenvalues are below the line tend to project simulation structure
finer than the resolution of the measurement, and thus induce
instabilities.
\par
The effectiveness of the regularization is measured with
the goodness-of-fit between the $\chi^2(\sin^2\theta_W)$ value
of the scan points and the parabolic function. In the basis
vector space of the error matrix, the $\chi^2$ of each scan point
is $\sum_n (\delta A_{\rm fb})^2_n / (\lambda_n + r_n)$,
where the index $n$ runs over all the eigenvector numbers and
$\delta A_{\rm fb}$ is the difference between the measured and
calculated values of $A_{\rm fb}$. 
The regularization function shown in Fig.~\ref{fig_covEigenval}
is defined and optimized as follows.
The shape of the regularization function is chosen so that
it selectively suppresses eigenvectors that project onto
noise rather than the uncertainties of the measurement.
To identify these eigenvectors, the expansion of the error
matrix is truncated one eigenvector at a time. Truncating
eigenvectors 14 and 13 from the error matrix significantly
improves the goodness-of-fit. There is no further improvement
with the truncation of lower numbered eigenvectors.
Consequently, the regularization terms for eigenvectors 13
and 14 are set to values significantly larger than the
eigenvalues so that the contributions of their constituents
to the $\chi^2$ are negligible. The regularization term for
eigenvector 12 is set to a value that is comparable with its
companion eigenvalue. For eigenvector numbers 11 and under,
the regularization terms are set to zero or values much smaller
than the eigenvalues so that their components in the $\chi^2$
are unaffected or negligibly affected by the regularization
terms. The optimum normalization level is determined via a
scan of level scale-factor values, starting from 0.
As the scale-factor
value increases, the goodness-of-fit rapidly improves then
enters a plateau region without significant improvement and
only a degradation of the measurement resolution. The optimum
is chosen to be slightly beyond the start of the plateau
region, where the $\overline{\sin}^2\theta_W$ parameter
is also stable in value.
\begin{figure}
\includegraphics
   [width=85mm]
   {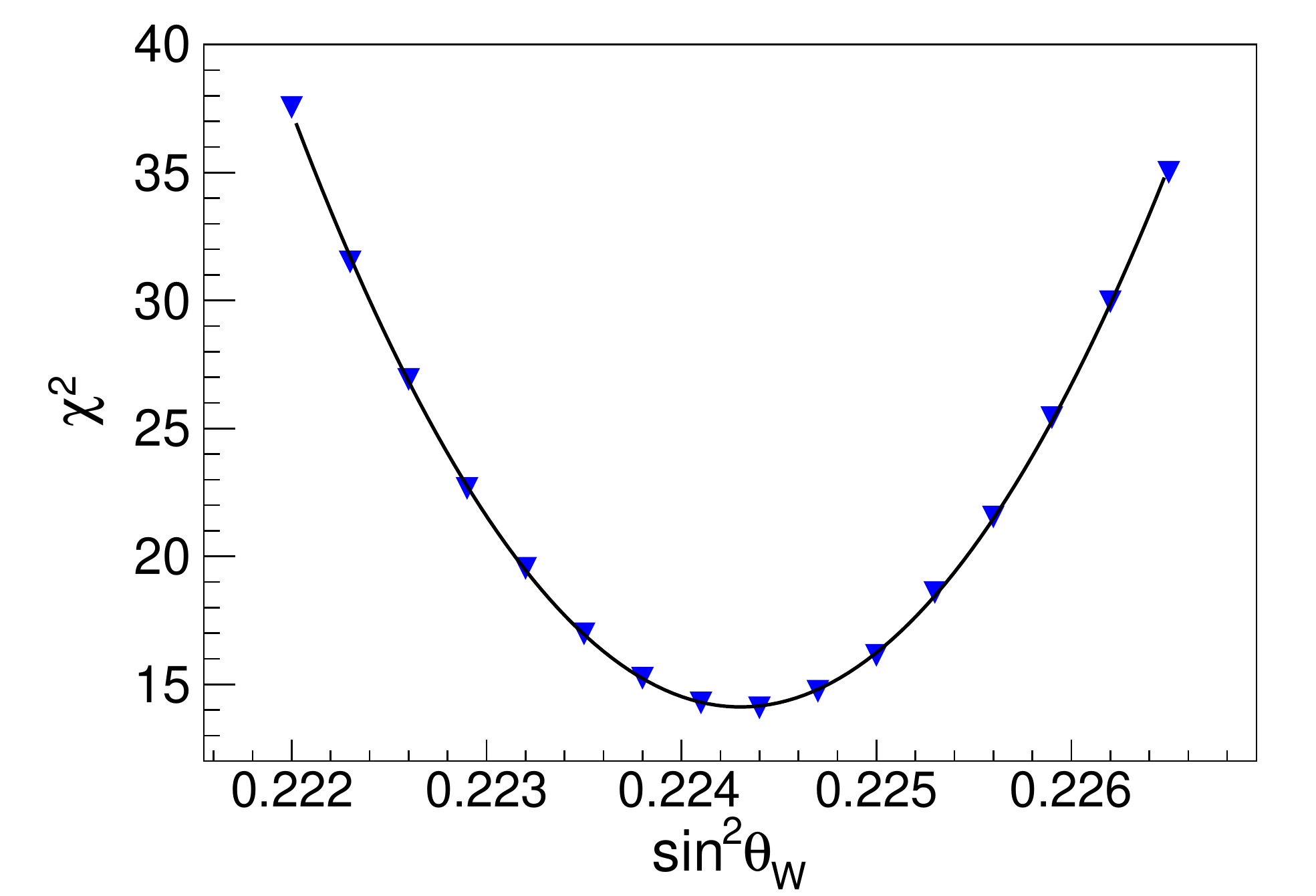}
\caption{\label{fig_scanPHboxsw2}
Values of $\chi^2$ as functions of scan points in the  $\sin^2\theta_W$
variable with the parabolic fit overlaid. The triangles are the comparisons
of the electron-pair $A_{\rm fb}$ measurement with the
\textsc{powheg-box} NLO calculations. The $A_{\rm fb}$ templates
of each scan point are calculated with the default PDF of NNPDF-3.0.
The solid curve is the fit of those points to the $\chi^2$
parabolic function.
}
\end{figure}
\par
As a cross check, the extraction of $\sin^2\theta_W$ is performed
using only CC or CP events for the measurement of $A_{\rm fb}$ and
its error matrix, and the default PDF of NNPDF-3.0 for the
calculation of templates. The extracted values
using only CC or CP events differ by about 0.6 standard deviations
of the statistical uncertainty. Since they are consistent with each
other, CC and CP events are hereafter combined.
An example template scan extraction of $\sin^2\theta_W$ from the
asymmetry of CC and CP events using $\chi^2$ values calculated
with the regularized error matrix, and then fit to the parabolic $\chi^2$
function of Eq.~(\ref{chiSqFit}) is shown in Fig.~\ref{fig_scanPHboxsw2}.
\par
The EBA-based tree and \textsc{powheg-box} NLO calculations of
$A_{\rm fb}$ use NNPDF-3.0 PDFs, an ensemble of probability-based
PDFs. Such ensembles are random samples drawn from the probability
density distribution of PDF parameters constrained by a
global fit to prior measurements. Thus, all information within
the probability density distribution is utilized. The predicted
value of an observable is the convolution of the probability density
distribution with the calculation. Consequently, the rms dispersion
about the mean is the associated PDF uncertainty \cite{GKpdfRewt}.
Typically, the PDF ensemble consists of equally likely samples.
The NNPDF-3.0 ensemble consists of 100 equally probable samples.
New measurements, if compatible with the measurements used to
constrain the PDFs, are incorporated into the ensemble without
regenerating it. This is accomplished by weighting the ensemble
PDFs, numbered 1 to $N$, with the likelihood of the new measurement
being consistent with the calculations:
\begin{equation}
   w_k = \frac{ \exp(-{\frac{1}{2}\chi^2_k}) }
	      {\sum_{l=1}^{N} \exp(-{\frac{1}{2}\chi^2_l}) } \;
\label{GKwtEqn}
\end{equation}
where $w_k$ is the weight for PDF number $k$, and $\chi^2_k$ is
the $\chi^2$ between the new measurement and the calculation using
that PDF \cite{GKpdfRewt,BayesPdfRewt}. These weights are denoted
as $w_k$ {\it weights} \cite{GKpdfRewt}.
\par
The $A_{\rm fb}$ measurement is used simultaneously to
extract $\sin^2\theta^{\rm lept}_{\rm eff}$ and to constrain PDFs
\cite{pdfGKweightMethod}.
Scan templates of $A_{\rm fb}$ are calculated for each ensemble
PDF, and its best-fit parameters, $\overline{\sin}^2\theta_W$,
$\bar{\chi}^2$ and $\bar{\sigma}$, are derived.
\begin{figure}
\includegraphics
   [width=85mm]
   {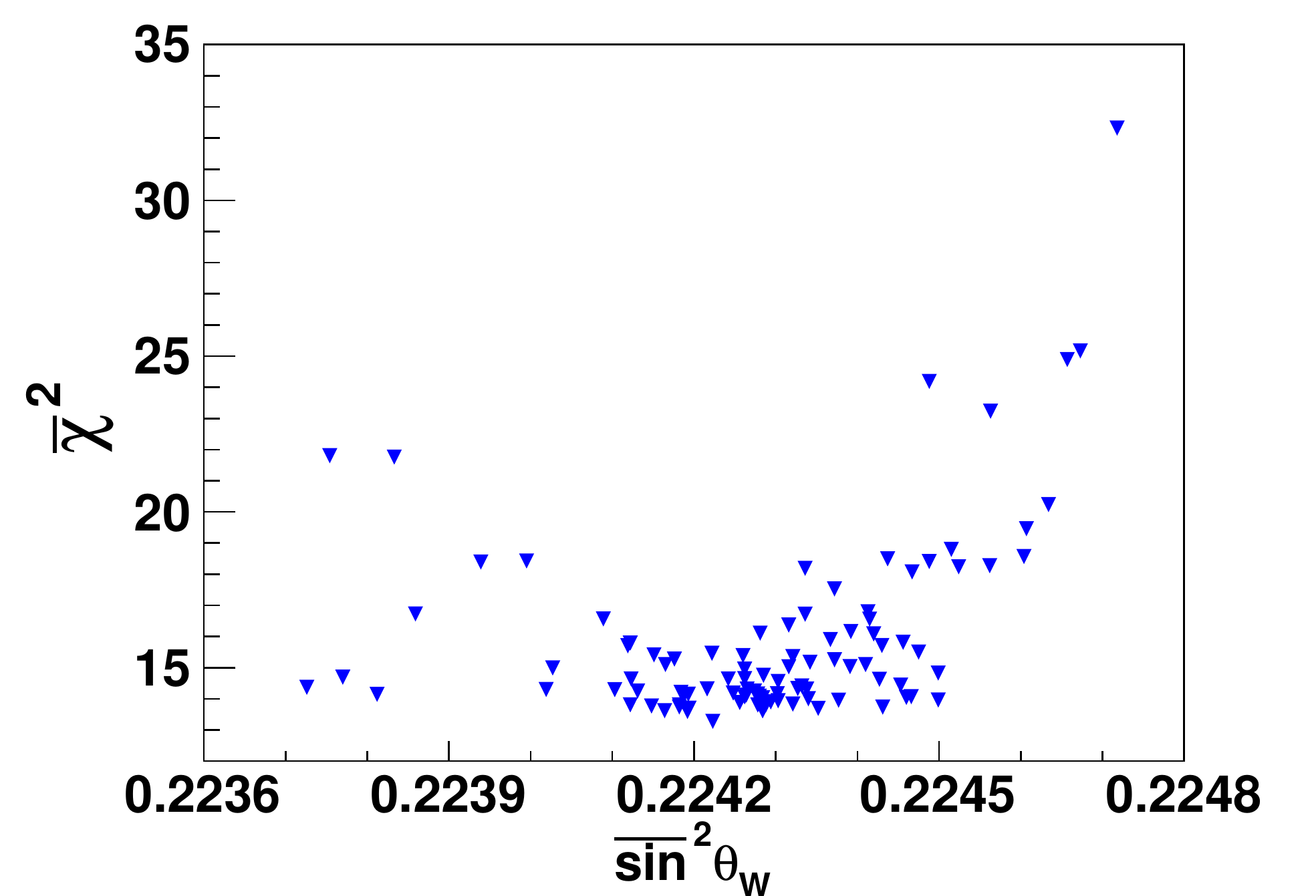}
\caption{\label{fig_eech2sw2}
$\bar{\chi}^2$ versus $\overline{\sin}^2\theta_W$ parameters
for the best-fit templates of the \textsc{powheg-box} NLO
calculation for each of the NNPDF-3.0 ensemble PDFs.
The $A_{\rm fb}$ measurement with electron pairs covers
15 mass bins.
}
\end{figure}
Figure~\ref{fig_eech2sw2} shows that the $A_{\rm fb}$ measurement
is compatible with those included in the NNPDF-3.0 fits of PDF
parameters. The results of the template scans are summarized
in Table~\ref{tblSW2results}.
\begin{table*}
\caption{\label{tblSW2results}
Extracted values of $\sin^2\theta^{\rm lept}_{\rm eff}$ and
$\sin^2\theta_W$ from the $A_{\rm fb}$ measurement using electron
pairs. For the tree and \textsc{powheg-box} entries,
the values are averages over the NNPDF-3.0 ensemble;
``weighted'' templates denote the $w_k$-weighted average; and
$\delta\sin^2\theta_W$ is the PDF uncertainty.
The \textsc{pythia} entry is the value from the scan
over non-EBA templates calculated by \textsc{pythia 6.4}
with CTEQ5L PDFs. The uncertainties of the electroweak-mixing
parameters are the measurement uncertainties $\bar{\sigma}$.
For the $\bar{\chi}^2$ column, the number in parenthesis is
the number of mass bins of the $A_{\rm fb}$ measurement.
}
\begin{ruledtabular}
\begin{tabular}{lcccc}
Template       & $\sin^2\theta^{\rm lept}_{\rm eff}$ & $\sin^2\theta_W$ 
						     & $\delta\sin^2\theta_W$
						     & $\bar{\chi}^2$  \\
(Measurement) &    &      &      &                                 \\ \hline
\textsc{powheg-box} NLO, default
			 & $0.23249 \pm 0.00049$ & $0.22429 \pm 0.00048$
						 & $\pm 0.00020$	 & $15.9\;(15)$ \\
\textsc{powheg-box} NLO, weighted
			 & $0.23248 \pm 0.00049$ & $0.22428 \pm 0.00048$
						 & $\pm 0.00018$	 & $15.4\;(15)$ \\
\textsc{resbos}     NLO  & $0.23249 \pm 0.00049$ & $0.22429 \pm 0.00047$
						 &  $-$ 		 & $21.3\;(15)$ \\
Tree LO, default
			 & $0.23252 \pm 0.00049$ & $0.22432 \pm 0.00047$
						 & $\pm 0.00021$	 & $22.4\;(15)$ \\
Tree LO, weighted
			 & $0.23250 \pm 0.00049$ & $0.22430 \pm 0.00047$
						 & $\pm 0.00021$	 & $21.5\;(15)$ \\
\textsc{pythia}          & $0.23207 \pm 0.00046$ &   $-$  & $-$          & $24.6\;(15)$ \\
(CDF 9 fb$^{-1}$ $A_{\rm fb}^{(\mu\mu)}$ \cite{cdfAfb9mmprd})
			 & $0.2315 \pm 0.0010$ & $0.2233 \pm 0.0009$
					       & $-$  		 	 & $21.1\;(16)$ \\
(CDF 2 fb$^{-1}$ $A_4^{(ee)}$ \cite{zA4ee21prd,*zA4ee21prdE})
			 & $0.2328 \pm 0.0010$ & $0.2246 \pm 0.0009$ &  $-$  & $-$ \\
(LEP-1 and SLD $A_{\rm FB}^{0,{\rm b}}$ \cite{LEPfinalZ})
			 & $0.23221 \pm 0.00029$ & $-$ & $-$ & $-$   \\
(SLD ${\cal A}_\ell$ \cite{LEPfinalZ})
			 & $0.23098 \pm 0.00026$ & $-$ & $-$ & $-$   \\		  
\end{tabular}
\end{ruledtabular}
\end{table*}
Included in the table for comparison are other measurements of
$\sin^2\theta^{\rm lept}_{\rm eff}$; the
CDF results are derived from EBA-based QCD templates. 
\par
The EBA-based \textsc{powheg-box} calculations of $A_{\rm fb}$ using
the $w_k$-weighted PDFs give the central value of $\sin^2\theta_W$.
The $\bar{\chi}^2$ values listed in Table~\ref{tblSW2results} indicate
that the \textsc{powheg-box} calculation provides the best description
of the $A_{\rm fb}$ measurement. For graphical comparisons of best-fit
$A_{\rm fb}$ templates, the difference relative to a reference
calculation is used: $A_{\rm fb} - A_{\rm fb}({\rm \textsc{pythia}})$
where the reference $A_{\rm fb}({\rm \textsc{pythia}})$ is the tuned
\textsc{pythia} calculation described in Sec.~\ref{AfbexpDatSim} on the
signal simulation.
\begin{figure}
\includegraphics
   [width=85mm]
   {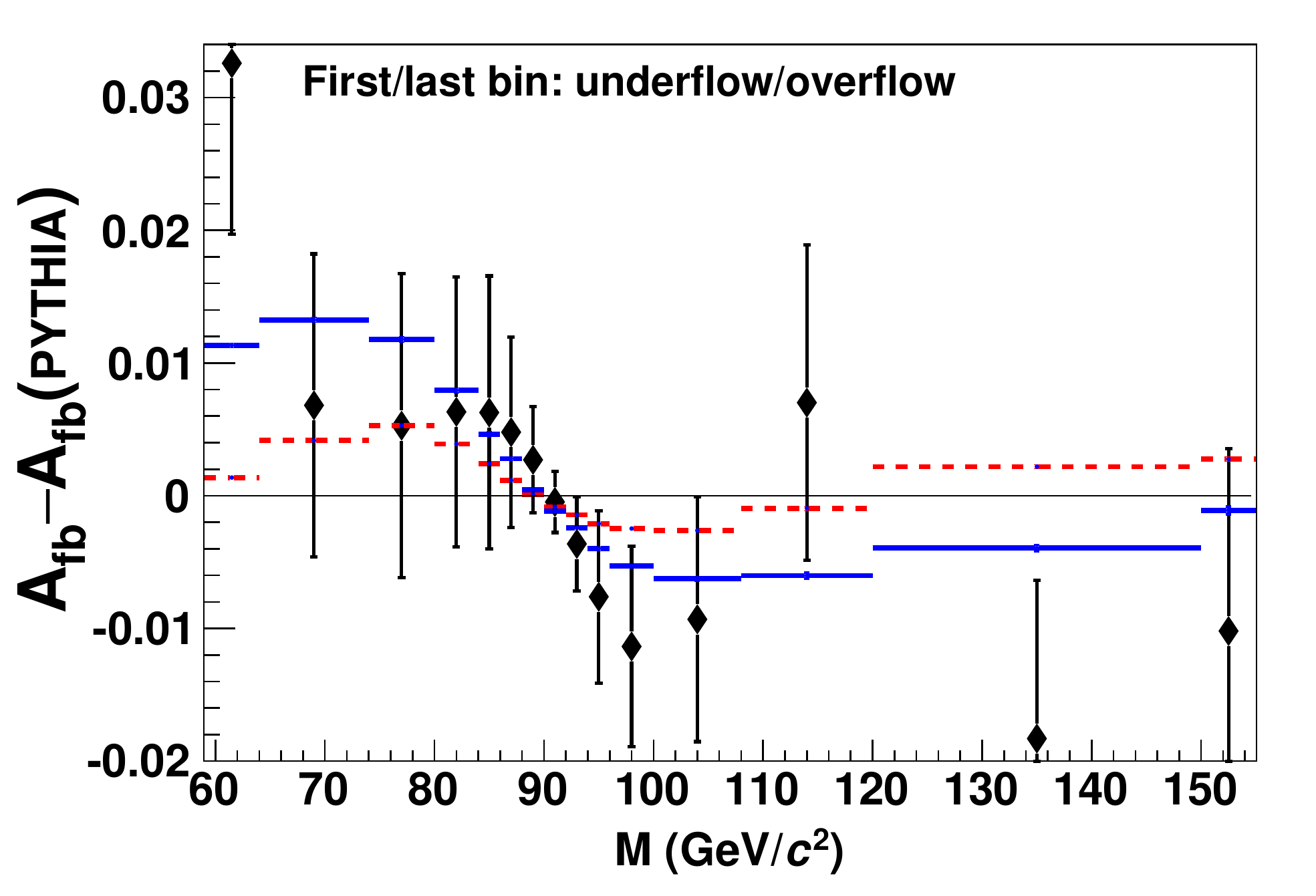}
\caption{\label{fig_afbDiff}
$A_{\rm fb} - A_{\rm fb}({\rm \textsc{pythia}})$ for $|y|<1.7$.
The diamonds represent the measurement using electron pairs,
and the uncertainties shown are the bin-by-bin unfolding estimates
which are correlated. There are no suppressed measurement values.
The solid bars represent the \textsc{powheg-box} calculation
with the default NNPDF-3.0 PDFs. The dashed
bars represent the \textsc{resbos} calculation with
CTEQ6.6 PDFs. Both calculations use $\sin^2 \theta_W = 0.2243$.
The horizontal line represents the reference \textsc{pythia}
calculation which uses CTEQ5L PDFs with
$\sin^2 \theta_{\rm eff}^{\rm lept} = 0.232$.
}
\end{figure}
Figure~\ref{fig_afbDiff} shows the difference
distributions for the measurement, the \textsc{powheg-box}
calculation, and the \textsc{resbos} calculation. The comparison
of \textsc{powheg-box} with NNPDF-3.0 PDFs to \textsc{resbos}
with CTEQ6.6 PDFs illustrates the nature of $A_{\rm fb}(M)$ as
a simultaneous probe of the electroweak-mixing parameter and the
PDFs. The NNPDF-3.0 PDFs include collider data from the LHC
while the CTEQ6.6 PDFs do not.

\section{\label{systUncerts}
Systematic Uncertainties}

The systematic uncertainties of the results derived from electron pairs
contain contributions from both the
measurement of $A_{\rm fb}$ and the template predictions of $A_{\rm fb}$
for various input values of $\sin^2\theta_W$. Both the experimental and
prediction-related systematic uncertainties are small compared to the
experimental statistical uncertainty. The $A_{\rm fb}$ templates of
the \textsc{powheg-box} calculations are used to estimate
systematic uncertainties on the $\sin^2\theta_W$ parameter from various
sources.

\subsection{\label{datSystUncerts}
Measurement}

The measurement uncertainties considered are from the energy
scale and resolution, and from the background estimates.
The bias-correction uncertainty from the PDFs, expected to be
a small secondary effect, is not included.
For the propagation of
uncertainties to the extracted value of $\sin^2\theta_W$,
the default PDF of the NNPDF-3.0 ensemble is used.
The total measurement systematic uncertainty is
$\pm 0.00003$.
\par
The energy scale and resolution of the simulation and data samples
are accurately calibrated (Sec.~\ref{CorrDatSim}) using
electron-pair mass distributions. In conjunction, the mass
distributions of the simulation have been tuned to agree
with those of the data, and the agreement between them,
presented in Figs. \ref{fig_mee1CCos} and \ref{fig_mee1CP} is good.
Since the energy scales
of the data and simulation are calibrated separately from the
underlying-physics scale, the potential effect of an offset
between the global scales of the simulation and data is investigated
as a systematic uncertainty. The electron-pair mass distributions in
the vicinity of the $Z$-boson mass peak are used to constrain shifts.
Scale shifts for the central and plug EM calorimeters are considered
separately. The resulting uncertainty from the energy scale is
$\pm 0.00003$. The potential effect of the limitations to the
energy-resolution model of the simulation is also investigated,
and the resulting uncertainty is estimated to be negligible.
\par
For the background systematic uncertainty, the normalization
uncertainties of the two largest backgrounds, QCD and
$Z\rightarrow\tau\tau$, are considered. They amount to about
three-quarters of the total background. The uncertainties of
their normalization values from the background fits described
in Sec.~\ref{EEBackgrounds} are propagated into uncertainties
on $\sin^2\theta_W$. They have a negligible impact on the
measurement.
For the $Z\rightarrow\tau\tau$ background of the CC topology,
the difference between the constrained and unconstrained fit
normalizations is assigned as a systematic uncertainty. The
systematic uncertainty from the background is $\pm 0.00002$.
\par
The bias correction uses the \textsc{pythia} calculation with
CTEQ5L PDFs. To evaluate whether a PDF systematic
uncertainty is needed, the following bias metric is used:
the difference in asymmetries calculated with the measurement
rapidity range of $|y|<1.7$ and with the reduced rapidity range
of $|y|<1.5$. The bias metric calculated with \textsc{pythia}
is shown in Fig.~\ref{fig_afbEWbias}, along with the
PDF uncertainties estimated using the tree-level calculation
of $A_{\rm fb}$ with the NNPDF-3.0 ensemble of PDFs. The PDF
uncertainties are small when compared to the statistical
uncertainties of the bias correction. In addition, the
\textsc{pythia} calculation of the bias-metric function is
compatible, relative to PDF uncertainties, with the
tree-level calculation using NNPDF-3.0 PDFs; the comparison
$\chi^2$ has a value of 11 for the 15 mass bins.
The PDF uncertainty to the \textsc{pythia} calculation is not
included with the measurement because its effects are
sufficiently small relative to the statistical uncertainties
of the bias correction, and because the prediction includes an
uncertainty for PDFs.

\subsection{\label{systUncertsPred}
Predictions}

The theoretical uncertainties considered are from the PDFs,
higher-order QCD effects, and the \textsc{zfitter} calculation.
The dominant uncertainty is the PDF uncertainty of
$\pm 0.00018$, and it is the $w_k$-weighted value of
$\delta \sin^2\theta_W$ from the \textsc{powheg-box} NLO entry of
Table~\ref{tblSW2results}. The total prediction uncertainty is
$\pm 0.00020$.
\par
The uncertainty of higher-order QCD effects is estimated
with the difference between the values of $\sin^2\theta_W$
in Table~\ref{tblSW2results} extracted with the tree
and \textsc{powheg-box} NLO templates based on the
$w_k$-weighted ensemble of NNPDF-3.0 PDFs. This uncertainty,
denoted as the ``QCD scale'' uncertainty, is $\pm 0.00002$.
Although the \textsc{powheg-box} prediction is a
fixed-order NLO QCD calculation at large boson $P_{\rm T}$,
it is a resummation calculation in the low-to-moderate
$P_{\rm T}$ region. The parton-showering algorithm of
\textsc{pythia} incorporates multi-order real emissions
of QCD radiation over all regions of the boson $P_{\rm T}$.
\par
The $\sin^2\theta^{\rm lept}_{\rm eff}$ result, because of its
direct relationship with $A_{\rm fb}$, is independent of the
standard-model based calculations specified in the appendix.
However, the choice of input parameter values may affect the
fit value of $\sin^2\theta_W$ or $M_W$. The effect of
measurement uncertainties from the top-quark mass $m_t$ and from
the contribution of the light quarks to the ``running''
electromagnetic fine-structure constant at the $Z$ mass
$\Delta \alpha_{\rm em}^{(5)}(M_Z^2)$ is investigated using these
uncertainties: $\pm 0.9$~GeV/$c^2$ \cite{topMassCDFD0} and
$\pm 0.0001$ \cite{alpemh5}, respectively. 
\begin{figure}
\includegraphics
   [width=85mm]
   {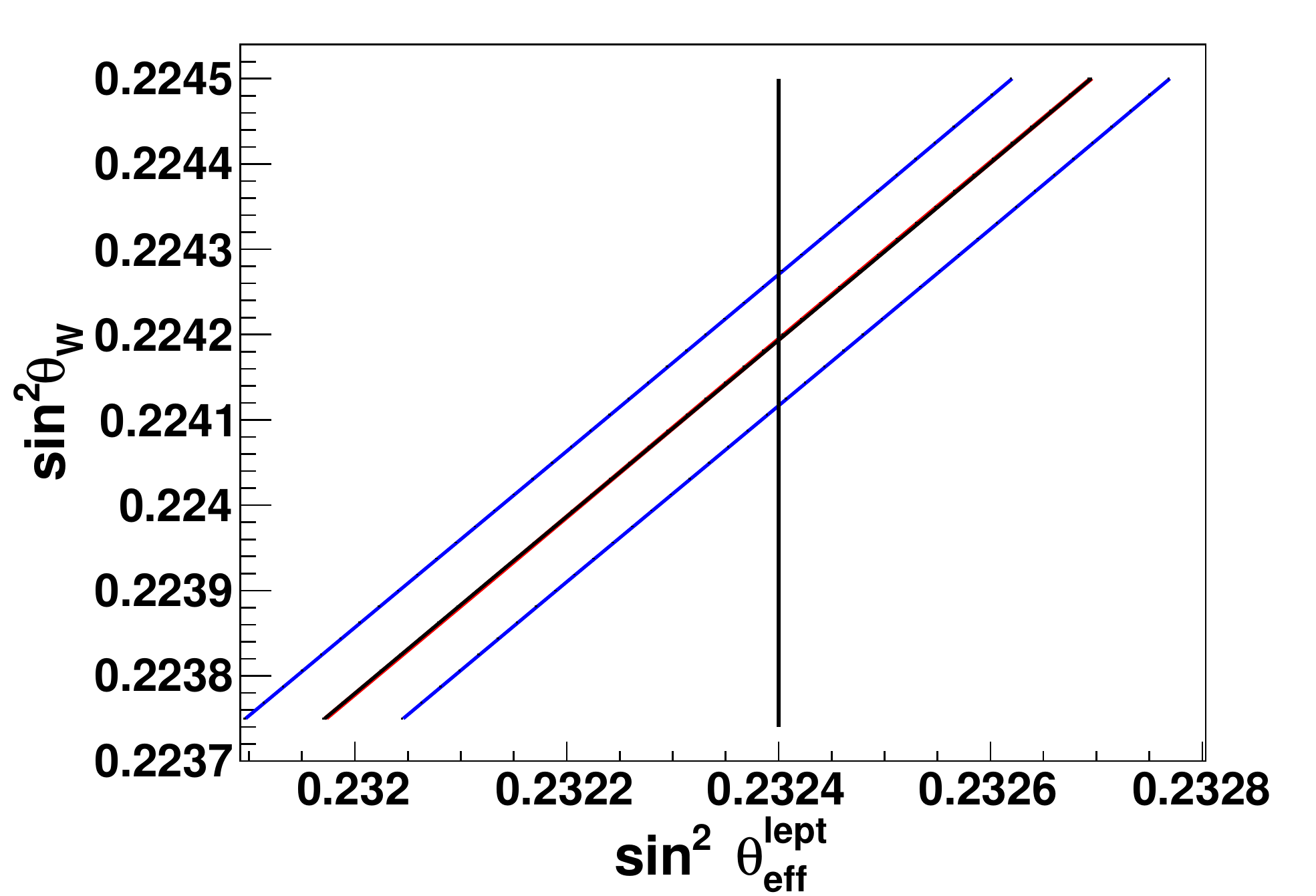}
\caption{\label{fig_zf643Syst}
$\sin^2\theta_W$ versus $\sin^2\theta^{\rm lept}_{\rm eff}$
relationships
from \textsc{zfitter} calculations. The default calculation
is the middle line of the group. The outermost lines are for
one standard-deviation shifts to the default value of the 
top-quark mass parameter ($173.2 \pm 0.9$) \cite{topMassCDFD0};
the lower line corresponds to a higher value of the top-quark
mass. The lines for one standard-deviation variations
of the $\Delta \alpha_{\rm em}^{(5)}(M_Z^2)$ parameter are close
to the default calculation and not easily distinguishable.  
The vertical line, an example reference value for
$\sin^2\theta^{\rm lept}_{\rm eff}$, is explained in the text.
}
\end{figure}
Figure~\ref{fig_zf643Syst} shows the relation between
$\sin^2\theta_W$ and $\sin^2\theta^{\rm lept}_{\rm eff}$ for the
default parameter values, and for one standard-deviation shifts
to the default values of the $m_t$ and $\Delta \alpha_{\rm em}^{(5)}(M_Z^2)$
parameters. Offsets from the default parameter curve to the one
standard-deviation curves along a reference value for
$\sin^2\theta^{\rm lept}_{\rm eff}$ (\mbox{e.g.}, the vertical
line in Fig.~\ref{fig_zf643Syst}) are used as systematic
uncertainties to $\sin^2\theta_W$ from the input parameters.
The uncertainty to $\sin^2\theta_W$ from
$\Delta \alpha_{\rm em}^{(5)}(M_Z^2)$ is negligible, and that
from $m_t$ is $\pm 0.00008$. This uncertainty,
denoted as the ``form factor'' uncertainty, is included in
systematic uncertainties for $\sin^2\theta_W$ and $M_W$.

\section{\label{finalResults}
Results}

The values for $\sin^2\theta^{\rm lept}_{\rm eff}$ and
$\sin^2\theta_W$ ($M_W$) extracted from this measurement
of $A_{\rm fb}$ are
\begin{eqnarray*}
  \sin^2 \theta^{\rm lept}_{\rm eff} & = &
	0.23248 \pm 0.00049 \pm 0.00019   \\
  \sin^2 \theta_W   & = &
	0.22428 \pm 0.00048 \pm 0.00020   \\
  M_W \; {\rm (indirect)}       & = & 
	80.313 \pm 0.025 \pm 0.010 \; {\rm GeV}/c^2 \:,
\end{eqnarray*}
where the first contribution to the uncertainties is statistical
and the second is systematic. All systematic uncertainties are
combined in quadrature.
\par
A summary of the sources and values of systematic
uncertainties is presented in Table~\ref{tblSystErrors}.
\begin{table}
\caption{\label{tblSystErrors}
Summary of the systematic uncertainties on the extraction of
the electroweak-mixing parameters
$\sin^2\theta^{\rm lept}_{\rm eff}$ and
$\sin^2\theta_W$
from the $A_{\rm fb}$ measurement with electron pairs.
}
\begin{ruledtabular}
\begin{tabular}{lcc}
Source  & $\sin^2\theta^{\rm lept}_{\rm eff}$ & $\sin^2\theta_W$ \\ \hline
Energy scale   & $\pm 0.00003$  & $\pm 0.00003$ \\
Backgrounds    & $\pm 0.00002$  & $\pm 0.00002$ \\
NNPDF-3.0 PDF  & $\pm 0.00019$  & $\pm 0.00018$ \\
QCD scale      & $\pm 0.00002$  & $\pm 0.00002$ \\
Form factor    & $-$            & $\pm 0.00008$ 
\end{tabular}
\end{ruledtabular}
\end{table}
The results of this section supersede those derived from the $A_4$
angular-distribution coefficient of $ee$-pairs from a sample
corresponding to 2.1~fb$^{-1}$ of collisions
\cite{zA4ee21prd,*zA4ee21prdE}.

\section{\label{cdfemuCombination}
CDF Result Combination}

The measurement of $A_{\rm fb}$ presented in this paper and
the previous CDF measurement using Drell-Yan $\mu^+\mu^-$ pairs
\cite{cdfAfb9mmprd} are used to extract the combined result
for the electroweak-mixing parameter. Both measurements are
fully corrected and use the full Tevatron Run~II data set.
Since they are defined for different regions of the
lepton-pair rapidity, $|y_{ee}| < 1.7$ and $|y_{\mu\mu}| < 1.0$,
each measurement is compared separately to $A_{\rm fb}$
templates calculated with the rapidity restriction of the
measurements, and the joint $\chi^2$ is used to extract the
combined values for electroweak-mixing parameters
$\sin^2\theta^{\rm lept}_{\rm eff}$ and $\sin^2 \theta_W$.

\subsection{Method}

The templates for both measurements are calculated using the
EBA-based \textsc{powheg-box} NLO framework and the NNPDF-3.0
PDF ensemble of this analysis. The corresponding tree-level
templates are also calculated. The $A_{\rm fb}$ templates for
both the $\mu\mu$- and $ee$-channel measurements
are calculated in the same \textsc{powheg-box} or tree-level
computational runs. Thus, they share common events and
scan-point values of the $\sin^2 \theta_W$ parameter.
\par
The method for the extraction of $\sin^2\theta^{\rm lept}_{\rm eff}$
from each measurement is unaltered. For each of the ensemble PDFs,
the parabolic fits to $\chi^2(\sin^2\theta_W)$ shown in
Eq.~(\ref{chiSqFit})
from each measurement are combined to obtain the values of
$\overline{\sin}^2\theta_W$, $\bar{\chi}^2$, and $\bar{\sigma}$.
\begin{figure}
\includegraphics
   [width=85mm]
   {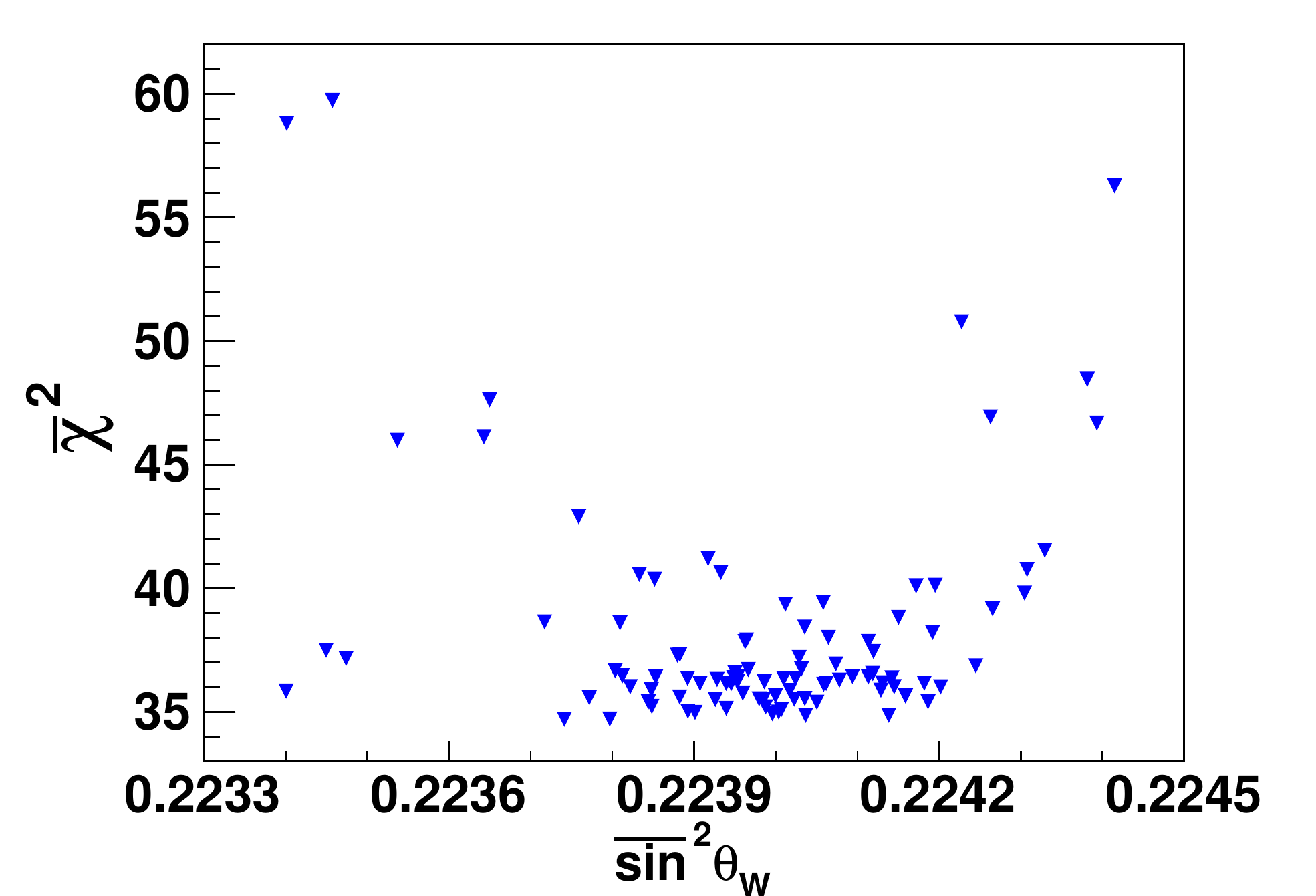}
\caption{\label{fig_emch2sw2}
$\bar{\chi}^2$ versus $\overline{\sin}^2\theta_W$ parameters
of the $\mu\mu$- and $ee$-channel combination. The prediction
templates are calculated
with \textsc{powheg-box} NLO and each of the NNPDF-3.0 ensemble
PDFs. The $\mu\mu$- and $ee$-channel $A_{\rm fb}$
measurements contain 16 and 15 mass bins, respectively.
}
\end{figure}
Figure~\ref{fig_emch2sw2} shows the
$\bar{\chi}^2$ and $\overline{\sin}^2\theta_W$ parameters
associated with each ensemble PDF. The corresponding table of
fit parameters is provided as supplemental material
\cite{[{See
ancilliary file
for the table of
fit parameter values from the joint $\chi^2$ of comparisons
of the individual muon and electron-channel $A_{\rm fb}$
measurements and templates calculated with
\textsc{powheg-box} NLO for each ensemble PDF}][{}]{supplemental}}.
\begin{table*}
\caption{\label{ctblSW2results}
Extracted values of $\sin^2\theta^{\rm lept}_{\rm eff}$ and
$\sin^2\theta_W$ after averaging over the NNPDF-3.0 ensembles.
The ``weighted'' templates denote the  $w_k$-weighted ensembles; and
$\delta\sin^2\theta_W$ is the PDF uncertainty.
The uncertainties of the electroweak-mixing
parameters are the measurement uncertainties $\bar{\sigma}$.
For the $\bar{\chi}^2$ column, the number in parenthesis is
the number of mass bins of the $A_{\rm fb}$ measurement.
The $ee$-channel values are from Table~\ref{tblSW2results}, and
the $\mu\mu$-channel values use the previous CDF measurement of
$A_{\rm fb}$ with $\mu^+\mu^-$ pairs \cite{cdfAfb9mmprd}.
}
\begin{ruledtabular}
\begin{tabular}{lccccc}
Template      & Channel
	            & $\sin^2\theta^{\rm lept}_{\rm eff}$ & $\sin^2\theta_W$
                                                 & $\delta\sin^2\theta_W$
                                                 & $\bar{\chi}^2$  \\ \hline
\textsc{powheg-box} NLO, default
                         & $\mu\mu$
                         & $0.23140 \pm 0.00086$ & $0.22316 \pm 0.00083$
                                                 & $\pm 0.00029$   & $21.0\;(16)$ \\
\textsc{powheg-box} NLO, weighted
			 & $\mu\mu$
                         & $0.23141 \pm 0.00086$ & $0.22317 \pm 0.00083$
                                                 & $\pm 0.00028$   & $20.7\;(16)$ \\
\textsc{powheg-box} NLO, default
                         & $ee$
                         & $0.23249 \pm 0.00049$ & $0.22429 \pm 0.00048$
                                                 & $\pm 0.00020$   & $15.9\;(15)$ \\
\textsc{powheg-box} NLO, weighted
                         & $ee$
                         & $0.23248 \pm 0.00049$ & $0.22428 \pm 0.00048$
                                                 & $\pm 0.00018$   & $15.4\;(15)$ \\
\textsc{powheg-box} NLO, default
			 & $ee+\mu\mu$
                         & $0.23222 \pm 0.00043$ & $0.22401 \pm 0.00041$
                                                 & $\pm 0.00021$   & $38.3\;(31)$ \\
\textsc{powheg-box} NLO, weighted
                         & $ee+\mu\mu$
                         & $0.23221 \pm 0.00043$ & $0.22400 \pm 0.00041$
                                                 & $\pm 0.00016$   & $35.9\;(31)$ \\
Tree LO, default
			 & $\mu\mu$
                         & $0.23154 \pm 0.00085$ & $0.22330 \pm 0.00082$
                                                 & $\pm 0.00031$   & $20.9\;(16)$ \\
Tree LO, weighted
			 & $\mu\mu$
                         & $0.23153 \pm 0.00085$ & $0.22329 \pm 0.00082$
                                                 & $\pm 0.00029$   & $20.5\;(16)$ \\
Tree LO, default
                         & $ee$
                         & $0.23252 \pm 0.00049$ & $0.22432 \pm 0.00047$
                                                 & $\pm 0.00021$   & $22.4\;(15)$ \\
Tree LO, weighted
                         & $ee$
                         & $0.23250 \pm 0.00049$ & $0.22430 \pm 0.00047$
                                                 & $\pm 0.00021$   & $21.5\;(15)$ \\
Tree LO, default
                         & $ee+\mu\mu$
                         & $0.23228 \pm 0.00042$ & $0.22407 \pm 0.00041$
                                                 & $\pm 0.00023$   & $44.4\;(31)$ \\
Tree LO, weighted
                         & $ee+\mu\mu$
                         & $0.23215 \pm 0.00043$ & $0.22393 \pm 0.00041$
                                                 & $\pm 0.00016$   & $37.4\;(31)$ \\
\end{tabular}
\end{ruledtabular}
\end{table*}
The ensemble-averaged values of the individual channels, along
with their combination, are shown in Table~\ref{ctblSW2results}.
The $w_k$-weighted averaging method with \textsc{powheg-box}
NLO calculations is selected for the central value of the
combination result.

\subsection{\label{systUncertsComb}
Systematic uncertainties}

The categories of systematic uncertainties for both the $\mu\mu$-
and $ee$-channel extractions of the electroweak-mixing parameters
are the same. Uncertainties associated with the measurements include
those on the electroweak-mixing parameter from the backgrounds and the
energy scales. Those associated with the predictions include
uncertainties from the PDFs
and higher-order QCD effects (QCD scale). The numerical
values for systematic uncertainties in this section are for the
$\sin^2 \theta_W$ parameter.
\par
The measurement uncertainties of the $\mu\mu$ and $ee$ channels
are uncorrelated, and thus the propagation of their uncertainties
to $\sin^2 \theta_W$ is uncorrelated. The combined
energy-scale and background uncertainties are $\pm 0.00002$ and
$\pm 0.00003$, respectively.
\par
As the prediction uncertainties of both channels are
correlated, the corresponding uncertainties of the combination
are derived from the fit parameters of the joint $\chi^2$.
The uncertainty due to the PDF is $\pm 0.00016$,
which is the $w_k$-weighted $\delta \sin^2\theta_W$ value from
the \textsc{powheg-box} NLO entry of Table~\ref{ctblSW2results}.
The uncertainty due to the QCD scale is $\pm 0.00007$, which is
the difference
between the $w_k$-weighted $\sin^2\theta_W$ values of the
\textsc{powheg-box} NLO and Tree entries from
Table~\ref{ctblSW2results}.

\subsection{\label{finalResultsComb}
Results}

The combination values for $\sin^2\theta^{\rm lept}_{\rm eff}$
and $\sin^2\theta_W$ ($M_W$) are
\begin{eqnarray*}
  \sin^2 \theta^{\rm lept}_{\rm eff} & = &
        0.23221 \pm 0.00043 \pm 0.00018   \\
  \sin^2 \theta_W   & = &
        0.22400 \pm 0.00041 \pm 0.00019   \\
  M_W \; {\rm (indirect)}       & = &
        80.328 \pm 0.021 \pm 0.010 \; {\rm GeV}/c^2 \:,
\end{eqnarray*}
where the first contribution to the uncertainties is statistical
and the second is systematic. All systematic uncertainties are
combined in quadrature, and the sources and values of these
uncertainties are summarized in Table~\ref{tblSystErrComb}.
\begin{table}
\caption{\label{tblSystErrComb}
Summary of the systematic uncertainties on the $\mu\mu$- and
$ee$-channel combination for the electroweak-mixing parameters
$\sin^2\theta^{\rm lept}_{\rm eff}$ and
$\sin^2\theta_W$.
}
\begin{ruledtabular}
\begin{tabular}{lcc}
Source  & $\sin^2\theta^{\rm lept}_{\rm eff}$ & $\sin^2\theta_W$ \\ \hline
Energy scale   & $\pm 0.00002$  & $\pm 0.00002$ \\
Backgrounds    & $\pm 0.00003$  & $\pm 0.00003$ \\
NNPDF-3.0 PDF  & $\pm 0.00016$  & $\pm 0.00016$ \\
QCD scale      & $\pm 0.00006$  & $\pm 0.00007$ \\
Form factor    & $-$            & $\pm 0.00008$
\end{tabular}
\end{ruledtabular}
\end{table}
The form-factor uncertainty, estimated in Sec.~\ref{finalResults},
is the uncertainty from the standard-model based calculation specified
in the appendix.
\par
\begin{figure}
\includegraphics
   [width=85mm]
   {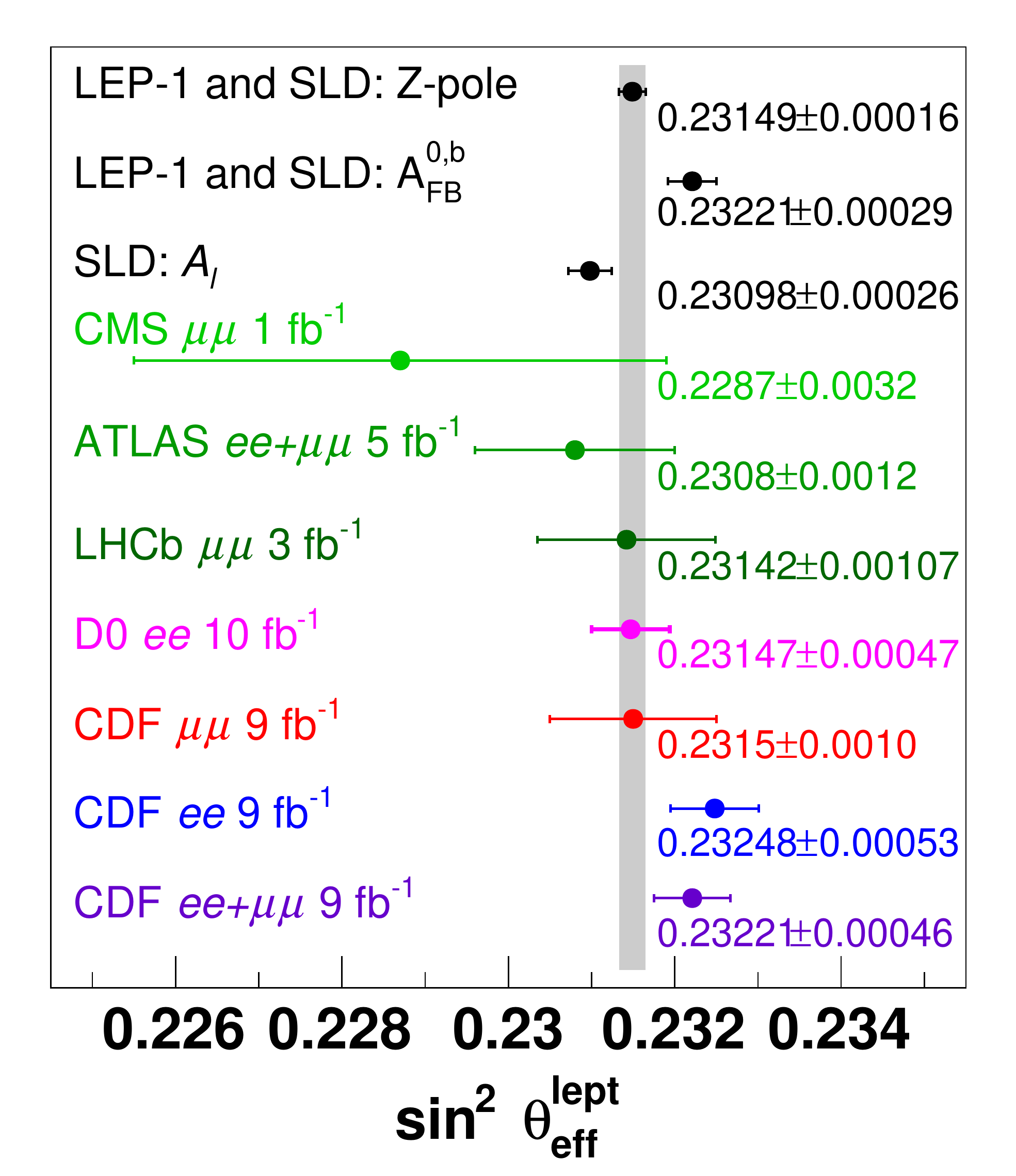}
\caption{\label{fig_compareSW2leff}
Comparison of experimental measurements of 
$\sin^2\theta^{\rm lept}_{\rm eff}$. 
The horizontal bars represent total uncertainties.
The CDF $\mu\mu$-channel, $ee$-channel, and combination results
are denoted as CDF $\mu\mu$ 9 fb$^{-1}$ \cite{cdfAfb9mmprd},
CDF $ee$ 9 fb$^{-1}$, and CDF $ee+\mu\mu$ 9 fb$^{-1}$, respectively.
The other measurements are
LEP-1 and SLD \cite{LEPfinalZ},
CMS \cite{CMSsw2eff1},
ATLAS \cite{ATLASsw2eff},
LHCb \cite{LHCBsw2eff}, and
D0 \cite{D0sw2e10} 
The LEP-1 and SLD $Z$ pole result is the combination of their six
measurements.
}
\end{figure}

 
The measurements of $\sin^2\theta^{\rm lept}_{\rm eff}$
are compared with previous results from the Tevatron, LHC,
LEP-1, and SLC in Fig.~\ref{fig_compareSW2leff}.
The hadron collider results are based on $A_{\rm fb}$ measurements.
The LEP-1 and SLD results on $\sin^2\theta^{\rm lept}_{\rm eff}$
are from these asymmetry measurements at the $Z$ pole \cite{LEPfinalZ}:
\begin{eqnarray*}
   A_{\rm FB}^{0,\ell}        & \rightarrow & 0.23099 \pm 0.00053 \\
   {\cal A}_{\ell}(P_{\tau})  & \rightarrow & 0.23159 \pm 0.00041 \\
   {\cal A}_{\ell}({\rm SLD}) & \rightarrow & 0.23098 \pm 0.00026 \\
   A_{\rm FB}^{0,{\rm b}}     & \rightarrow & 0.23221 \pm 0.00029 \\
   A_{\rm FB}^{0,{\rm c}}     & \rightarrow & 0.23220 \pm 0.00081 \\
   Q_{\rm FB}^{{\rm had}}     & \rightarrow & 0.2324  \pm 0.0012 
\end{eqnarray*}
The $Q_{\rm FB}^{{\rm had}}$ measurement is based on the
hadronic-charge asymmetry from all-hadronic final states.
\par
\begin{figure}
\includegraphics
   [width=85mm]
   {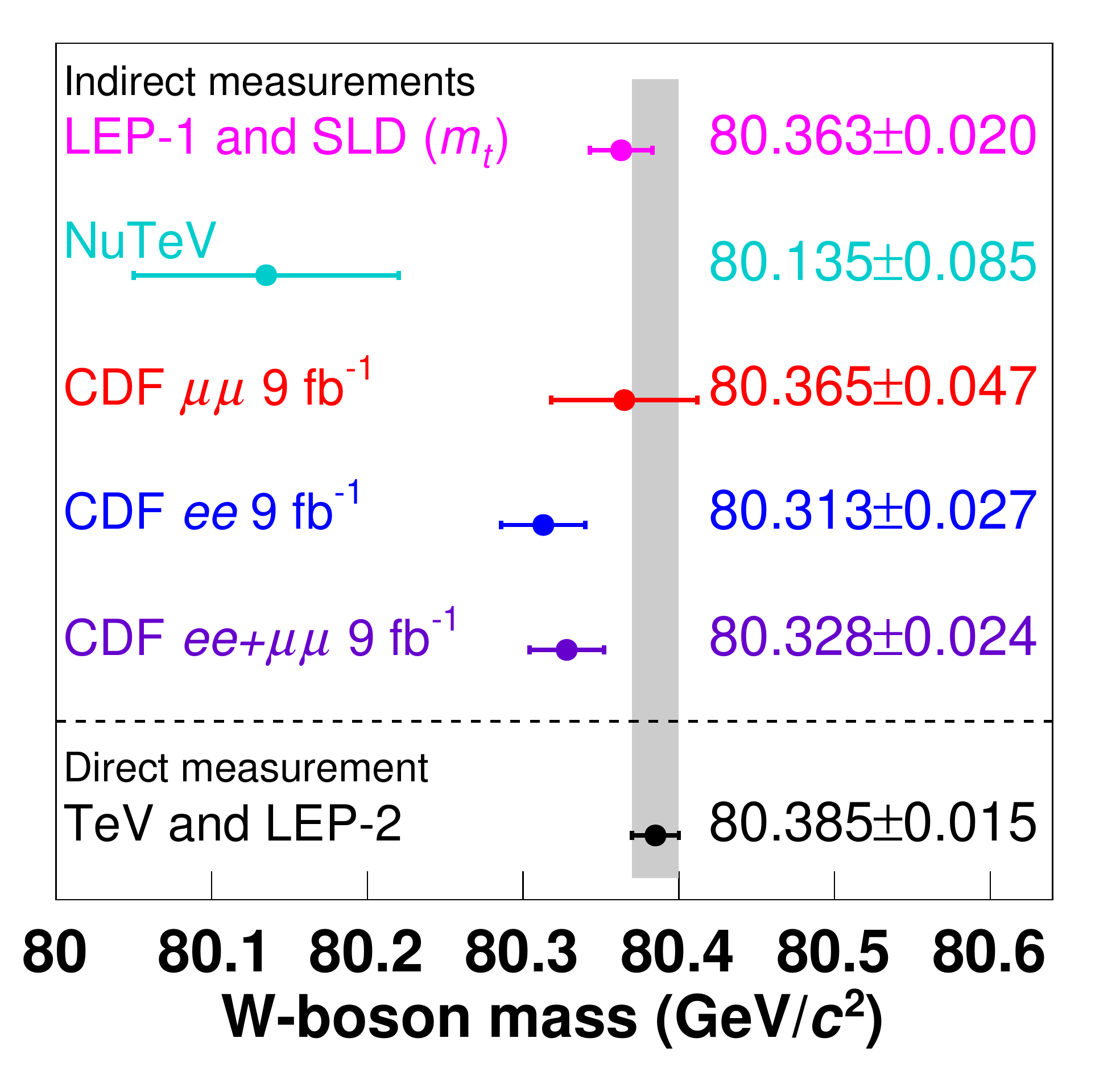}
\caption{\label{fig_compareMW}
Comparison of experimental determinations of the $W$-boson mass.
The horizontal bars represent total uncertainties. The CDF
$\mu\mu$-channel, $ee$-channel, and combination results are
denoted as
CDF $\mu\mu$ 9 fb$^{-1}$ \cite{cdfAfb9mmprd},
CDF $ee$ 9 fb$^{-1}$, and CDF $ee+\mu\mu$ 9 fb$^{-1}$, respectively.
The other indirect measurements are from
LEP-1 and SLD \cite{LEPfinalZ,LEPfinalZ2}, which include the
Tevatron top-quark mass measurement \cite{topMassCDFD0}, and 
NuTeV \cite{NuTev1,*NuTev2}. The direct measurement is from
the Tevatron and LEP-2 \cite{tevWmassCDFD0}.
}
\end{figure}

The $W$-boson mass inference is compared in
Fig.~\ref{fig_compareMW} with previous direct and indirect
measurements from the Tevatron, NuTeV, LEP-1, SLD, and LEP-2.
The direct measurement is from the Tevatron and
LEP-2 \cite{tevWmassCDFD0}. 
The previous indirect measurement from the Tevatron is derived
from the CDF measurement of $A_{\rm fb}$ with muon pairs and it
uses the same EBA-based method of inference.
The indirect measurement of $\sin^2\theta_W$ from LEP-1 and SLD,
$0.22332 \pm 0.00039$, is from the standard-model fit to all
$Z$-pole measurements \cite{LEPfinalZ,LEPfinalZ2} described in
Appendix~F of Ref.~\cite{LEPfinalZ2}. The following input
parameters to \textsc{zfitter}, the Higgs-boson mass $m_H$, the
$Z$-boson mass $M_Z$, the QCD coupling at the $Z$ pole $\alpha_s(M_Z^2)$,
and the QED correction $\Delta \alpha_{\rm em}^{(5)}(M_Z^2)$, are varied
simultaneously within the constraints of the LEP-1 and SLD data,
while the
top-quark mass $m_t$ is constrained to the directly measured value
from the Tevatron, $173.2 \pm 0.9$ GeV/$c^2$ \cite{topMassCDFD0}.
The NuTeV value is an inference based on the on-shell
$\sin^2\theta_W$ parameter extracted from the measurement of the
ratios of the neutral-to-charged current $\nu$ and
$\bar{\nu}$ cross sections at Fermilab \cite{NuTev1,*NuTev2}.

\section{\label{theEndSummary}
Summary}

The angular distribution of Drell-Yan lepton pairs provides
information on the electroweak-mixing parameter $\sin^2\theta_W$.
The electron forward-backward asymmetry in the polar-angle
distribution $\cos\vartheta$ is governed by the
$A_4\cos\vartheta$ term, whose $A_4$ coefficient is directly
related to the $\sin^2 \theta^{\rm lept}_{\rm eff}$ mixing
parameter at the lepton vertex, and indirectly to $\sin^2\theta_W$.
The effective-leptonic parameter $\sin^2 \theta^{\rm lept}_{\rm eff}$
is derived from the measurement of the forward-backward
asymmetry $A_{\rm fb}(M)$ based on the entire CDF Run~II
sample of electron pairs, reconstructed in 9.4~fb$^{-1}$ of
integrated luminosity from $p\bar{p}$ collisions at a
center-of-momentum energy of 1.96~TeV.
Calculations of $A_{\rm fb}(M)$ with different values of
the electroweak-mixing parameter are compared with the
measurement to determine the value of the parameter that
best describes the data. The calculations include QCD
radiative corrections and virtual electroweak radiative
corrections.
\par
For the $ee$-channel measurement of $A_{\rm fb}$ presented
in this paper, the best-fit values from the comparisons
are
\begin{eqnarray*}
  \sin^2 \theta^{\rm lept}_{\rm eff} & = & 0.23248 \pm 0.00053, \\
  \sin^2 \theta_W & = & 0.22428 \pm 0.00051, \; {\rm and} \\   
  M_W ({\rm indirect}) & = & 80.313 \pm 0.027 \;{\rm GeV}/c^2 \, .
\end{eqnarray*}
Each uncertainty includes statistical and systematic
contributions. The inferred value of $\sin^2\theta_W$ ($M_W$) is
based on the standard-model calculations specified in the
appendix. When this measurement of $A_{\rm fb}$ and the previous
CDF measurement based on muon pairs \cite{cdfAfb9mmprd} are used
jointly in fits, the corresponding best-fit values are
\begin{eqnarray*}
  \sin^2 \theta^{\rm lept}_{\rm eff} & = & 0.23221 \pm 0.00046, \\
  \sin^2 \theta_W & = & 0.22400 \pm 0.00045, \; {\rm and} \\
  M_W ({\rm indirect}) & = & 80.328 \pm 0.024 \;{\rm GeV}/c^2 \, .
\end{eqnarray*}
Both results are consistent with LEP-1 and SLD measurements at the
$Z$-boson pole. The value of $\sin^2 \theta^{\rm lept}_{\rm eff}$
is also consistent with the previous results from the
Tevatron~\cite{D0sw2e,cdfAfb9mmprd}.

\begin{acknowledgments}
\input{cdf_ack_050516.itex}
\end{acknowledgments}

\appendix*
\section{\label{appendixZFITTER}
ZFITTER}

\par
The input parameters to the \textsc{zfitter} radiative-correction
calculation are particle masses, the electromagnetic fine-structure
constant $\alpha_{\rm em}$, the Fermi constant $G_F$, the
strong-interaction coupling
at the $Z$ mass $\alpha_s(M_Z^2)$, and the contribution of the
light quarks to the ``running''  $\alpha_{\rm em}$ at the $Z$ mass
$\Delta \alpha_{\rm em}^{(5)}(M_Z^2)$.
The scale-dependent couplings are
$\alpha_s(M_Z^2)=0.118 \pm 0.001$~\cite{alphaS09}
and $\Delta \alpha_{\rm em}^{(5)}(M_Z^2)=0.0275 \pm 0.0001$ \cite{alpemh5}.
The mass parameters are
$M_Z = 91.1875 \pm 0.0021$ GeV/$c^2$~\cite{LEPfinalZ,LEPfinalZ2},
$m_t = 173.2 \pm 0.9$ GeV/$c^2$ (top quark) \cite{topMassCDFD0}, and
$m_H = 125$ GeV/$c^2$ (Higgs boson).
Form factors and the $Z$-boson total decay-width $\Gamma_Z$ are
calculated.
The central values of the parameters provide the context of the
\textsc{zfitter} standard-model calculations.
\par
\textsc{zfitter} uses the on-shell renormalization scheme
scheme~\cite{OnShellScheme}, where particle masses are on-shell and
\begin{equation}
  \sin^2 \theta_W = 1 - M_W^2/M_Z^2
\label{eqnOnShellScheme}
\end{equation}
holds to all orders of perturbation theory by definition.
If both $G_F$ and $m_H$ are specified, $\sin\theta_W$ is not
independent, and related to $G_F$ and $m_H$ by
standard-model constraints
from radiative corrections. To vary the $\sin\theta_W$ ($M_W$)
parameter, the value of $G_F$ is not constrained. The value of
the $M_W$ is varied over 80.0--80.5 GeV/$c^2$, and
for each value, \textsc{zfitter} calculates $G_F$ and the
form factors. Each set of calculations corresponds to a family
of physics models with standard-model like couplings where
$\sin^2\theta_W$ and the $G_F$ coupling are defined by the
$M_W$ parameter.
The Higgs-boson mass constraint $m_H=125$~GeV/$c^2$ keeps the
form factors within the vicinity of standard-model fit
values from LEP-1 and SLD~\cite{LEPfinalZ,LEPfinalZ2}.
\par
The primary purpose of \textsc{zfitter} is to provide tables of
form factors for each model. As the form factors are calculated
in the massless-fermion approximation, they only depend on the fermion
weak isospin and charge, and are distinguished via three indices:
$e$ (electron type), $u$ (up-quark type), and $d$ (down-quark type).
\par
For the $ee \rightarrow Z \rightarrow q\bar{q}$ process,
the \textsc{zfitter} scattering-amplitude ansatz is
\begin{eqnarray*}
A_q &=&  \frac{i}{4} \:
         \frac{\sqrt{2} G_F M_Z^2}
              {\hat{s} - (M_Z^2 - i\,\hat{s} \Gamma_Z/M_Z)} \:
         4T_3^e T_3^q \: \rho_{eq}                   \nonumber \\
    & & [
        \langle \bar{e}| \gamma^\mu (1+\gamma_5) |e \rangle
        \langle \bar{q}| \gamma_\mu (1+\gamma_5) |q \rangle  + \nonumber \\
    & & -4|Q_e| \kappa_e\sin^2 \theta_W \:
         \langle \bar{e}| \gamma^\mu |e \rangle
         \langle \bar{q}| \gamma_\mu (1+\gamma_5) |q \rangle + \nonumber \\
    & & -4|Q_q| \kappa_q\sin^2 \theta_W \:
         \langle \bar{e}| \gamma^\mu (1+\gamma_5) |e \rangle
         \langle \bar{q}| \gamma_\mu |q \rangle  +             \nonumber \\
    & &  16|Q_e Q_q| \kappa_{eq}\sin^4 \theta_W
         \langle \bar{e}| \gamma^\mu |e \rangle
         \langle \bar{q}| \gamma_\mu |q \rangle ] \:,
\end{eqnarray*}
where $q = u$ or $d$, the $\rho_{eq}$, $\kappa_e$, $\kappa_q$,
and $\kappa_{eq}$
are complex-valued form factors, the bilinear $\gamma$ matrix
terms are covariantly contracted, and
$\frac{1}{2}(1+\gamma_5)$ is the left-handed helicity projector in
the \textsc{zfitter} convention. 
The $\kappa_e$ form factors of the $A_u$ and $A_d$ amplitudes are
not equivalent; however, at $\hat{s} = M_Z^2$, they are numerically
equal.
\par
The $\rho_{eq}$, $\kappa_e$, and $\kappa_q$ form factors are incorporated
into QCD calculations as corrections to the Born-level $g_A^f$ and $g_V^f$
couplings,
\begin{eqnarray*}
  g_V^f & \rightarrow & \sqrt{\rho_{eq}}\,
                       ( T_3^f - 2Q_f \kappa_f \: \sin^2\theta_W )
                       \: {\rm and} 
                       \nonumber \\
  g_A^f & \rightarrow & \sqrt{\rho_{eq}} \, T_3^f ,
\end{eqnarray*}
where $f = e$ or $q$. The resulting current-current amplitude is
similar to $A_q$, but the $\sin^4 \theta_W$ term contains
$\kappa_e \kappa_q$. This difference is eliminated by adding
the $\sin^4 \theta_W$ term of $A_q$ with the replacement of
$\kappa_{eq}$ with $\kappa_{eq} - \kappa_e \kappa_q$
to the current-current amplitude.
Implementation details are provided in Ref. \cite{zA4ee21prd,*zA4ee21prdE}.

\bibliography{cdfAfb9ee}

\end{document}